\documentclass[compact,diff]{jkpaper}
\usepackage{tikz}
\usetikzlibrary{arrows.meta}
\usetikzlibrary{decorations.pathreplacing}
\usepackage{tikz-cd}
\usepackage{bigints}
\usepackage{aas_macros}
\usepackage{soul}
\usepackage{subcaption}
\usepackage{hhline}
\usepackage{tabularx}

\makeatletter
\newcommand\footnoteref[1]{\protected@xdef\@thefnmark{\ref{#1}}\@footnotemark}
\makeatother

\def\cA{{\mathcal{A}}}
\def\cG{{\mathcal{G}}}
\def\cE{{\mathcal{E}}}

\def\cN{{\mathcal{N}}}
\def\cM{{\mathcal{M}}}

\def\tA{\Tilde{A}}
\def\dt{\text{d}}
\def\B{\mathcal{B}}

\def\volC{ \epsilon _{\p \Sigma } }

\renewcommand{\Tilde}{\widetilde}

\def\l{\left(}
\def\r{\right)}
\def\p{\partial}

\def\d{\delta}

\definecolor{MyRed}{rgb}{0.5,0,0}
\definecolor{MyBrown}{RGB}{146,81,27}

\newauthornote{\francesco}{F}{MyRed}
\newauthornote{\Bi}{Bilyana}{magenta}
\newauthornote{\goncalo}{Goncalo}{blue}
\newauthornote{\pah}{[P:]}{violet}

\def\ra{{\text{dr}}}
\def\dr{{\text{dr}}}
\def\Ag{{\mathcal{A}}}
\def\rf{\Phi}

\newcommand{\OIST}{\raisebox{-0.08em}{\includegraphics[height=0.8em]{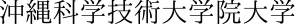}}}

\title{Soft edges: the many links between soft and edge modes }
\author{Gonçalo Araujo-Regado\texorpdfstring{\textsuperscript{a}}{}, Philipp A.\ H\"ohn\texorpdfstring{\textsuperscript{b}}{}, Francesco Sartini,\texorpdfstring{\textsuperscript{c}}{} and Bilyana Tomova\texorpdfstring{\textsuperscript{d}}{}}
\institution{Qubits and Spacetime Unit,\texorpdfstring{\\}{ }Okinawa Institute of Science and Technology \emph{(}\OIST\emph{)},\texorpdfstring{\\}{ }1919-1 Tancha, Onna-son, Kunigami-gun, Okinawa, Japan 904-0495
}

 \email{
\begin{tabular}{rlrl}
     \textsuperscript{a}\hspace*{-1em} &\emaillink{goncalo.araujo@oist.jp} 
     &\textsuperscript{b}\hspace*{-1em}   &  \emaillink{philipp.hoehn@oist.jp} \\
     \textsuperscript{c}\hspace*{-1em} & \emaillink{francesco.sartini@oist.jp}
     &\textsuperscript{d}\hspace*{-1em}   & \emaillink{b.tomova@oist.jp}  \\ 
\end{tabular}
 }

\abstr{
Boundaries in gauge theory and gravity give rise to symmetries and charges at both finite and asymptotic distance.
Due to their structural similarities, it is often held that soft modes are some kind of asymptotic limit of edge modes. Here, we show in Maxwell theory that there is an arguably more interesting relationship between the \emph{asymptotic} symmetries and their charges, on one hand, and their \emph{finite-distance} counterparts, on the other, without the need of a limit. Key to this observation is to embed the finite region in the global spacetime and identify edge modes as  dynamical $\rm{U}(1)$-reference frames for dressing subregion variables. 
Distinguishing \emph{intrinsic} and \emph{extrinsic} frames, according to whether they are built from field content in- or outside the region, we find that non-trivial corner symmetries arise only for extrinsic frames. Further, the asymptotic-to-finite relation requires asymptotically charged ones (like Wilson lines).
Such frames, called \emph{soft edges}, extend to asymptotia and, in fact, realize the corner charge algebra in multiple ways, for example, by ``pulling in'' the asymptotic one from infinity, or physically through the addition of asymptotic \emph{soft} and \emph{hard} radiation.  Realizing an infinite-dimensional  algebra 
requires 
a new set of \emph{soft boundary conditions}, relying on the distinction between extrinsic and intrinsic data. We identify the subregion Goldstone mode as the relational observable between extrinsic and intrinsic frames and clarify the meaning of vacuum degeneracy. We also connect the asymptotic memory effect with a more operational \emph{quasi-local} one.  A main conclusion is that the relationship between asymptotia and finite distance is \emph{frame-dependent}; each choice of soft edge mode probes distinct cross-boundary data of the global theory. 
}

\begin{document}
\maketitleandtoc

\section{Introduction}
\label{sec:intro}

\newcolumntype{M}[1]{>{\centering\arraybackslash}m{#1}}

Gauge theories play a fundamental role in modern theoretical physics, providing the frameworks for describing interactions between elementary particles and for gravitational physics. These theories, which are grounded in the principles of symmetry and invariance, exhibit fascinating properties when boundaries, both finite and asymptotic, are introduced. In such cases, a set of physical symmetries, resembling boundary-supported gauge transformations, and charges generating them can emerge. This has triggered significant efforts to understand the physical meaning and implications of these boundary symmetries. 

In the context of asymptotic boundaries, this has led to the extensively studied infrared (IR) triangle in gauge theory and gravity \cite{PhysRev.140.B516, Strominger:2013lka, Kapec:2014zla, Strominger:2014pwa, He:2014cra, Campiglia:2015qka, Pasterski:2015zua,Pasterski:2015tva,Pasterski:2016qvg,Strominger:2017zoo,Laddha:2017vfh, Pasterski:2017kqt,Pate:2017vwa, Kapec:2017tkm, He:2019jjk, Raclariu:2021zjz,Pasterski:2021rjz}. The IR triangle encompasses the interplay between so-called memory effects, soft theorems, and asymptotic symmetries. These elements are not merely theoretical curiosities, they hold significant physical implications. The memory effect captures the permanent imprint left on a system by passing waves, while soft theorems dictate the behaviour of low-energy quanta in scattering experiments. Together, they highlight the physical reality of the third vertex, containing asymptotic symmetries, which are often also referred to as large gauge transformations, and their associated charges. The low-energy quanta tying the IR triangle together are called \emph{soft modes} and are the low to zero frequency modes of the respective theory at (or near) the asymptotic boundary; for example, they can be soft photons or gravitons. Originally introduced in the context of asymptotically flat spacetimes, the concept of IR triangles now echoes across various areas of physics, conjecturally appearing whenever a massless particle is present \cite{Strominger:2017zoo}.

Boundaries at a finite distance, on the other hand, are considered when partitioning a global theory into subregions with which one would like to associate subsystems. This is a prerequisite for meaningfully defining entropies and other quantum information-theoretic or thermodynamic quantities in gauge theory or gravity. Owing to the gauge constraints of the theory, one finds that a gauge-invariant description of the bounded region necessitates the appearance of boundary degrees of freedom, called \emph{edge modes} \cite{Donnelly:2014fua,Donnelly:2015hxa,Donnelly:2016auv,Geiller:2017xad,Blommaert:2018oue,Geiller:2019bti,Riello:2021lfl,Gomes:2018dxs,Freidel:2020xyx,Freidel:2020svx,Freidel:2020ayo,Carrozza:2021gju,Carrozza:2022xut,Francois:2021jrk, Gomes:2024coh}, which also contribute to entanglement entropies \cite{Donnelly:2011hn,Donnelly:2014gva,Donnelly:2014fua,Donnelly:2015hxa,Wong:2017pdm,Ball:2024hqe,Frenkel:2023yuw,Fliss:2024don}. While the interpretation of the term edge mode varies across the literature (we will come back to this), they are commonly associated with the appearance of physical charges and symmetries on the boundary. 
The identification of such edge degrees of freedom has catalyzed the development of the \textit{corner program} \cite{Freidel:2015gpa,Wieland:2017cmf,Freidel:2019ees,Freidel:2019ofr,Freidel:2020xyx,Freidel:2020svx,Freidel:2020ayo,Donnelly:2022kfs,Donnelly:2020xgu,Ciambelli:2021nmv,Freidel:2023bnj,Ciambelli:2022vot,Ciambelli:2023mir,Klinger:2023tgi,AliAhmad:2024vdw, Langenscheidt:2024nyw}, which aims to quantize gravity quasi-locally by understanding the symmetry groups localized at the entangling surfaces (corners) of spacetime regions and finding their appropriate unitary representations.

The appearance of physical boundary symmetries and charges at both finite and asymptotic distance naturally leads to the question of whether there is a relation between these symmetries, as well as between soft and edge modes. Previous works bringing these concepts in relation suggest in various guises that one can view soft modes and asymptotic symmetries as some kind of infinite-distance limit of edge modes and corner symmetries, respectively \cite{Wieland:2020gno,Chen:2023tvj,Chen:2024kuq,He:2024ddb,He:2024skc,Ball:2024hqe}. While this is part of the story, we will demonstrate in this work that there is an arguably more interesting relationship between \emph{asymptotic} soft modes and symmetries, on the one hand, and \emph{finite-distance} edge modes and corner symmetries, on the other. We will see that the finite corner symmetries and charges can be viewed as finite-distance incarnations of their asymptotic counterparts without taking an infinite-volume limit. One can, of course, still take such a limit in line with earlier works to directly map the finite into the asymptotic symmetry structures, but the crux is that this is not needed in order to establish a relationship. In fact, we will argue that the corner symmetry structure enjoys an enhancement at finite distance, not seen in its asymptotic limit. By this we mean that there are many more ways to generate physical corner symmetries from global operators, other than the asymptotic charges. Of particular physical significance are the addition of asymptotic soft and hard radiation modes. This enlarged set of generators is a direct consequence of subregion/complement relationality.

The first key to the observation that there are links between the asymptotic and finite-distance symmetry structures is the identification of edge modes as dynamical reference frames \cite{Carrozza:2021gju,Carrozza:2022xut,Kabel:2023jve,Giesel:2024xtb,Gomes:2024coh} (see also \cite{Gomes:2016mwl,Gomes:2018dxs,Gomes:2021nwt,Riello:2020zbk,Riello:2021lfl}). These are frames transforming under the gauge group of the theory and are thus, in non-gravitational gauge theories, not associated with spatiotemporal symmetries; dynamical refers here to the frames being field-dependent and hence subject to the equations of motion.\footnote{In the quantum theory they become what are known as internal quantum reference frames \cite{Giacomini:2017zju,Hoehn:2019fsy,Hoehn:2023ehz,delaHamette:2021oex,AliAhmad:2021adn,Castro-Ruiz:2021vnq,DeVuyst:2024pop,Fewster:2024pur,Giacomini:2021gei,Vanrietvelde:2018pgb,Kabel:2024lzr}.} Roughly speaking, being non-invariant, they constitute the degrees of freedom one fixes when fixing a gauge, or that can be used to dress other non-invariant degrees of freedom to form composite gauge-invariant observables, often called dressed or relational observables.\footnote{The terms dressed and relational observables are used in different parts of the literature, but they are the same type of observables \cite{Carrozza:2021gju,Goeller:2022rsx}.}
We mentioned before that the interpretation of the term ``edge mode'' varies across the literature, in some cases referring to invariant, in others to non-invariant boundary degrees of freedom. The identification as dynamical reference frames refers to the latter. However, there are relations between the different interpretations and we shall later clarify what they are, proposing a more encompassing use of the term.

The second key to this observations is our distinction of two types of dynamical reference frames on a subregion boundary, namely \emph{intrinsic} and \emph{extrinsic frames}, which also entails a distinction of intrinsic and extrinsic edge modes. Intrinsic frames are built out of the field content within the subregion of interest.
Extrinsic frames, by contrast, are (typically non-locally) built out of the field content in the \emph{complement} of the subregion of interest, and it has been shown that the corner symmetries and charges correspond to \emph{reorientation symmetries} of such  frames \cite{Carrozza:2021gju,Carrozza:2022xut}. The inclusion of extrinsic frames thus requires one to depart from setups considering the finite region in isolation and to explore instead how its embedding into a global spacetime affects its symmetry structures. This is precisely what we will exploit  in order to study the interplay of the symmetries of the subregion with those of its complement.

While extrinsic frames therefore have to be added by hand to the subregional phase space (extended phase space), they are intrinsic to the global theory. For example, in gauge theories, extrinsic edge mode frames can be realized via Wilson lines connecting the asymptotic with the finite  boundary \cite{Carrozza:2021gju}, whereas in gravity, a similar construction with  geodesics yields an example realization \cite{Carrozza:2022xut}. Crucially, the dressed observables built with extrinsic edge mode frames thus encode gauge-invariant cross-boundary information, and this extends all the way to the asymptotics when the frame is charged under large gauge transformations as in the case of boundary-anchored Wilson lines or geodesics. 

As we will see, it is these large-gauge charged extrinsic frames that constitute what we call \emph{soft edges}, linking the asymptotic soft mode   with the finite-distance edge mode physics. These asymptotically extended dynamical frames ``pull in'' the asymptotic symmetries and their charges to finite distance and incarnate them as the corner symmetries and charges. Given that the latter symmetries amount to reorientations of the edge frame \cite{Carrozza:2021gju,Carrozza:2022xut}, this means that the asymptotic symmetries induce such reorientations (though we will see that they can also be realized in other ways).

One might wonder whether extrinsic frames are \emph{necessary} to understand the meaning of finite-distance corner symmetries and charges, or whether one could also realize them with intrinsic frames, thereby justifying to consider the finite region in isolation. It will turn out that extrinsic frames indeed are necessary. While intrinsic and extrinsic frames share essentially all properties and the corner terms of the symplectic structure can be equally formulated with either, they differ in one crucial point: intrinsic frames admit no reorientation symmetries and so cannot give rise to boundary charges. This will clarify that the edge modes and symmetries introduced by Donnelly and Freidel in their seminal work on finite regions \cite{Donnelly:2016auv} can necessarily only be realized via extrinsic frames in the global theory. Non-trivial corner symmetries thus always correspond to changes in the relation between the finite region and its complement, refining the discussion in \cite{Carrozza:2021gju,Carrozza:2022xut}. Furthermore, challenges to the presence of corner symmetries, such as those raised in \cite{Gomes:2018dxs,Riello:2021lfl,Riello:2020zbk}, can be traced to the implicit choice of intrinsic frames in these works. Our work can  be seen as reconciling these diverging views on edge modes and corner symmetries by placing them consistently in one picture.

These intrinsic edge mode frames, also become essential when departing from the \emph{kinematical} setup of a spatial subregion and its causal diamond and moving to the \emph{dynamical} setting of the subregion tracing out a timelike tube in spacetime, as earlier done in \cite{Geiller:2019bti,Carrozza:2021gju,Carrozza:2022xut,Ball:2024hqe,Canfora:2024awy,Maleknejad:2023nyh}. In the kinematical extended setting, it is sometimes argued, e.g.\ in \cite{Donnelly:2016auv}, that one obtains an infinite-dimensional corner symmetry group independently of boundary conditions.
By contrast, in the dynamical setting, it was pointed out \cite{Carrozza:2021gju,Carrozza:2022xut} that the corner symmetry group depends on the class of boundary conditions one must impose in order to obtain a well-defined subregional phase space, as the symmetries must preserve those boundary conditions, even within the extended phase space. For example, in Maxwell theory, which we will focus on in the main body, none of the standard Dirichlet, Neumann or Robin boundary conditions admit any non-trivial corner symmetries \cite{Carrozza:2021gju}.\footnote{This is different in Yang-Mills theory \cite[Sec.~7.4]{Carrozza:2021gju} and general relativity \cite[Sec.~6.3]{Carrozza:2022xut}, where one obtains a finite-dimensional symmetry consistent with standard boundary conditions, and in Chern-Simons theory \cite[Sec.~7.3]{Carrozza:2021gju}, where one does obtain infinite-dimensional symmetries.} Importantly, these boundary conditions were formulated in terms of the \emph{extrinsic} frame dressed boundary observables. 

Are there any boundary conditions that \emph{do} lead to a full infinite-dimensional corner symmetry algebra as the finite-distance imprint of the asymptotic symmetries? One of the core results of the present work is to identify a new class of boundary conditions, which we shall call \emph{soft boundary conditions}, that achieves this. This class is characterized by leaving the corner symplectic pair free, which instead forces one to fix the time derivative of the normal electric field (and thus the charges), and further splits into Dirichlet, Neumann and Robin boundary conditions for the remaining radiative data, which belong to the \emph{intrinsic} frame-dressed observables.
They are a generalization of the recently introduced ``dynamical edge mode conditions'' in \cite{Ball:2024hqe}, which correspond to our soft Neumann case.
Hence, while the symmetries depend on boundary conditions, there exist ones that give rise to infinite-dimensional charge algebras also at finite distance.

This construction further permits us to propose an arguably operationally more interesting \emph{quasi-local} (i.e.\ finite-distance) version of the \emph{memory effect} in Maxwell theory.  In contrast to the asymptotic version, which is usually formulated at null infinity, it turns out to be more appropriate at finite distance to formulate it in terms of the timelike tube traced out by the evolution of a spatial subregion along timelike geodesics, as opposed to the null boundary of its causal diamond. 
We will show that, using our soft edge frames, one can relate the quasi-local and asymptotic memory effects, however, in ways that will not permit observers to infer any information about the asymptotic one with quasi-local measurements only.

In this construction, we identify the \emph{Goldstone mode} at finite distance as the \emph{difference} between the extrinsic and intrinsic frames, i.e.\ as the relational observable describing the intrinsic and extrinsic frames relative to one another. The split between purely radiative and non-radiative (with respect to the subregion) gauge-invariant observables thereby takes the form of a change of frame transformation between the extrinsic and intrinsic frames.
The quasi-local vacuum transitions, on the other hand, are generated by the corner symmetries, i.e.\ reorientations of the extrinsic frame. The interpretation of this is that the many quasi-local vacua correspond to the multitude of possible relations between the subregion and the extrinsic frame and thus more generally between the subregion and its complement. The many quasi-local vacua correspond to global states of the theory that all share the same radiative data within the subregion, but differ outside.

A main result of our work is that the relation between the asymptotic and finite-distance physics is \emph{frame-dependent}. This pertains in particular to symmetries, and in the non-abelian case to charges and memory effect as well. 
For extrinsic frames that are not charged under large gauge transformations (such as Wilson lines anchored to charged matter in the complement), there is no relation between the finite corner and the asymptotic symmetries. For the large-gauge charged soft edge frames, on the other hand, this relation depends upon the choice of such frame. For example, a given system of Wilson lines between asymptotic and finite boundary constitutes one choice of soft edge frame field and changing the system of Wilson lines means changing the frame. With each such frame choice, one has access to a particular subset of gauge-invariant cross-boundary observables, but never to all of them. This means that with distinct extrinsic frames, we are probing the finite subregion with distinct gauge-invariant degrees of freedom. This is a gauge theory manifestation of \emph{subsystem relativity} \cite{AliAhmad:2021adn,delaHamette:2021oex,Hoehn:2023ehz,DeVuyst:2024pop,DeVuyst:2024uvd,Castro-Ruiz:2021vnq}. In particular, since the subregion theories are defined in terms of boundary conditions involving a choice of extrinsic frame, we have that the regional theories are edge-frame-dependent. There are therefore many relations between asymptotic and finite-distance symmetries and charges and each is physically meaningful.

Along the way of our discussion, we attempt to bring clarity into the literature by carefully distinguishing different concepts and streamlining terminology. In different parts of the literature, different setups are invoked to address boundaries and physical symmetries. Broadly, they differ by two main criteria. The first, and more obvious one is the distinction between:
\begin{enumerate}
    \item \emph{Gauge-fixed} description and definition of the theory in terms of \emph{gauge-fixed} observables. See, for instance, \cite{Harlow:2019yfa,  Ball:2024hqe,Blommaert:2018oue,Canfora:2024awy,Bub:2024nan}.  
    \item Manifestly \emph{gauge-invariant} description and definition of the theory in terms of \emph{relational/dressed} observables ($A^\dr$). In some cases, including this manuscript, these observables are built out of the \emph{bare} fields ($A$) and the \emph{edge mode} fields living on the boundary, see, for instance, \cite{Donnelly:2016auv,Geiller:2019bti,Carrozza:2021gju}. In another approach, these originate in an intrinsic regional split between gauge-invariant and -variant data \cite{Riello:2021lfl,Gomes:2018dxs,Riello:2020zbk,Gomes:2016mwl}. 
\end{enumerate}
As we will see, the former is just a particular description of the latter, corresponding to a choice of gauge-fixing condition on the manifestly gauge-invariant phase space. However, there is a lot of physical insight and clarity to be gained by working within the second framework, which is one of the main messages of this work.

While both approaches can successfully treat gauge transformations with support on the subregion boundary (but without support on the asymptotic boundary) as unphysical, confusion can easily arise in the gauge-fixed approach. This often leads to misleading statements such as ``boundaries break gauge invariance'' or ``turn gauge into physical degrees of freedom''. As we will see, within this approach, gauge transformations on the finite boundary must be considered as gauge-fixed. On the other hand, the gauge-invariant formulation naturally gives the correct boundary supported physical symmetries of a subregion, without any gauge ambiguity. Moreover, it allows us to develop a structured understanding of how to deal with a crucial feature of gauge theories, namely the existence of shared observables between a subregion and its complement. When defining the subregion theory, we must specify which of these observables we choose to include in our description (if any). This leads to the second important distinction between descriptions of a subregion in gauge theories, that we present and highlight in our work:
\begin{enumerate}
    \item \emph{Unextended} phase space, which only includes observables built out of local operators in the subregion. Gauge-invariant observables are built with an \emph{intrinsic} reference frame.
    \item \emph{Extended} phase space, which, in addition, includes \emph{some} cross-boundary observables. Gauge-invariant observables are built with an \emph{extrinsic} reference frame. There exist many inequivalent extended phase spaces, depending on which shared extrinsic frame we include. 
\end{enumerate} 
In this work, we will explicitly establish the relation between the different approaches and unify them in one picture. This will include establishing the equivalence between the gauge-fixed and gauge-invariant formulations of both the extended and unextended phase spaces by explicitly writing down the maps between them in section \ref{subsec:equivalence}. (For the unextended phase space, such a map is also discussed in \cite{Gomes:2018dxs}.) This will also highlight that the variables in apparently same-looking symplectic structures of different approaches actually have a distinct physical meaning, overcoming previously diverging conclusions in the literature.

The distinction between extended and unextended phase spaces is important, because of the crucial differences summarized in table \ref{table: summary} below.

\begingroup
\renewcommand{\arraystretch}{1.75}
    
    \begin{table}[h!]
        \centering
        \begin{tabular}{|M{55mm}||M{35mm}|M{45mm}|}
            \hline
            & \textbf{Unextended} & \textbf{Extended} \\
            \hhline{|=||=|=|}
            Type of edge mode & Intrinsic & Extrinsic \\
            \hline 
            Shared subregion-complement \newline observables & None & Some\\
            \hline
            Gauge-invariant corner factor \newline (finite Goldstone mode) & No & Yes \\
            \hline
            Corner symmetries & No & Yes (with soft boundary conditions) \\
            \hline
            Imprint of asymptotic charges & No & Yes (if soft edge)\\
            \hline
            Finite vacuum transition related \newline to asymptotic vacuum transition & No & Yes (if soft edge)\\
            \hline
            Frame dependence & No & Yes\\
            \hline
        \end{tabular}
        \caption{Summary of structures present in the unextended and extended phase spaces, respectively. All of the concepts have a manifestation in both gauge-fixed and manifestly gauge-invariant descriptions.}
        \label{table: summary}
    \end{table}

\endgroup

For simplicity and explicitness, we shall focus on classical Maxwell theory in the remainder of this article. Its gauge group consists of a suitable set of compactly supported $\rm{U}(1)$-valued spacetime functions, and its (bulk) action is defined in terms of the connection $A$ as
\begin{equation} \label{eq:Maxwell} 
S = -\frac{1}{2} \int _{\mathcal{M} } \, \star F \wedge F , \q F= \dt A\,. \end{equation} 
As our discussion will invoke a number of variable splits, this restriction to the Maxwell case will become convenient at multiple stages of the discussion. We anticipate, however, that our core insights will carry over to Yang-Mills theory and gravity. Our introductory discussion has thus been kept more general and in the main body we aim to acknowledge whenever conclusions depend on the abelian nature of the theory.

For the convenience of the reader, after summarizing our main notations, we shall provide an outline of the results of this paper in section~\ref{sec:out_res} for a quick overview. Thereafter, in section~\ref{sec:ref_frame}, we explain the concept of dynamical reference frames in Maxwell theory, and in section~\ref{sec:BC} we discuss the various forms and formulations of phase spaces, as well as their relations. In section~\ref{sec:Soft_bc}, we introduce the new soft boundary conditions and contrast them with traditional choices. In section~\ref{sec:sol_space}, we offer an explicit implementation of our construction in Minkowski space, while in section~\ref{sec:gold_soft} we discuss the quasi-local memory effect, Goldstone and soft modes, and their relation with their asymptotic counterparts. Before concluding, we then show in section~\ref{sec:charge action} that, for soft edge frames, the asymptotic charge action agrees with the finite-distance one, and we present other asymptotic operators which generate subregional corner symmetries. Several details have been moved to appendices.

\section*{Main Notations}
\addcontentsline{toc}{section}{Main Notations}%
\label{sec:notations}
In this section, we collect the notations for the relevant objects we use throughout the paper, for reader convenience.

\paragraph{Spacetime and symplectic structures}\ \\[-10pt]

\begin{tabular}{ |c|c|c| c| } 
\hline
 & Subregion theory & Global theory  & Complement region theory \\
\hline
Interior & $\mathcal{R}$ & $\mathcal{M}$ & $\bar{\mathcal{R}} $ \\
Boundary & $\Gamma $ & $\mathcal{B}$ & $\Gamma  \cup \mathcal{B} $ \\
Cauchy Slice & $\Sigma $ & $ \mathbf{\Sigma} = \Sigma \cup \bar \Sigma $ & $ \bar \Sigma $ \\
Presymplectic form & $\Omega $ & $\mathbf{\Omega }$ & $\bar{\Omega }$ \\
Presymplectic current & $\omega $ & $\mathbf{\omega }$ & $\bar{\omega }$ \\
Presymplectic potential & $\Theta $ & $\mathbf{\Theta }$ & $\bar{\Theta }$ \\
Any other quantity  & $X$ & $\mathbf{X}$ & $\bar{X}$ \\
\hline 
\end{tabular}

\paragraph{Spacetime and field space symbols}\ \\[-10pt]

\begin{tabular}{ll} 
$\dt $ & Spacetime exterior derivative\\
$\wedge$ & Spacetime wedge product\\
$|_{\Gamma}\;\; |_{\p\Sigma}$, ...& Pullbacks to (respectively) $\Gamma$, $\p \Sigma$, etc...\\
$\epsilon_{\Gamma}\;\; \epsilon_{\p\Sigma}$, ... & Volume form on (respectively)  $\Gamma$, $\p \Sigma$, etc...\\
$\d $ & (Pre)symplectic space exterior derivative\\
$\curlywedge$ & (Pre)symplectic space wedge product \\
$\cdot$ & (Pre)symplectic space interior product \\
\end{tabular}

{ \noindent When both the spacetime and field space wedge product are present, we keep the second implicit, e.g.
\[
\delta A \wedge \delta \star F = \sqrt{-g} \,\varepsilon_{\nu\rho\sigma\lambda}\, g^{\nu\sigma} g^{\rho \lambda} \l \delta A_\mu \curlywedge \delta F^{\sigma \lambda}\r \dt x^\mu \wedge \dt x^\nu \wedge \dt x^\rho\,.
\]
}

\paragraph{Equalities}\ \\[-10pt]

\begin{tabular}{ll} 
$=$ & Off-shell equality\\
$\hat =$ & Off-shell equality with boundary conditions (postselected space) \\
$\approx$ & On-shell equality (global solution space)\\
$\hat \approx$ & On-shell equality with boundary conditions (postselected space)\\
\end{tabular}

\paragraph{Dynamical variables}\ \\[-10pt]

\begin{minipage}{\textwidth}
\hspace{-20pt}
   \begin{tabular}{ll} 
$A$     &   Bare connection\\[4pt]
$\Ag$   &   gauge-fixed connection \\[4pt]
$U[A]$  &   Reference frame \\[4pt]
$\rf[A] = i \ln U [A]$&    Frame phase (edge mode) \\[4pt]
$A^{\ra} =A -\dt \rf$&     Frame-dressed connection \\[4pt]   
$\cG[A^\dr]=0$&      Dressing condition \\[4pt]
$\cG[\Ag]=0$&        gauge-fixing condition \\[4pt]
$ \phi$ & Global Goldstone mode \\[4pt]
\end{tabular}\vline
\begin{tabular}{ll} 
$ \varphi $& Subregion Goldstone mode  \\[4pt]
$ \varphi^+, \,  \varphi^- $& Values of $\varphi$ at $t^\pm $ respectively \\[4pt]
$ \phi ^+ , \, \phi ^- $& Values of $\phi$ at $ \scri ^+ _\pm  $ respectively  \\[4pt]
$\Tilde{A}^\dr= A^{\ra} - \dt \varphi$& Radiative field  in the subregion \\[4pt]
$\p _a N =\int \,  \dt  u \,F_{ua} \big| _{\scri ^+}$& Asymptotic soft mode\\[4pt]
$\cN_a =\int \,  \dt t \,F_{ta} \big| _{\Gamma }$& Finite distance memory\\[4pt]
$ \l \star F\big|_{\p \Sigma } , \,  \rf \r$& Edge pair \\[4pt]
$ \l \star F\big|_{\p \Sigma }  , \,  \varphi \r$ & Goldstone pair
\end{tabular}
\end{minipage}

\paragraph{Charges}\ \\[-10pt]

\begin{tabular}{ll} 
 $\mathbf{Q} = \lim _{r \to \infty } r^2 F^{rt} \big|_{i^+}$ & Total electric charge density at $\scri ^+ _+$ \\ [4pt] 
$\mathcal{Q} = F^{tr} \big|_{\p \Sigma} $ &  Radial electric field at the subregion boundary \\[4pt]
$ \mathcal{Q} ^\pm = \mathcal{Q} \big|_{t=t^\pm} $ & Radial electric field at initial ($t^-$) and final slice  ($t^+$) 
\\ [4pt]
$Q_\rho = \int _{\p \Sigma } \, \epsilon _{\p \Sigma }\,  \rho \, \mathcal{Q} $ & Charge generating subregion frame reorientations parametrized by $\rho $ \\ [4pt]
$\mathbf{Q} _\alpha   = \int _{\scri ^+ _-} \, \epsilon _{\scri ^+ _- } \, \alpha \l D^2 N - \mathbf{Q}  \r $ &  Charge generating global large gauge transformations parametrized by $\alpha $
\end{tabular}

\paragraph{Symmetry transformation parameters}\ \\[-10pt]

\begin{tabular}{ll} 
 $\lambda$  & Small gauge -- $\lambda |_\B = 0 $  \\[4pt]
$\alpha$ &  Large gauge -- $\alpha |_\B \neq 0$ \\[4pt]
$ \rho$ & Frame reorientations 
\end{tabular}

\paragraph{Metric and coordinate choice} \ \\[-10pt] 

\begin{tabular}{ll} 
$t$ & Timelike function, where constant $t$ slices are $\Sigma $ \\[4pt] 
$r, x^a $ & Coordinates on $\Sigma $ ($x^a$ angular coordinates on $\p\Sigma$) \\[4pt] 
$g_{tt}\big|_\Gamma=-1$, $g_{t r }\big|_\Gamma=g_{ta}\big|_\Gamma = 0 $, $g_{rr}\big|_\Gamma=1$, $g_{ra}\big|_\Gamma=0$ & Coordinate conditions on $\Gamma$ 
\end{tabular}

\vspace{1mm}
\paragraph{Coordinates on $\Gamma\cong\p\Sigma\times I$ with $I\subset\mathbb{R}$}\ \\[-10pt] 

\begin{tabular}{ll} 
 $x^a$ , $g_{ab}$, $\Tilde{\nabla}^a $ & Coordinates,  induced metric on $\p \Sigma  $ and its covariant derivative \\[4pt] 
 $\dt t  = n_\mu$ & Future-pointing time-like vector, with which we contract the symplectic flux density \\[4pt] 
 $\dt r=m_\mu$ & Outward-pointing vector normal to $\Gamma$\\[4pt]
$\star _{\p \Sigma } $ & Hodge star on $\p \Sigma $ \\[4pt]
$\dt _{\p \Sigma }$ & Exterior derivative on $\p \Sigma $ \\[4pt]
$\Delta ^{-1} _{\p \Sigma } $ & Green's function of Laplace operator on $\p \Sigma $ 
\end{tabular}

\begin{tabular}{ll} 

$\gamma _{ab}, \, \sqrt{\gamma}$ &  Metric and determinant on unit round sphere\\[4pt]
$ D_a $ & Covariant derivative with respect to $\gamma _{ab}$   \\[4pt]
\end{tabular}

 Whenever we are in Minkowski spacetime $(r,x^a)$ are understood to be the standard spherical coordinates, and $\p _t$ is a Minkowski Killing vector field. 


\section{Outline of Results}\label{sec:out_res}

In this section, we summarise our construction and the key messages in a concise form. In the process, we highlight connections with various concepts in the literature on edge and soft modes, respectively. The goal is to bring clarity into how all these concepts, including our new ones, fit together, and especially how the asymptotic and finite-distance physics are related. We refer to the relevant sections in the bulk of the paper in which these issues are explored in detail. 

\paragraph{Basic Setup and Philosophy:} 

Our setup involves a lab in the bulk of an asymptotically flat universe. We imagine we have some detectors covering some codimension-2 surface ($\p\Sigma$) which evolve in time along some worldlines, thus tracing out a timelike tube $\Gamma$. This construction enables a dynamical setting, where we can investigate whether kinematical symmetries persist on the solution space and study how various boundary conditions influence these symmetries. This, in turn, allows us to explore vacuum transitions across two Cauchy surfaces of the subregion, clarifying their connection to corner symmetries.

\begin{figure}[h]
    \centering
    \tikzset{every picture/.style={line width=0.75pt}} 
    
    \begin{tikzpicture}[x=0.75pt,y=0.75pt,yscale=-1,xscale=1]

\draw   (330.9,40.6) -- (462.8,170.8) -- (330.9,301) -- (199,170.8) -- cycle ;
\draw    (301.06,212.96) .. controls (311.78,195.23) and (282.61,167.14) .. (296.77,151.58) .. controls (310.93,136.02) and (306.64,127.81) .. (304.92,112.24) ;
\draw [shift={(295.69,176.12)}, rotate = 70.06] [color={rgb, 255:red, 0; green, 0; blue, 0 }  ][line width=0.75]    (10.93,-3.29) .. controls (6.95,-1.4) and (3.31,-0.3) .. (0,0) .. controls (3.31,0.3) and (6.95,1.4) .. (10.93,3.29)   ;
\draw [shift={(306.82,127.24)}, rotate = 91.47] [color={rgb, 255:red, 0; green, 0; blue, 0 }  ][line width=0.75]    (10.93,-3.29) .. controls (6.95,-1.4) and (3.31,-0.3) .. (0,0) .. controls (3.31,0.3) and (6.95,1.4) .. (10.93,3.29)   ;
\draw    (357.26,212.96) .. controls (359.4,194.8) and (371.42,174.49) .. (363.69,157.63) .. controls (355.97,140.77) and (355.11,131.7) .. (364.55,112.24) ;
\draw [shift={(365.33,179.44)}, rotate = 102.31] [color={rgb, 255:red, 0; green, 0; blue, 0 }  ][line width=0.75]    (10.93,-3.29) .. controls (6.95,-1.4) and (3.31,-0.3) .. (0,0) .. controls (3.31,0.3) and (6.95,1.4) .. (10.93,3.29)   ;
\draw [shift={(358.45,128.76)}, rotate = 97.11] [color={rgb, 255:red, 0; green, 0; blue, 0 }  ][line width=0.75]    (10.93,-3.29) .. controls (6.95,-1.4) and (3.31,-0.3) .. (0,0) .. controls (3.31,0.3) and (6.95,1.4) .. (10.93,3.29)   ;
\draw   (304.92,112.24) .. controls (304.92,107.47) and (318.27,103.6) .. (334.74,103.6) .. controls (351.2,103.6) and (364.55,107.47) .. (364.55,112.24) .. controls (364.55,117.02) and (351.2,120.89) .. (334.74,120.89) .. controls (318.27,120.89) and (304.92,117.02) .. (304.92,112.24) -- cycle ;
\draw   (301.06,212.96) .. controls (301.06,208.18) and (313.64,204.31) .. (329.16,204.31) .. controls (344.68,204.31) and (357.26,208.18) .. (357.26,212.96) .. controls (357.26,217.73) and (344.68,221.6) .. (329.16,221.6) .. controls (313.64,221.6) and (301.06,217.73) .. (301.06,212.96) -- cycle ;
\draw  [color={rgb, 255:red, 155; green, 155; blue, 155 }  ,draw opacity=1 ] (301.95,191.71) .. controls (302.95,185.74) and (317.84,178.39) .. (335.22,175.28) .. controls (352.61,172.17) and (365.9,174.49) .. (364.9,180.45) .. controls (363.91,186.42) and (349.02,193.77) .. (331.63,196.88) .. controls (314.25,199.99) and (300.96,197.68) .. (301.95,191.71) -- cycle ;
\draw  [color={rgb, 255:red, 74; green, 144; blue, 226 }  ,draw opacity=1 ] (294.66,156.28) .. controls (299.64,150.24) and (318.87,146.77) .. (337.61,148.52) .. controls (356.34,150.27) and (367.48,156.58) .. (362.5,162.62) .. controls (357.51,168.66) and (338.28,172.14) .. (319.55,170.38) .. controls (300.81,168.63) and (289.67,162.32) .. (294.66,156.28) -- cycle ;
\draw  [color={rgb, 255:red, 155; green, 155; blue, 155 }  ,draw opacity=1 ] (309.28,126.35) .. controls (314.82,123.55) and (329.73,125.42) .. (342.57,130.53) .. controls (355.41,135.64) and (361.33,142.05) .. (355.79,144.85) .. controls (350.25,147.64) and (335.35,145.77) .. (322.5,140.66) .. controls (309.66,135.55) and (303.74,129.14) .. (309.28,126.35) -- cycle ;
\draw [color={rgb, 255:red, 155; green, 155; blue, 155 }  ,draw opacity=1 ]   (311.35,219.44) .. controls (315.64,200.77) and (308.78,179.15) .. (307.49,165.76) .. controls (306.21,152.36) and (313.07,136.97) .. (319.08,119.68) ;
\draw [shift={(310.87,186.43)}, rotate = 81.53] [color={rgb, 255:red, 155; green, 155; blue, 155 }  ,draw opacity=1 ][line width=0.75]    (10.93,-3.29) .. controls (6.95,-1.4) and (3.31,-0.3) .. (0,0) .. controls (3.31,0.3) and (6.95,1.4) .. (10.93,3.29)   ;
\draw [shift={(313.09,136.41)}, rotate = 108.47] [color={rgb, 255:red, 155; green, 155; blue, 155 }  ,draw opacity=1 ][line width=0.75]    (10.93,-3.29) .. controls (6.95,-1.4) and (3.31,-0.3) .. (0,0) .. controls (3.31,0.3) and (6.95,1.4) .. (10.93,3.29)   ;
\draw [color={rgb, 255:red, 155; green, 155; blue, 155 }  ,draw opacity=1 ]   (329.16,221.6) .. controls (333.45,202.93) and (337.52,179.15) .. (331.52,166.19) .. controls (325.51,153.22) and (332.8,138.52) .. (334.74,120.89) ;
\draw [shift={(334.46,187.64)}, rotate = 94.12] [color={rgb, 255:red, 155; green, 155; blue, 155 }  ,draw opacity=1 ][line width=0.75]    (10.93,-3.29) .. controls (6.95,-1.4) and (3.31,-0.3) .. (0,0) .. controls (3.31,0.3) and (6.95,1.4) .. (10.93,3.29)   ;
\draw [shift={(331.69,137.67)}, rotate = 101.75] [color={rgb, 255:red, 155; green, 155; blue, 155 }  ,draw opacity=1 ][line width=0.75]    (10.93,-3.29) .. controls (6.95,-1.4) and (3.31,-0.3) .. (0,0) .. controls (3.31,0.3) and (6.95,1.4) .. (10.93,3.29)   ;
\draw [color={rgb, 255:red, 155; green, 155; blue, 155 }  ,draw opacity=1 ]   (345.89,220.3) .. controls (350.18,201.63) and (358.55,177.86) .. (352.54,164.89) .. controls (346.53,151.92) and (349.57,137.29) .. (351.5,119.65) ;
\draw [shift={(353.74,186.64)}, rotate = 100.3] [color={rgb, 255:red, 155; green, 155; blue, 155 }  ,draw opacity=1 ][line width=0.75]    (10.93,-3.29) .. controls (6.95,-1.4) and (3.31,-0.3) .. (0,0) .. controls (3.31,0.3) and (6.95,1.4) .. (10.93,3.29)   ;
\draw [shift={(349.57,136.42)}, rotate = 94.36] [color={rgb, 255:red, 155; green, 155; blue, 155 }  ,draw opacity=1 ][line width=0.75]    (10.93,-3.29) .. controls (6.95,-1.4) and (3.31,-0.3) .. (0,0) .. controls (3.31,0.3) and (6.95,1.4) .. (10.93,3.29)   ;
\draw  [color={rgb, 255:red, 208; green, 2; blue, 27 }  ,draw opacity=1 ] (275.92,106.22) .. controls (274.58,108.72) and (273.31,111.1) .. (273.94,111.74) .. controls (274.58,112.38) and (276.96,111.1) .. (279.46,109.76) .. controls (281.96,108.42) and (284.34,107.14) .. (284.98,107.78) .. controls (285.62,108.42) and (284.34,110.8) .. (283,113.3) .. controls (281.65,115.79) and (280.38,118.17) .. (281.02,118.81) .. controls (281.66,119.45) and (284.03,118.17) .. (286.53,116.83) .. controls (289.03,115.49) and (291.41,114.21) .. (292.05,114.85) .. controls (292.69,115.49) and (291.41,117.87) .. (290.07,120.37) .. controls (288.72,122.86) and (287.45,125.24) .. (288.09,125.88) .. controls (288.73,126.52) and (291.11,125.25) .. (293.6,123.9) .. controls (295.43,122.92) and (297.2,121.97) .. (298.27,121.78) ;
\draw  [color={rgb, 255:red, 208; green, 2; blue, 27 }  ,draw opacity=1 ] (377.92,207.22) .. controls (376.58,209.72) and (375.31,212.1) .. (375.94,212.74) .. controls (376.58,213.38) and (378.96,212.1) .. (381.46,210.76) .. controls (383.96,209.42) and (386.34,208.14) .. (386.98,208.78) .. controls (387.62,209.42) and (386.34,211.8) .. (385,214.3) .. controls (383.65,216.79) and (382.38,219.17) .. (383.02,219.81) .. controls (383.66,220.45) and (386.03,219.17) .. (388.53,217.83) .. controls (391.03,216.49) and (393.41,215.21) .. (394.05,215.85) .. controls (394.69,216.49) and (393.41,218.87) .. (392.07,221.37) .. controls (390.72,223.86) and (389.45,226.24) .. (390.09,226.88) .. controls (390.73,227.52) and (393.11,226.25) .. (395.6,224.9) .. controls (397.43,223.92) and (399.2,222.97) .. (400.27,222.78) ;
\draw  [color={rgb, 255:red, 208; green, 2; blue, 27 }  ,draw opacity=1 ] (264.92,117.22) .. controls (263.58,119.72) and (262.31,122.1) .. (262.94,122.74) .. controls (263.58,123.38) and (265.96,122.1) .. (268.46,120.76) .. controls (270.96,119.42) and (273.34,118.14) .. (273.98,118.78) .. controls (274.62,119.42) and (273.34,121.8) .. (272,124.3) .. controls (270.65,126.79) and (269.38,129.17) .. (270.02,129.81) .. controls (270.66,130.45) and (273.03,129.17) .. (275.53,127.83) .. controls (278.03,126.49) and (280.41,125.21) .. (281.05,125.85) .. controls (281.69,126.49) and (280.41,128.87) .. (279.07,131.37) .. controls (277.72,133.86) and (276.45,136.24) .. (277.09,136.88) .. controls (277.73,137.52) and (280.11,136.25) .. (282.6,134.9) .. controls (284.43,133.92) and (286.2,132.97) .. (287.27,132.78) ;
\draw  [color={rgb, 255:red, 208; green, 2; blue, 27 }  ,draw opacity=1 ] (365.92,219.22) .. controls (364.58,221.72) and (363.31,224.1) .. (363.94,224.74) .. controls (364.58,225.38) and (366.96,224.1) .. (369.46,222.76) .. controls (371.96,221.42) and (374.34,220.14) .. (374.98,220.78) .. controls (375.62,221.42) and (374.34,223.8) .. (373,226.3) .. controls (371.65,228.79) and (370.38,231.17) .. (371.02,231.81) .. controls (371.66,232.45) and (374.03,231.17) .. (376.53,229.83) .. controls (379.03,228.49) and (381.41,227.21) .. (382.05,227.85) .. controls (382.69,228.49) and (381.41,230.87) .. (380.07,233.37) .. controls (378.72,235.86) and (377.45,238.24) .. (378.09,238.88) .. controls (378.73,239.52) and (381.11,238.25) .. (383.6,236.9) .. controls (385.43,235.92) and (387.2,234.97) .. (388.27,234.78) ;
\draw  [color={rgb, 255:red, 208; green, 2; blue, 27 }  ,draw opacity=1 ] (280.88,204.22) .. controls (278.38,202.88) and (276,201.61) .. (275.36,202.24) .. controls (274.72,202.88) and (276,205.26) .. (277.34,207.76) .. controls (278.68,210.26) and (279.96,212.64) .. (279.32,213.28) .. controls (278.68,213.92) and (276.3,212.64) .. (273.8,211.3) .. controls (271.31,209.95) and (268.93,208.68) .. (268.29,209.32) .. controls (267.65,209.96) and (268.93,212.33) .. (270.27,214.83) .. controls (271.61,217.33) and (272.89,219.71) .. (272.25,220.35) .. controls (271.61,220.99) and (269.23,219.71) .. (266.73,218.37) .. controls (264.24,217.02) and (261.86,215.75) .. (261.22,216.39) .. controls (260.58,217.03) and (261.85,219.41) .. (263.2,221.9) .. controls (264.18,223.73) and (265.13,225.5) .. (265.32,226.57) ;
\draw  [color={rgb, 255:red, 208; green, 2; blue, 27 }  ,draw opacity=1 ] (293.88,216.22) .. controls (291.38,214.88) and (289,213.61) .. (288.36,214.24) .. controls (287.72,214.88) and (289,217.26) .. (290.34,219.76) .. controls (291.68,222.26) and (292.96,224.64) .. (292.32,225.28) .. controls (291.68,225.92) and (289.3,224.64) .. (286.8,223.3) .. controls (284.31,221.95) and (281.93,220.68) .. (281.29,221.32) .. controls (280.65,221.96) and (281.93,224.33) .. (283.27,226.83) .. controls (284.61,229.33) and (285.89,231.71) .. (285.25,232.35) .. controls (284.61,232.99) and (282.23,231.71) .. (279.73,230.37) .. controls (277.24,229.02) and (274.86,227.75) .. (274.22,228.39) .. controls (273.58,229.03) and (274.85,231.41) .. (276.2,233.9) .. controls (277.18,235.73) and (278.13,237.5) .. (278.32,238.57) ;
\draw  [color={rgb, 255:red, 208; green, 2; blue, 27 }  ,draw opacity=1 ] (391.88,106.22) .. controls (389.38,104.88) and (387,103.61) .. (386.36,104.24) .. controls (385.72,104.88) and (387,107.26) .. (388.34,109.76) .. controls (389.68,112.26) and (390.96,114.64) .. (390.32,115.28) .. controls (389.68,115.92) and (387.3,114.64) .. (384.8,113.3) .. controls (382.31,111.95) and (379.93,110.68) .. (379.29,111.32) .. controls (378.65,111.96) and (379.93,114.33) .. (381.27,116.83) .. controls (382.61,119.33) and (383.89,121.71) .. (383.25,122.35) .. controls (382.61,122.99) and (380.23,121.71) .. (377.73,120.37) .. controls (375.24,119.02) and (372.86,117.75) .. (372.22,118.39) .. controls (371.58,119.03) and (372.85,121.41) .. (374.2,123.9) .. controls (375.18,125.73) and (376.13,127.5) .. (376.32,128.57) ;
\draw  [color={rgb, 255:red, 208; green, 2; blue, 27 }  ,draw opacity=1 ] (403.88,118.22) .. controls (401.38,116.88) and (399,115.61) .. (398.36,116.24) .. controls (397.72,116.88) and (399,119.26) .. (400.34,121.76) .. controls (401.68,124.26) and (402.96,126.64) .. (402.32,127.28) .. controls (401.68,127.92) and (399.3,126.64) .. (396.8,125.3) .. controls (394.31,123.95) and (391.93,122.68) .. (391.29,123.32) .. controls (390.65,123.96) and (391.93,126.33) .. (393.27,128.83) .. controls (394.61,131.33) and (395.89,133.71) .. (395.25,134.35) .. controls (394.61,134.99) and (392.23,133.71) .. (389.73,132.37) .. controls (387.24,131.02) and (384.86,129.75) .. (384.22,130.39) .. controls (383.58,131.03) and (384.85,133.41) .. (386.2,135.9) .. controls (387.18,137.73) and (388.13,139.5) .. (388.32,140.57) ;

\draw (397,80.4) node [anchor=north west][inner sep=0.75pt]    {$\mathscr{I}^{+}$};
\draw (397,249.4) node [anchor=north west][inner sep=0.75pt]    {$\mathscr{I}^{-}$};
\draw (470,163.4) node [anchor=north west][inner sep=0.75pt]    {$i ^{0}$};
\draw (322,17.4) node [anchor=north west][inner sep=0.75pt]    {$i ^{+}$};
\draw (323,304.4) node [anchor=north west][inner sep=0.75pt]    {$i ^{-}$};
\draw (279.38,152.42) node [anchor=north west][inner sep=0.75pt]    {$\Gamma $};
\draw (362.57,123.1) node [anchor=north west][inner sep=0.75pt]    {$t$};
\draw (369.5,149.89) node [anchor=north west][inner sep=0.75pt]    {$\textcolor[rgb]{0.29,0.56,0.89}{\partial \Sigma }$};

\end{tikzpicture}
    \caption{Finite lab in the bulk of an asymptotically flat spacetime. Radiation from asymptotic infinity passes through the walls of the lab. The worldvolume of the lab's wall ($\Gamma$) is foliated into slices ($\partial\Sigma$) of equal proper time ($t$) of the detectors (which follow some prescribed worldlines in spacetime). Later, we will also allow for matter outside of the subregion.}
    \label{lab}
\end{figure}
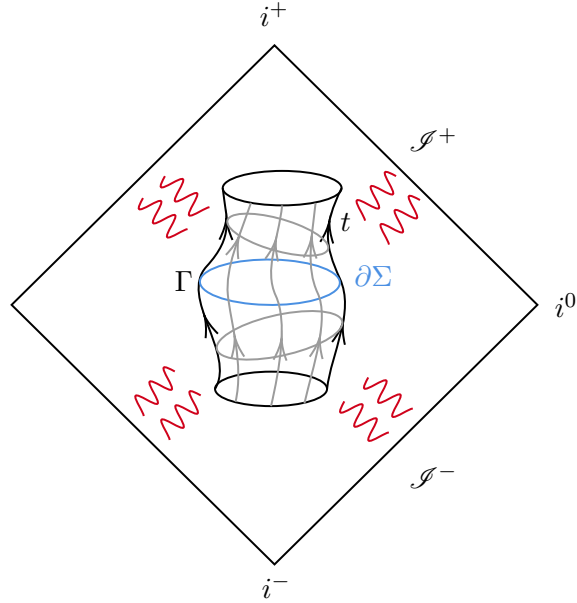

For now, we take this lab to have a finite proper lifetime, as shown in figure \ref{lab}. Later we will also discuss the limit of infinite temporal extension, to connect with the asymptotic memory effect. 

\paragraph{Edge Modes as Dynamical Reference Frames:}
In the literature, the term ``edge mode'' is used to mean different things: for example, in \cite{Donnelly:2015hxa,Donnelly:2014fua} it is the boundary value of the component of the electric field perpendicular to $\Gamma$ -- $\star F|_{\partial\Sigma}$, while in \cite{Geiller:2019bti,Carrozza:2021gju,Freidel:2023bnj,Riello:2021lfl,Gomes:2016mwl} it is its conjugate variable at the level of the \emph{pre}symplectic form\footnote{The \emph{pre} is important here. At the level of the symplectic form, after having quotiented by degenerate directions, the conjugate variable to the perpendicular electric field is a gauge-invariant observable which we call the \emph{finite Goldstone mode}. This only exists in the extended phase space, as it turns out.} -- a boundary field $\Phi$ which covariantly transforms under boundary-supported gauge transformations.  In this paper, we \emph{always} mean the latter. In \cite{Ball:2024hqe,Ball:2024xhf}, on the other hand, the term edge mode is used to refer to what, in this paper, we call the Goldstone mode, as explained shortly.

Originally, when edge modes (in the form $\Phi$) were postulated in gauge theories \cite{Donnelly:2016auv,Geiller:2017xad,Blommaert:2018oue,Geiller:2019bti,Freidel:2020xyx,Freidel:2020svx,Freidel:2020ayo}, they were defined as new degrees of freedom which one simply added to the regional theory, in order to have manifest boundary gauge invariance, with their meaning and origin otherwise left open. However, it was later understood \cite{Carrozza:2021gju,Carrozza:2022xut} that they have a transparent interpretation once we take seriously the fact that the subregion is embedded in a larger, global theory. They are, in fact, already contained in the global theory. Namely, there, it was shown that they can be thought of as non-local functionals of the connection (and possibly other dynamical fields). By contrast, the view of edge modes as additional degrees of freedom was disputed, for example, in \cite{Riello:2020zbk,Riello:2021lfl,Gomes:2018dxs}.

In order to reconcile these diverging views on edge modes in a single picture, we will distinguish between two types of edge modes: 
\begin{enumerate}
    \item \emph{Intrinsic} edge mode: built out of dynamical degrees of freedom living on $\Gamma$. There is one canonical such choice.
    \item \emph{Extrinsic} edge mode: built out of dynamical degrees of freedom living in the complement. There are many inequivalent choices.
\end{enumerate}
This distinction, tied to the distinction between \textit{unextended} and \emph{extended} phase space,
turns out to be crucial and has perhaps not been sufficiently appreciated in the literature. In most works, e.g.\ see \cite{Donnelly:2016auv,Carrozza:2021gju,Carrozza:2022xut,Geiller:2017xad,Geiller:2019bti,Freidel:2020xyx,Freidel:2023bnj,Ciambelli:2021nmv}, an edge mode of the extrinsic type is at least implicitly assumed.\footnote{In most works, these were not identified as having support in the complement of the finite region, mostly because the latter was considered in isolation. However, as edge modes in these works were added ``by hand'' to the regional phase space, they were still extrinsic to the other regional degrees of freedom. It was only pointed out in \cite{Carrozza:2021gju,Carrozza:2022xut} that these edge modes can be realized via frames with support in the complement in both gauge theory and gravity.} However, the existence of the intrinsic edge mode clarifies many aspects of the relational structure of gauge theories, as well as the absence of boundary symmetries in \cite{Riello:2020zbk,Riello:2021lfl,Gomes:2018dxs}. By distinguishing between intrinsic and extrinsic frames, our analysis provides a framework that explains why different works have arrived at seemingly contrasting conclusions regarding edge modes: they were, in fact, examining distinct types of edge modes tied to different frame choices.

A key requirement on such functionals (for both intrinsic and extrinsic types) is that they transform covariantly under the action of the gauge group:
\begin{align}
    \Phi \mapsto \Phi + \alpha|_\Gamma\,,
\end{align}
where $\alpha$ parameterizes a gauge transformation with non-trivial support on $\Gamma$. We give a proper definition in section \ref{subsec:dyn_ref_fr}. The bottom line is that exponentiated, $U=e^{i\Phi}$, the edge mode is a $\rm{U}(1)$-valued boundary function that transforms covariantly under gauge transformations, $U\mapsto g(\alpha)\,U$, where $g=e^{i\alpha}\in\rm{U}(1)$. Now reference frames often take value in the group they transform under (e.g.\ tetrads are Lorentz group-valued), and so $U$ is a realization of the more general structure of a  \emph{dynamical reference frame} \cite{Carrozza:2021gju,Carrozza:2022xut,Goeller:2022rsx,delaHamette:2021oex,Hoehn:2023ehz}: it is a complete $\rm{U}(1)$-frame at each point of $\Gamma$. $U$ is thus a dynamical frame field for the gauge group of the theory and thus here not associated with a spacetime symmetry. We call the values of $U(x)$ and $\Phi(x)$ the (local) \emph{orientation} and \emph{phase} of the reference frame at each point of $\Gamma$, respectively. This holds for both the extrinsic and intrinsic edge mode frames. We give explicit constructions of each case in section \ref{subsec:extr_intr_ref_fr}. For distinction, we shall henceforth label the intrinsic edge mode frame by $\Tilde U$ and its phase by $\Tilde\Phi$. 

The canonical intrinsic edge mode arises as the exact term in the unique Hodge decomposition on the corner:
\be
\label{eq:Hodge_intro}
A|_{\p \Sigma}= \Tilde A^\dr +\dt \Tilde \Phi\,,
\ee
where the intrinsic edge mode $\Tilde \Phi$ is thus determined up to an overall constant. The notation $\Tilde A^\dr$ for the transverse non-exact, mode will be clarified shortly.\footnote{This split is akin to the usual Helmholtz decomposition in space, even though here we are performing the decomposition on a boundary of space only, via the Hodge decomposition. Borrowing the notation, $\Tilde A^\dr$ is denoted as the radiative (or transverse) field. It is indeed a gauge-invariant observable related to the photon polarizations in the subregion. In contrast, the edge mode, corresponding to the longitudinal component, is not gauge-invariant, marking a key distinction from the (truly non-local) Goldstone mode introduced later.} 
Crucially, the Hodge decomposition depends on the value of the connection over the entire entangling surface, and only on its value there, making the edge mode quasi-local.\footnote{In \cite{Riello:2020zbk,Riello:2021lfl} a choice of intrinsic edge mode has been implicitly made in the construction, albeit a different one, involving a \emph{bulk} functional of $A$ based on the Helmholtz decomposition. While both choices yield a valid unextended regional phase space, we view our choice as the canonical one, in the sense that the edge mode need only be defined on $\Gamma$ for the gauge-invariant definition of the subregion theory.}

Within the extrinsic type, there is a special class of edge modes which ``pull in'' the asymptotic soft physics into the subregion theory. The \emph{soft edges} in the title refer to edge modes of such type as they are the key ingredient in relating soft and edge physics. In this paper, we use a particular realization of such edge modes, built out of Wilson lines connecting $\Gamma$ to the asymptotic infinity \cite{Carrozza:2021gju}, as shown in figure \ref{dynamical frame}. There are other realizations of asymptotically charged frames upon which we comment in appendix \ref{app:examples}.

\begin{figure}[h!]
    \centering
    
\tikzset{every picture/.style={line width=0.75pt}} 

\begin{tikzpicture}[x=0.75pt,y=0.75pt,yscale=-1,xscale=1]

\draw [color={rgb, 255:red, 155; green, 155; blue, 155 }  ,draw opacity=1 ]   (307.2,111.6) .. controls (357.2,103.6) and (367.2,131.6) .. (412.2,145.69) ;
\draw [color={rgb, 255:red, 155; green, 155; blue, 155 }  ,draw opacity=1 ]   (306.2,129.6) .. controls (356.2,121.6) and (365.2,145.6) .. (412.2,145.69) ;
\draw [color={rgb, 255:red, 155; green, 155; blue, 155 }  ,draw opacity=1 ]   (312.2,148.6) .. controls (362.2,140.6) and (358.2,157.6) .. (412.2,145.69) ;
\draw [color={rgb, 255:red, 155; green, 155; blue, 155 }  ,draw opacity=1 ]   (318.2,164.4) .. controls (362.2,174.4) and (372.2,163.6) .. (412.2,145.69) ;
\draw [color={rgb, 255:red, 155; green, 155; blue, 155 }  ,draw opacity=1 ]   (313.2,185.4) .. controls (349.2,190.4) and (376.2,172.6) .. (412.2,145.69) ;
\draw    (310.2,193.6) .. controls (324.2,163.6) and (317.2,163.6) .. (312.2,148.6) .. controls (307.2,133.6) and (300.2,119.6) .. (311.2,100.6) ;
\draw   (290.99,265.7) -- (291.06,24.04) -- (412.2,145.69) -- cycle ;

\draw (296,140.4) node [anchor=north west][inner sep=0.75pt]    {$\Gamma $};
\draw (417,136.4) node [anchor=north west][inner sep=0.75pt]    {$i ^{0}$};
\draw (356,65.4) node [anchor=north west][inner sep=0.75pt]    {$\mathscr{I}^{+}$};
\draw (357,210.4) node [anchor=north west][inner sep=0.75pt]    {$\mathscr{I}^{-}$};

\end{tikzpicture}

    \caption{Example of a dynamical reference frame built out of spacelike (or null) Wilson lines connecting $\Gamma$ to asymptotic infinity. They can end anywhere on $i^0$ (including $\scri^+_-$ and $\scri^-_+$). They cannot, however, end on $\scri^+$ or $\scri^-$ as this would lead to the wrong transformation under small gauge transformations.}
    \label{dynamical frame}
\end{figure}
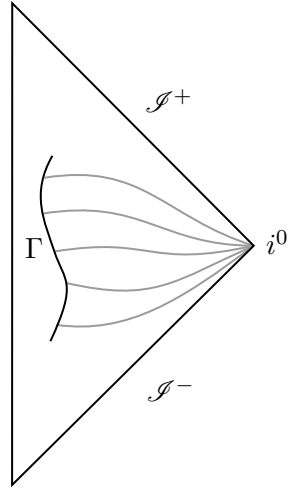

Taking this perspective, it is clear that the manifestly gauge-invariant approach does not amount to adding new degrees of freedom to the theory, as sometimes stated. On the one hand, extrinsic edge modes $\Phi$, as composite fields of the complement, were always there in the phase space of the global theory and only appear as ``new'' from the point of view of the subregion \cite{Carrozza:2021gju,Carrozza:2022xut}. This is why they correspond to a phase space \emph{extension} from the perspective of the subregion. On the other hand, the intrinsic edge mode $\Tilde\Phi$, as a composite field built entirely from $\Gamma$, was always there even in the purely subregional data; it does not  correspond to a phase space extension. Thus, one does not need to extend to have an ``edge mode''. We will see shortly, however, that without extension the regional theory will encode no information about its relation with the complement and this will be reflected in the absence of corner symmetries.

By dressing the bare fields with the edge mode, we build manifestly gauge-invariant observables. For example:
\begin{align}
    A^\dr:=A-\dt \Phi \label{dressed}\,.
\end{align}
These are also manifestly \emph{relational observables} \cite{Carrozza:2021gju,Goeller:2022rsx}, as they gauge-invariantly encode the (Lie algebra) value of the field $A$ when the frame (phase) $\Phi=\text{const}$, as we explain in section \ref{subsec:dyn_ref_fr}. Since $\Phi(x)$ fully parameterizes the gauge orbit at $x\in\Gamma$ (i.e.\ of the local structure group), fixing its value is equivalent to a gauge-fixing condition, something we will revisit later. For distinction, we henceforth label the intrinsically dressed connection as $\Tilde{A}^{\rm dr}$, identifying the non-exact piece in the Hodge decomposition \eqref{eq:Hodge_intro} as the corresponding dressed observable.

\paragraph{Properties of Intrinsic vs Extrinsic Frames:} 
Intrinsic and extrinsic edge modes are equally good for the purpose of having manifest gauge invariance at $\Gamma$. However, they are physically very different. The phase space built with the intrinsic edge mode is the \emph{unextended} one and only includes observables with support entirely within the subregion, while the phase space built with the extrinsic edge mode is an \emph{extended} one and includes some non-local observables with support shared between the subregion and the complement. We explore this issue in detail in section \ref{subsec:extensions}. 

For now, we note that the core difference lies in whether or not they give rise to a non-trivial corner symmetry algebra.
Through the inclusion of some shared non-local observables, the extended phase space acquires physical symmetries. In addition to gauge covariance, the extrinsic frames also transform under another class of transformations, which can be realized as certain changes of the complement data at spacelike separation, with the effect that (see \cite{Donnelly:2016auv,Geiller:2019bti,Carrozza:2021gju}):\footnote{See \cite{delaHamette:2021oex,Hoehn:2023ehz,DeVuyst:2024pop} and \cite{Carrozza:2022xut,Goeller:2022rsx,DeVuyst:2024pop} for their quantum and gravity realizations, respectively.} 
\be
\delta A=0\,,\q \delta \Phi =-\rho\,,\label{eq:reor}
\ee
with $\rho:\p\Sigma\to \mathbb{R}$ field-independent. These are called \emph{frame reorientations}, as they change the orientation and phase of the frame $U$, while leaving the remaining regional data invariant. 
When these transformations survive on the space of solutions, they become physical symmetries and constitute the corner symmetries when restricted to $\p\Sigma$ \cite{Carrozza:2021gju}. Under frame reorientations,  $A^{\rm dr}\mapsto A^{\rm dr}+\dd{\rho}$ and so it is clear that reorientations commute with gauge transformations. We discuss the various possibilities  of generating frame reorientations in detail in section \ref{subsec:fr_reor} and we clarify in section \ref{sec:Soft_bc} why $\rho$ has to be a function of a boundary cut only. These frame reorientations alter the gauge-invariant relationship between the connection and the frame (and thereby the subregion and its complement), by acting solely on the frame. This highlights the relational nature of these transformations.

By contrast, as we will see in section~\ref{subsec:fr_reor}, no such transformations are possible for the intrinsic frame $\Tilde\Phi$, essentially because it is built from the intrinsic field content of the region, so that in \eqref{eq:Hodge_intro} one cannot vary the frame without varying the connection. This means that the unextended phase space has \emph{no} corner symmetries, recovering the observations in \cite{Riello:2020zbk,Riello:2021lfl}.

\paragraph{Gauge-fixed vs Gauge-invariant Approaches:}

In section \ref{subsec:equivalence} we discuss the equivalence of these two approaches: the former simply yields a specific description of the latter. However, in the gauge-fixed description certain aspects appear less transparent due to removing part of the kinematical data, such as for example the interpretation of the physical corner symmetries, which has led to incorrect statements in the literature. 

The two types of edge modes (intrinsic and extrinsic) lead to different gauge-invariant phase spaces, and this distinction is reflected in the corresponding gauge-fixed descriptions, as illustrated in section \ref{subsec:equivalence}. In the unextended (intrinsic) phase space, frame reorientations do not exist and this is mirrored in the gauge-fixed version, where no residual symmetries remain. In contrast, gauge-fixing in the extended (extrinsic) phase space does not remove reorientations; instead, these manifest themselves as physical, boundary-supported \emph{gauge-looking} transformations. But, crucially, they are \emph{not} gauge transformations, as sometimes claimed in the literature.\footnote{The mechanism behind this is the same as in the Page-Wootters formalism for extracting a relational dynamics from a time reparametrization-invariant system. The Page-Wootters conditioning is nothing but a gauge-fixing of the clock quantum reference frame \cite{Hoehn:2019fsy} and the relational dynamics corresponds to the reorientation of that clock. It is a physical symmetry whose action on the evolving degrees of freedom however appears the same as a gauge transformation upon conditioning \cite{DeVuyst:2024pop,DeVuyst:2024uvd}.} Due to their non-local character, these transformations act physically on the global solution space and affect dynamical information. They are, thus, the imprint on the subregion physics of an action at the level of the global solution space. 

To illustrate this, let us consider a gauge condition fixing the value of the frame phase (e.g.\ $\Phi=0$).\footnote{The 1-dimensionality of the gauge (structure) group implies that a complete gauge-fixing is provided by fixing a single scalar, transforming covariantly under the gauge group, such as the frame phase.} While we initially defined reorientations as acting solely on the frame \eqref{eq:reor}, this is true only up to gauge transformations, which act on all fields. On the gauge-fixed section where $\Phi = 0$, the corresponding representation of reorientations is thanks to \eqref{dressed} given by:
\be
\label{eq:reor_gf}
\d \Phi =0 \,,\q \d A = \dt \rho\,.
\ee
If we disregard the frame and focus exclusively on the \textit{system} ($A$), starting from a gauge-fixed perspective, we might (incorrectly) conclude that \eqref{eq:reor_gf} is a gauge transformation. It only appears as such. In reality, it is a symmetry that modifies the relationship between the system and the frame.\footnote{In analogy with particles subject to translation invariance, this would be akin to altering the relative distance between a reference particle and the others \cite{Carrozza:2021gju}.}

Starting from a gauge-fixing viewpoint can obscure the inherent relationality of these transformations, making it challenging to recognize their physical meaning and sometimes leading to the statement that boundaries ``break gauge invariance'' or ``turn gauge into physical degrees of freedom''.  This underscores the value of a gauge-invariant approach, which facilitates a more comprehensive understanding of the interplay between boundary conditions, gauge transformations, symmetries and the overall structure of the theory.

\paragraph{The Finite Goldstone Mode:}

The relationship between the intrinsic ($\Tilde{\Phi}$) and extrinsic ($\Phi$) edge modes encodes the same physics as the split between \emph{radiative} and \emph{Goldstone} modes.\footnote{Similarly to how we have introduced the intrinsic edge mode, the same $\Tilde A^\dr_t$ can be obtained through a Hodge decomposition of the dressed observable $A^\dr$, instead of the bare connection $A$. The important difference is that here both the \emph{transversal} and \emph{longitudinal} modes are gauge-invariant. This turns out to play a role in cavity QED \cite{TBH}.} In particular, the relational observable $\varphi:=\Tilde{\Phi}-\Phi$, describing the intrinsic and extrinsic frames relative to one another (and thereby encoding the transformation between them), is nothing but the exact piece of the boundary-induced extrinsic dressed observable: $A^\dr\big|_{\p\Sigma}$. We thus write:
\be
    A^\dr\big|_\Gamma=\Tilde{A}^\dr\big|_\Gamma + \dt\varphi \,\label{eq:extr_intr_rel}\, , 
\ee
and call $\varphi$ the \emph{finite Goldstone mode}. This terminology is justified by the fact that, under the frame reorientations defined in \eqref{eq:reor}, we have that:
\be
    \delta\Tilde A^\dr\big|_\Gamma=0\,, \q \delta\varphi=\rho \, . 
\ee
Therefore, $\varphi$ is a gauge-invariant observable (unlike the frame $\Phi$), which parameterizes the orbit of the physical symmetry group. In addition, it parameterizes the local vacuum solutions ($F=0$) in a gauge-invariant, frame-dependent and non-local fashion.\footnote{We will see in section \ref{subsec:sym_form} that it corresponds to a zero-energy mode around the vacuum solution.} This means it acts as a Goldstone mode for the phenomenon of \emph{relational spontaneous symmetry breaking}, whereby the breaking of the subregional symmetry group is achieved via transformations (at spacelike separation) in the complement, while the subregion stays in the vacuum as far as any local observables are concerned. The different vacua thus correspond to distinct relations between the subregion and its complement, where only the data defining the extrinsic frame in the complement varies. This finite Goldstone mode will be a key player throughout the paper;  we discuss it in sections \ref{subsec:framechange}, \ref{subsubsec:splitting} and \ref{subsec:GM}.

\paragraph{Various (Pre-)Symplectic Forms:}

The presymplectic form before gauge-fixing reads \cite{Donnelly:2016auv,Geiller:2019bti,Carrozza:2021gju}:
\be
    \Omega=\int_\Sigma\delta A\wedge\delta\star F - \int_{\partial\Sigma}\delta\Phi\,\delta\star F\,. \label{eq:edge form}
\ee
We call the corner  variables here, $(\Phi,\star F\big|_{\p\Sigma})$, the \emph{edge mode pair}.\footnote{We have chosen this nomenclature to accommodate the different uses in the literature with regards to what the ``edge mode'' is alluded to.
As both variables are defined on the corner and conjugates of one another, this terminology is justified. In \cite{Ball:2024hqe,Ball:2024xhf} the term ``edge mode'' is used to denote what here we called the Goldstone mode $\varphi$. These are different types of quantities, so having distinct terminology is important.} As discussed, it is our choice which edge mode $\Phi$ (extrinsic or intrinsic) to take. This simply corresponds to a choice of observables to include in our phase space and thus applies both to the extended and unextended phase space.

For an extrinsic edge mode (like the soft edge), using \eqref{eq:extr_intr_rel}, we get the following factorization of the phase space:\footnote{This factorization should be understood upon quotienting by degenerate directions. Here, we have implicitly extended the frame $\Phi$ into the bulk in some way, like in figure \ref{edge extension}, and extend the definition of dressed observables accordingly. Then the symplectic form reads:
\[
    \Omega\approx\int_\Sigma\delta A^\dr\wedge \delta\star F\,.
\]
$A^\dr$ extended in the bulk labels the gauge orbits.We are free to extend the frame in any way we like. It just amounts to a different description of the same physics, by choosing a different label for the same gauge orbit.}
\be
    \Omega_\text{ext}&\approx \Omega_\text{int}+\Omega_\text{Gold}\,,\\
    &\approx \int_\Sigma \delta \Tilde{A}^\dr\wedge \delta\star F + \int_{\p\Sigma}\delta\varphi\, \delta\star F\,,\label{eq:gold form}
\ee
where we have used suggestive notation to denote the extrinsic and intrinsic symplectic forms. We call the corner term here, $(\varphi,\star F\big|_{\p\Sigma})$, the \emph{Goldstone mode pair}. In other words, an extended phase space always includes a Goldstone-like corner term in its symplectic form. This leads to the slogan:
\be
    \text{Extrinsic }=\text{ Intrinsic }\cross \text{ Goldstone}\,.
\ee

Despite their similar form, the two corner terms (in \eqref{eq:edge form} and \eqref{eq:gold form}) are fundamentally different, as they involve different kinds of objects: $\Phi$ is a gauge-covariant reference frame, while $\varphi$ is a gauge-invariant relational observable. Crucially, $\star F|_{\p\Sigma}$ is conjugate to both.
 The form in \eqref{eq:gold form} is what is analogous to the asymptotic symplectic structure at $\scri^+$, which includes the asymptotic soft factor as its corner piece \cite{Strominger:2017zoo}.

\paragraph{Soft Boundary Conditions:}

In order to have a well-defined subregional theory, we need to specify boundary conditions such that the symplectic flux vanishes on $\Gamma$.\footnote{We can think of these as defining the type of walls we have in our lab. In fact, some of the classes of boundary conditions discussed here even have an interpretation as an infinitesimally thin electric/magnetic conductor.} We are interested in this question for the extended phase space, built out of extrinsic edge modes of the soft edge type. The boundary conditions for the unextended phase space agree in part with those of the extended phase space. The difference relates to the additional cross-boundary observables included in the phase space. 

It turns out that a crucial role is played by $A^\dr_t\big|_\Gamma$, where $t$ measures the time along the detector worldlines. Its value is physical and must \emph{always} be specified as a boundary condition in order to evolve the initial data.\footnote{This is in contrast with $\Tilde{A}^\dr_t\big|_\Gamma$, which is fixed by the initial data in the extended phase space. Different values of $A^\dr_t\big|_\Gamma$ lead to different vacuum transitions for the same initial data.} We explain this in section \ref{subsec:in_dat_ev}. This observation leads to a clear understanding of which boundary conditions refer to the (intrinsic) radiative modes, localized on $\Gamma$, and which to the complement-supported observables we added to the phase space, which are not radiative for the subregion. 

Because boundary conditions are part of the definition of a theory, only symmetry transformations which preserve them are allowed. This leads to a classification of boundary conditions in terms of the sub-algebra of the asymptotic algebra which survives as physical symmetries for the subregion:
\begin{enumerate}
    \item \emph{Soft Boundary Conditions} - full infinite-dimensional symmetry algebra:
    \begin{align}\label{softcornercharge}
        \{Q_{\partial\Sigma}\}=Q[\rho]:=\int_{\partial\Sigma}\rho\star F,\;\;\;\rho:\partial\Sigma\to \mathbb{R}\,.
    \end{align}

These are obtained by imposing boundary conditions on the intrinsic radiative modes (here we give Dirichlet and Neumann as examples), as well as minimal conditions on the Goldstone sector which do not eliminate it from the phase space.  As explained in section \ref{sec:Soft_bc}, it turns out that only the divergence-free parts of $\Tilde{A}^\dr_a $ (here called $\p_a h$) and $F^{ra}$ (here called $ \p_a \mathfrak{h} $) need to be fixed, as summarized in Table \ref{table: soft bcs}. These are related to the magnetic field orthogonal and tangential to the corner, respectively, as we illustrate in section \ref{sec:Soft_bc}. 

\begingroup
\renewcommand{\arraystretch}{1.7}
    
    \begin{table}[h!]
        \centering
        \begin{tabularx}{0.5\textwidth}{|>{\centering\arraybackslash}X|>{\centering\arraybackslash}X|}
            \hline
            \multicolumn{2}{|c|}{$\d A^\dr _t  \big |_{\Gamma } = 0$} \\
            \multicolumn{2}{|c|}{$\d \p _t \l \sqrt{g}  F^{rt}  \r \big |_{\Gamma } =  0$} \\
            \hline
            (Soft) Dirichlet & (Soft) Neumann\\
            \hline 
            $\delta \p_a h=0$& $\delta \p_a \mathfrak{h} =0$\\
            \hline
        \end{tabularx}
        \caption{Table showing the two basic examples of \emph{soft} boundary conditions. The first two rows are the key to the different character of these boundary conditions compared to the traditional ones below. The last row corresponds to a simple Dirichlet/Neumann choice for the \emph{radiative} modes of the dressed observables. The index $a$ runs over the coordinates on $\partial\Sigma$ and $t$ is the proper-time along the detector worldlines.}
        \label{table: soft bcs}
    \end{table}

\endgroup

This classification of the boundary conditions allowing for the imprint of the full infinite-dimensional asymptotic symmetry algebra on the subregion theory is one of the key results of this paper. We point out that the soft Neumann case was recently considered in \cite{Ball:2024hqe,Ball:2024xhf,Canfora:2024awy}, where it was proposed as an example of boundary conditions allowing for corner charges.\footnote{They called them ``dynamical edge mode'' boundary conditions. We also note that the soft Dirichlet case corresponds to what can be called ``Dirichlet up to large gauge'' boundary conditions.} In this paper, we derive the above classification in section \ref{sec:Soft_bc}. 

\item \emph{Traditional Boundary Conditions} - no symmetries:
    \begin{align}
        \{Q_{\partial\Sigma}\}=\{\emptyset\}\,.
    \end{align}

    \begingroup
    \renewcommand{\arraystretch}{1.7}
    \begin{table}[h!]
        \centering
        \begin{tabularx}{0.6\textwidth}{|>{\centering\arraybackslash}X|>{\centering\arraybackslash}X|}
        \hline
           (Traditional) Dirichlet  &  (Traditional) Neumann\\
        \hline
            $\delta A^\dr\big|_\Gamma=0$ & $\delta\star F\big|_\Gamma=0$\\
        \hline
        \end{tabularx}
        \caption{Table showing the two standard examples of boundary conditions fully fixing either $A^\dr\big|_\Gamma$ or $\star F\big|_\Gamma$.}
        \label{table: traditional}
    \end{table}
    \endgroup

    Both of the traditional type of boundary conditions summarized in table~\ref{table: traditional} fix the gauge-invariant corner piece of the symplectic form, which eliminates the possibility of frame reorientations. This is because these conditions do not discriminate between radiative and Goldstone modes, as we explain in section \ref{sec:Soft_bc}. As we will see later, they are partially redundant. 
\end{enumerate}

This classification justifies our choice of the soft boundary conditions from here onwards.\footnote{We note here that the boundary conditions on the global theory are also of the soft type and they reduce to the usual class of boundary conditions on $i^0$ considered in the asymptotic literature. See section \ref{sec:sol_space}.} The other cases simply do not allow for the imprint of the asymptotic symmetries into the subregional theory. This reconciles a debate in the literature about whether, on the one hand, a non-trivial symmetry algebra is always possible irrespective of boundary conditions \cite{Donnelly:2016auv}, or whether, on the other, it depends on boundary conditions and there are any at all which allow a non-trivial algebra \cite{Carrozza:2021gju}.

\paragraph{Embedding and Postselection:}

Taking into account that the subregional theory is embedded in a larger, global theory, the question arises how the regional phase space relates to the global phase space of solutions. The  effect of putting boundary conditions on $\Gamma$ corresponds to a \emph{postselection} on this global phase space to the subset of solutions consistent with those boundary conditions \cite{Carrozza:2021gju,Carrozza:2022xut}. 

We note that the postselection induced by the first line in table \ref{table: soft bcs} is of a non-local nature with respect to the complement initial data. It is also \emph{in addition} to the postselection conditions arising from the intrinsic, unextended subregion phase space (bottom line in table \ref{table: soft bcs}). This makes sense, since, in the extended phase space, we are keeping track of strictly \emph{more} observables, namely also the Goldstone sector, which encodes the frame-dependent set of cross-boundary data.

\paragraph{Subregion Symmetries:}

We now have all the ingredients to understand what the subregion corner symmetries actually mean. By symmetries here and throughout this work, we mean integrable phase space variations, tangential to the solution space (including boundary conditions) and transversal to the gauge orbits.\footnote{We do not necessarily require the stronger condition of having charges preserved under time evolution. In fact, depending on boundary conditions, their charges may have a non-trivial time evolution which is, however, constant on the phase space. In other words, they Poisson commute with the Hamiltonian up to a phase space constant that is determined by the boundary conditions and may or may not vanish. We will come back to this in section~\ref{sec:Soft_bc}.} Subregion symmetries thus refer to integrable flows on the \emph{regional} phase space; within the global theory this means the variation must be tangential to the postselected subspace (though it need not necessarily be integrable there). When working in a manifestly gauge-invariant description involving edge modes, the boundary-supported non-degenerate and integrable directions of the symplectic form in \eqref{eq:edge form} correspond to \emph{frame reorientations}, in which the bare field $A$ remains unchanged, while the extrinsic edge mode transforms as in equation \eqref{eq:reor}. While this was already noted in \cite{Carrozza:2021gju}, we can now refine that observation.

Notice already that, even at a kinematical level, no such transformations are possible for the intrinsic edge mode. This has two consequences. First, the finite-distance Goldstone mode thus transforms as $\varphi\mapsto\varphi+\rho$ under extrinsic edge frame reorientations, while $\Tilde{A}^{\rm dr}$ remains invariant. Second, the unextended phase space has \emph{no corner symmetries} (cf.\ section~\ref{subsec:extensions}), consistent with the discussion in \cite{Riello:2021lfl} (it may have other integrable symmetries that are not purely boundary supported). Corner symmetries only arise when we extend the phase space by certain non-local shared observables, like those associated with the extrinsic soft edge modes in figure \ref{dynamical frame}. 

There are two ways to achieve \eqref{eq:reor}. We either:
\begin{enumerate}
    \item modify the initial data ``locally'' in the complement,\footnote{This changes the value of the edge mode $\Phi$ on $\Gamma$. The requirement that it preserves the boundary conditions on $\Gamma$ is quite strong. There are, however, allowed transformations as discussed explicitly in section \ref{sec:gold_soft}.} or
    \item act with a large gauge transformation at the asymptotic boundary.
\end{enumerate}
Both types of transformation involve an action on dynamical fields living in a direction spacelike to the subregion. We discuss both cases in section \ref{subsec:fr_reor}. The conclusion is that, for soft edge modes:\footnote{This holds for any choice of boundary conditions on $\Gamma$, including those of traditional type in table \ref{table: traditional}. In that case, the post-selection is stronger, as explained in section \ref{sec:traditional bcs}, and no large gauge transformations are allowed on the post-selected global phase space, consistent with the lack of symmetries for the subregion theory.} 
\be
    \text{Large-gauge Transformations }\subseteq \text{Subregion Corner Symmetries} \label{symmetry inclusion}\,,
\ee

More precisely, in section \ref{sec:charge action} we show that:
\be
\label{eq:charge_action_post}
\{Q_\text{soft}\;,\; \cdot\;\}_\text{postselected}=\{Q_{\p\Sigma}\;,\; \cdot \; \}_\text{subregion}\,,
\ee
where $Q_\text{soft}$ are the usual soft charges at the \emph{asymptotic} boundary, $Q_{\p\Sigma}$ are the finite-distance corner charges given in \eqref{softcornercharge}, and the subscripts clarify on which phase spaces the respective Poisson brackets are evaluated. The equality holds on observables defined on the subregion extended phase space. Hence, we can think of the finite-distance corner charges as ``pulled in'' incarnations of the asymptotic soft charges. The entity behind this mechanism is the \emph{soft edge} mode $\Phi$. 

An important lesson from this analysis is that \eqref{symmetry inclusion} is a strict inclusion for the non-trivial case of soft boundary conditions on $\Gamma$. The notion of symmetry is enhanced for subregions, since transformations of the first kind mentioned above might not be symmetries of the global theory,\footnote{Being quite arbitrary transformations of initial data in the complement, they need not necessarily be integrable. They must be tangential to the global space of solutions and transversal to its gauge orbits, however.} yet their action on the subregion phase space is indistinguishable from a large gauge transformation.

In particular, we will show in section \ref{sec:charge action} that for the phase space built from a certain kind of soft edge modes, well-suited to the geometry of Minkowski spacetime, we can generate a subregional symmetry via the addition of asymptotic \emph{soft} photon modes or even \emph{hard} photon modes, that are causally disconnected from the subregion. These latter two types of physical transformations do not correspond to symmetries of the global theory, yet they are operationally meaningful ways of generating subregional symmetries from the action of global observables.

We see that \emph{how} to generate corner symmetries becomes a tractable question once we know how to construct the edge mode from complement observables. The answer is, of course, \emph{frame-dependent}. Even though any extrinsic edge mode may support the same full set of corner symmetry transformations, their generators as operators in the global theory may be radically different in each case. It is in this sense that corner symmetries are \emph{relational} in nature.

\paragraph{Finite-Distance Memory, Goldstone Modes and Relation to the Asymptotics:}

The electromagnetic memory effect consists in the momentum kick experienced by a charged inertial particle upon the passage of electromagnetic radiation. It is a purely (spatially) \emph{local} effect, approximated by the following integral along the particle's unperturbed world-line, $\tilde x(s) _0$ (meaning what would be the world-line without the effect of $F_{\mu \nu }$), parametrized by its proper time $s$:
\be\label{eq:local memory}
    \mathcal{N}_a=\int_{s_i}^{s_f}\dt s\;  F_{\mu a}\frac{\dt \tilde x _0 ^\mu}{\dt s}\, . 
\ee
Evaluating this for a particle whose worldline approaches $\scri^+$ tangentially over an infinite time extent connects it to the photon's soft mode. Under ``no magnetic charges'' boundary conditions, this soft mode coincides with the asymptotic vacuum transition $\Delta\phi$ \cite{Strominger:2017zoo}, thereby realizing one edge of the soft IR triangle correspondence.

While this asymptotic memory effect, similarly to amplitudes in scattering theory, is usually taken to be a good description of the impact of an electromagnetic wave sufficiently far from the source of radiation, our framework permits us to propose a somewhat more operational quasi-local version of the memory effect, i.e.\ for finite distance and finite time extent. 

Our setup already consists of a timelike tube, making it natural to describe memory within this framework. This is particularly suitable when the tube contains timelike geodesics, allowing us to treat ‘‘memory-recording'' as small perturbations to inertial particles with worldlines confined to $\Gamma$. Besides enabling a quasi-local description of memory, our framework offers flexibility for example by deforming the tube boundary towards a null surface or smoothly extending to the null asymptotic boundary by sending the radius to infinity. The latter will recover traditional results, as explained in section \ref{subsec:big_edges}.

Additionally, the timelike tube framework enables a dynamic treatment of vacuum transitions between two Cauchy surfaces for the subregion, a feature not afforded by a causal diamond.
In particular, we relate the quasi-local memory effect of \eqref{eq:local memory} (for a particle whose worldline lies on $\Gamma$) to the inherently \emph{frame-dependent} and \emph{non-local} transition of the regional vacuum:
\be
\label{eq:memory_Gm_finite}
    \mathcal{N}_a=\p_a\l\Delta\varphi-\int_{t^-}^{t^+}\dt t\; A^\dr_t\big|_\Gamma\r\,,
\ee
where $\varphi$ is the finite-distance Goldstone mode and this formula is valid for solutions with vanishing magnetic field at $t^\pm$ on $\Gamma$.\footnote{We consider more general cases in section \ref{sec:gold_soft}.} The term involving $A^\dr_t\big|_\Gamma$ is fixed by the soft boundary conditions, leading to an equality between a (spatially) local observable on $\Gamma$ 
and a non-local observable $\varphi$ (and non-local background field $A^{\rm dr}_t\big|_\Gamma$) which encodes the relationship between the subregion and the complement. This equality holds for any choice of extrinsic frame $\Phi$. 

The relationship between finite and asymptotic physics is brought to life if we take a soft edge mode as our extrinsic frame. In section \ref{sec:gold_soft} we consider a particular example using null Wilson lines to derive an explicit equation relating the finite Goldstone mode $\varphi$ to the asymptotic Goldstone mode $\phi$, given in  \eqref{eq:v_scri+}. This equality holds at the level of the postselected phase space and imprints the action of certain observables in the complement on the subregion relational vacuum. For example, its infinite time limit yields (we will properly explain this equation  in section~\ref{sec:gold_soft}):
\be
\varphi ^\pm   = \l\phi -\frac{1}{2} N\r \pm N    + \int _{S^2} \Delta _{S^2} ^{-1}  \l r_s^2  \mathcal{Q} ^\pm  + J^{\pm }  - \mathbf{Q} \r \, ,
\ee
where $N$ is the asymptotic memory, $\mathcal{Q}^\pm$ the initial and final radial electric fields on $\Gamma$,  $J^\pm$ the matter contributions in the complement and  $\mathbf{Q}$ the leading radial electric field on $\scri ^+_+$. The various quantities are depicted in figure \ref{fig:Memory_1}.

\begin{minipage}{.95\textwidth}
    \centering
    \vspace{5pt}
    \includegraphics[width=0.7\linewidth]{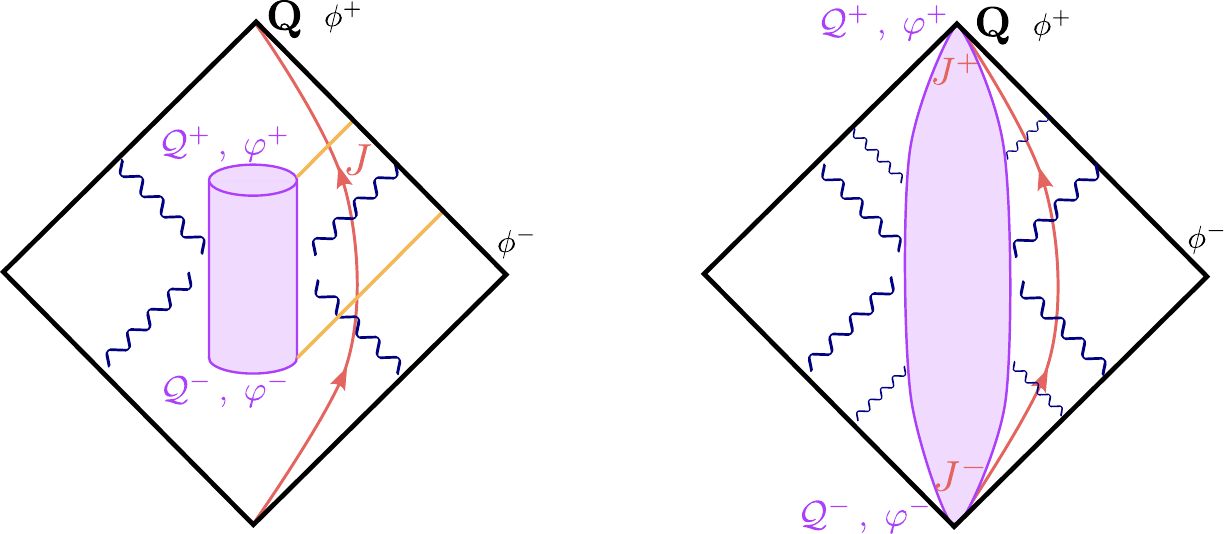}
    \captionof{figure}{\footnotesize On the left we show a finitely extended timelike tube, with the null Wilson lines from section  \ref{subsec:memory} highlighted in yellow. The electric fields $\mathcal Q^\pm$ and the Goldstone modes $\varphi^\pm$ are defined on the initial and final slices. These are positioned \textit{inside} $i^\pm$ in the limit $t\to \pm \infty$, shown on the right. Due to the compactification, we can't depict a detailed view of $i^\pm$, but it is important to emphasise that while $\mathbf Q$, $\phi^+$ are defined on $\scri^+_+$, the regional quantities $\mathcal Q^\pm$, $\varphi^\pm$ remain at finite distances, with the matter contributions $J^\pm$ lying somewhere in between. }
\label{fig:Memory_1}
\end{minipage}\\

To summarize, the frame dependence resides both in the relationship between quasi-local memory and vacuum transitions, as well as in the relationship between regional and global ‘‘vacua''. Conversely, both the asymptotic and finite memory (recorded in the momentum kicks) are edge-frame-independent concepts as they involve integrals of the frame-independent observable $F$.\footnote{This will be different in Yang-Mills theory, where $F$ is not gauge-invariant and thus too will have to be frame-dressed.} While asymptotic effects leave a finite-distance imprint via the soft edge frame, this imprint is fundamentally inaccessible from \emph{local} observations alone. In other words, quasi-local memory effects, measured at finite distances, bear a theoretical link to asymptotic phenomena, but this link remains elusive to direct measurement.

This ‘‘censorship'' arises because the same non-locality of frames that enables the finite-to-asymptotic link also relies on complementary data. Measurements restricted to a finite region cannot capture the full nature of the asymptotically extended dynamical frame or its properties in the complement, which remain inaccessible by definition. Without this additional information, inferring complete asymptotic details from quasi-local observables is impossible.

Nevertheless, this theoretical link between finite and asymptotic regimes remains essential for understanding edge modes and corner symmetries. Finite-distance observables may reflect aspects of the asymptotic structure, but operationally the latter cannot be decoded from them.

\paragraph{Frame Dependence:}

A key aspect highlighted by our construction is the fact that the subregional theory thus defined is \emph{frame-dependent}. 

This arises because the boundary conditions depend on our choice of edge mode field $\Phi$.\footnote{As of now, it might seem that this is not the case for the traditional Neumann boundary conditions. However, as explained before and in section \ref{subsec:in_dat_ev}, the frame dependence is still there through the fact that we still need to fix $A^\dr_t|_\Gamma$ \emph{on top of} fixing $\star F|_\Gamma$.} This, in turn, means that for different choices of dynamical reference frame, there will be different postselections on the global phase space and, thus, different theories inside the lab. Each frame choice, via the dressed observables, captures a \emph{different} set of cross-boundary gauge-invariant data of the global theory in the regional phase space. This frame dependence will be a recurrent theme in most of our results. Notice that the finite Goldstone mode $\varphi$, and thus the notion of vacuum itself, is also frame-dependent.

\paragraph{Going Agnostic:}

Throughout the paper, we adopt, as much as possible, an approach to phase space and boundary conditions that remains agnostic to the nature of the Cauchy slice (whether finitely or infinitely extended, or whether spacelike or null). This is advantageous for developing a unified framework that both encompasses and bridges finite-distance and asymptotic regimes.

This approach also shows that many results from the asymptotic symmetry literature can in principle be reformulated relationally. In the standard formulation, large gauge transformations are taken to be physical, thus not requiring gauge-fixing. This means that the global phase space is, in fact, an extended one relative to the purely radiative data. Indeed, this extension is constituted by the Goldstone mode and its conjugate momentum. One could equivalently formulate this as an asymptotically extended phase space with manifest large gauge invariance by introducing asymptotic extrinsic edge modes and reinterpreting the Goldstone mode, as we do at finite distance, as the relational observable between intrinsic and extrinsic edge modes. Just like at finite distance, the usual asymptotic symmetries are then reorientations of the asymptotic extrinsic edge modes and the standard formulation would be recovered by gauge-fixing the extrinsic edge mode. Of course, the price to pay is that one would have to make sense of the asymptotic extrinsic edge mode, which in spacetime extensions such as in \cite{Chen:2023tvj,Chen:2024kuq} would become meaningful.

\section{Dynamical Reference Frames in Maxwell Theory}
\label{sec:ref_frame}

In this section, we start by reviewing the key features of dynamical reference frames in gauge theory, following the constructions outlined in \cite{Carrozza:2021gju} and adding a couple of new observations along the way. The dynamical reference frames we are interested in are gauge frames transforming covariantly under the $\rm{U}(1)$ gauge symmetry of the theory, as opposed to some spatiotemporal symmetry. We also elaborate on the two important classes of reference frames which will be key players in the rest of the paper, namely the \emph{extrinsic} and \emph{intrinsic} ones. The terminology refers to whether they are constructed from data \emph{outside} or \emph{inside} the subregion, respectively.

Since we aim to use dynamical frames to realize edge modes on the boundary of the subregion, we first focus our discussion on defining \emph{boundary} or \emph{edge reference frames}. These are reference frames sensitive to the action of the gauge group on a given boundary only. This amounts to defining a frame field  $U[A]$ on a codimension-1 timelike surface, rather than on a codimension-0 spacetime region. We also explore extensions of frames into the bulk of the region of spacetime delimited by this boundary.
However, 
the primary focus of our paper lies in the relational structure at boundaries that gives rise to symmetries and charges.

\subsection{Dynamical Reference Frames}\label{subsec:dyn_ref_fr}

A defining feature of a reference frame is that it transforms under some symmetry group $G$. Its purpose in life is to define a description of a physical situation at hand. Typically, the frame is taken to be external to the physical situation.\footnote{Our use of external/internal here should not be confused with our use of extrinsic/intrinsic. External frames are non-dynamical, while what we call extrinsic and intrinsic frames are both dynamical internal frames.} For example, in standard quantum field theory in Minkowski space, we resort to Lorentz frames that are external to the fields and whose dynamics is not explicitly included. However, the frame could be internal too, i.e.\ part of the physical system to be described, and this is in fact what happens whenever there is a gauge symmetry at play. 

Whenever one is constructing dressed or relational observables, one is implicitly invoking an internal, i.e.\ dynamical reference frame constituted by the ``dressing'' degrees of freedom \cite{Carrozza:2021gju,Carrozza:2022xut,Goeller:2022rsx}.\footnote{In the quantum theory they become what are known as quantum reference frames \cite{Giacomini:2017zju,Hoehn:2019fsy,Hoehn:2023ehz,delaHamette:2021oex,AliAhmad:2021adn,Castro-Ruiz:2021vnq,DeVuyst:2024pop,Fewster:2024pur,Giacomini:2021gei,Vanrietvelde:2018pgb,Kabel:2024lzr,Kabel:2023jve,Giesel:2024xtb}.}  These are field-dependent reference frames that transform under the gauge group of the theory; in gravity, these transform under bulk diffeomorphisms, while in gauge theory they transform under structure-group-valued spacetime functions. In our case of Maxwell theory, this means they transform under $\rm{U}(1)$-valued spacetime functions. Thus, these are not necessarily associated with spatiotemporal symmetries and rather serve to deparametrize gauge orbits. Some examples are: the position of a particle in a translation-invariant theory \cite{Vanrietvelde:2018pgb}, the value of four scalar fields (possibly non-locally constructed from the fundamental fields including the metric) as a local dynamical coordinate system in gravity \cite{Goeller:2022rsx}, the color of one of the quarks inside the proton, or, as we explain in detail below, Wilson lines in Maxwell theory \cite{Carrozza:2021gju}. 

This is a manifestation of the broader physical insight that gauge-invariant observables are relational in nature \cite{Rovelli:2013fga,Rovelli:2020mpk,Gomes:2021tjm,Rovelli:1990pi}. Indeed, the dynamical frames give rise to an internal description of the physics, one that is gauge-invariant and relational: it describes the ``bare'' fields relative to the ``dressing'' frame field and in general there are many choices of such ``dressing'' frame fields. As we will see, one can change between the descriptions relative to different such internal frames just like one can between different Lorentz frames in Minkowski quantum field theory \cite{Carrozza:2021gju,Goeller:2022rsx}. In our case of gauge theory where the frame is not a spatiotemporal one, this relational description is not a spatiotemporal one either, e.g.\ it does not localize the fields to be described in spacetime (this is done via a standard external spacetime frame). Instead, it describes the bare field conditional on the frame being in a certain (local) orientation.

We now explain in some more detail the properties we want a reference frame field in pure Maxwell theory to obey; below we will construct examples that realize this. A general feature is that it is \emph{non-locally} constructed from the fundamental fields, which here is the connection $A$.\footnote{We can easily extend this construction to include dynamical matter fields/particles. Then, we can choose to construct non-local functionals of both $A$ and matter $\psi$: $U[A,\psi](x)$. These will, in general, not be group $G$-valued. 
This is something we will in fact do, when we include matter outside the subregion to discuss the memory effect in section \ref{sec:gold_soft}.}  
We denote it by $U[A](x)$, itself a field in spacetime; its value at $x$ will henceforth be referred to as its \emph{local orientation}. In this paper, for simplicity, we focus on a special class of reference frames: those whose orientations are valued in the group under which they transform. This will suffice for our purposes (for more general frames in gauge theory, see \cite[Sec.~4]{Carrozza:2021gju}). Such group-valued frames are common; for example, tetrad frames are Lorentz-group-valued, the position of a particle is translation-group-valued, and here, this means that the local orientations take value in the structure group $G=\rm{U}(1)$. Furthermore, we demand that, by construction, this field transforms covariantly under small gauge transformations:
\be\label{eq:covariance}
A \mapsto A+\dt \lambda \q \Rightarrow\q U[A] \mapsto e^{i\lambda} U[A]\,.
\ee
In this way, the gauge group acts on itself.
Such a non-local functional is a complete reference frame field, in the sense that it fully covers the group configuration space \cite{Carrozza:2021gju}.\footnote{The presence of the Gribov obstruction, arising from the non-uniqueness of gauge-fixing, might complicate the construction of global frames across field space. However, in Maxwell's theory, this obstruction is easily overcome due to the abelian nature of the gauge group and the simple topology of space, for the subregions of interest in this work. Such issues are anticipated to be more complex with non-abelian gauge groups or more intricate spacetime topologies.}

Constructing gauge-invariant observables is now straightforward. In our case, in which the frame itself takes values in the gauge group, it has a natural action on any functional of the fields and in particular on the connection: 
\be\label{eq:u_la_tr}
 U[A]^{-1} \rhd A := A  + i \dt \l \ln U[A] \r=:A-\dt \Phi\,,
\ee
where by $-i\ln U[A]$ we mean the \emph{local phase} of the $\rm{U}(1)$ group element, which from now on we label, suggestively, as $\Phi(x)$.
We denote the new (non-locally constructed) field containing both the bare connection and the phase of the frame by:
\begin{equation}\label{eq:A_dr_def}
 A^{\ra } := U[A]^{-1} \rhd A   =  A -\dt \rf  \,.
\end{equation}

This is an example of a \emph{frame-dressed} or \emph{relational observable}. It is clear that the covariance property of the frame \eqref{eq:covariance} implies $\Phi\mapsto\Phi+\lambda$ and so the small gauge invariance of the dressed field $A^{\rm dr}$. It is a relational observable giving the value of $A$ when the frame is in local orientation $U[A](x)=\mathds{1}$ or, equivalently, when the frame phase $\Phi$ is $0$. This can be seen by gauge-fixing the phase to $0$, which yields $A^{\rm dr}=A$ in that gauge. Since $A^{\rm dr}$ is gauge-invariant, it takes that value in any gauge.

We note that, in abelian theories, as here, the frame acts trivially on the field strength tensor $F^\ra :=U[A]^{-1}\rhd F=F$, but this is not true in non-abelian theories \cite{Donnelly:2016auv,Carrozza:2021gju}.

An important general feature of relational observables is that they know about the built-in redundancy of the fundamental fields. This is manifested through relationships between dressed observables, making them not independent. For example, for the components of the dressed connection, we can express this as:
\begin{align}
    \mathcal{G}[A^\dr]=0,
\end{align}
where $\mathcal{G}$ is a (generically) linear function on the set of dressed observables. We call this a \emph{dressing condition} and we will see explicit examples shortly. It is crucial not to confuse this with a gauge-fixing condition. We are working entirely at the level of gauge-invariant observables, so this condition holds \emph{in any gauge}.\footnote{What is true is that if we were to gauge-fix our frame to the value $\Phi_\text{g.f.}=0$, then the gauge-fixed field $A_\text{g.f.}$ would satisfy a condition like: $\mathcal{G}[A_\text{g.f}]=0$. But we are not forced to choose such a gauge-fixing condition.} We will revisit this distinction in section \ref{subsec:equivalence}, where we show that the two approaches (gauge-fixing and dressing), despite being conceptually different are equivalent.

We will now focus on two important classes of the non-local functionals $U[A](x)$ and $\Phi(x)$: \textit{extrinsic} and \textit{intrinsic}.  In doing so, we will emphasize the key distinctions between these two categories and their implications.

\subsection{Extrinsic vs Intrinsic Reference Frames}
\label{subsec:extr_intr_ref_fr}

In this paper, our interest lies in defining a gauge-invariant theory for a subregion of an asymptotically flat spacetime, as in figure \ref{lab}. To do that, we will need to define a reference frame field living on the timelike boundary $\Gamma$, whose phase will act as our edge mode for the subregion.

We have a choice between defining the reference frame via:
\begin{enumerate}
    \item A functional built non-locally from dynamical fields \emph{outside} of the subregion. We call this an \emph{extrinsic} frame.
    \item A functional built non-locally from dynamical fields \emph{on the boundary} itself. We call this an \emph{intrinsic} frame.
\end{enumerate}

\subsubsection{Dressing via Wilson Lines to the Asymptotic Boundary}\label{subsubsec:wilson}

We first describe an example of an \emph{extrinsic} frame, built from Wilson lines.

A Wilson line is a group-valued object that is covariantly (with respect to the gauge group) constant along a given path $x(\tau)$ in spacetime:
\begin{equation} \label{eq:Wilson}
      \l \p  _{\mu } U   + i A_{\mu } U \r  \frac{ \dt x^{\mu } }{\dt \tau } = 0  , \quad U_0 = U(0) = \mathds{1},
\end{equation}
where 
$\tau \in [0,1]$ parameterizes the path. Notice that we have the freedom to set this initial condition to whatever group element we like. The solution to the above equation is:
\begin{equation} \label{eq:im_frame}
    U(x, x_0(x))  =\mathcal{P} \exp \l  i \int _{ x_0 (x) } ^x A_{\mu} \frac{\dt x^{\mu } }{\dt \tau } \, \dt \tau  \r,
\end{equation}
where $\mathcal{P}$ stands for path-ordering, which, for the abelian case studied here, is irrelevant. Note that $U$ is group-valued, as desired. 

We have suggestively written the initial point of the path as a function of the final one. This is because we do not have just one path, but rather a congruence of smooth paths, as in figure \ref{dynamical frame}. With those paths we can define $U(x,x_0(x))$ everywhere in the bulk of the spacetime in an unambiguous way -- it is not only a functional of $A$, but also a field on the spacetime.

\begin{figure}[h]
     \centering
     \includegraphics[width=0.35\linewidth]{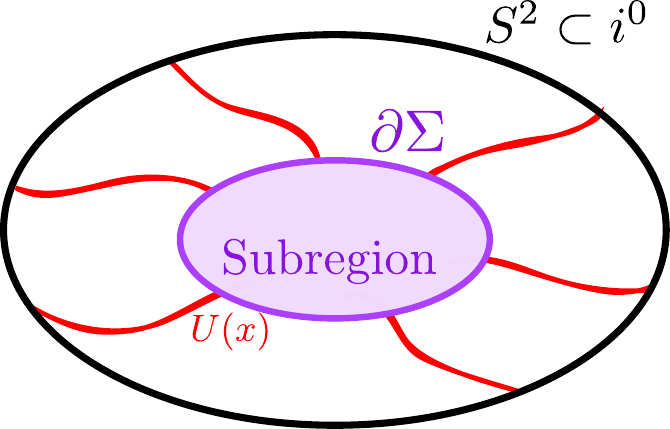}
     \caption{In this picture we illustrate the global and subregion Cauchy slices. As we can see, the latter is embedded in the former and it is bounded by the surface $\p \Sigma \subset \Gamma$. The reference frame $U(x)$ that is used to describe the subregion is constructed from Wilson lines (depicted in red) that stretch from the subregion all the way to the global boundary $i^0$. The congruence of paths provides a 1:1 map between $\p \Sigma$ and a spherical cut of $i^0$. 
     }
     \label{extrinsic}
 \end{figure}

We will be interested in Wilson lines that attach to the asymptotic boundary of space, $i ^0$ \cite{Carrozza:2021gju}. If we fix the system of Wilson lines and thereby the boundary anchor points $x_0(x)$, we are entitled to view $U[A](x,x_0(x))=:U[A](x)$ as a (non-locally constructed) $\rm{U}(1)$-valued field at $x$. These have the right covariance property of \eqref{eq:covariance} under small gauge transformations.\footnote{Of course, we could have taken the initial point to be anywhere in the bulk of spacetime if we had added, for example, a matter insertion at the same point in order to cancel the small gauge transformation there. Considering Wilson lines with bulk endpoints as edge modes was also proposed in \cite{Riello:2020zbk}, however that construction is not small gauge-invariant.} The Wilson lines can be in general spacelike or null.\footnote{In appendix \ref{app:examples}, we also discuss an example with timelike Wilson lines. There, we argue that those do not provide complete frames, and thus require a second dressing condition.} 
For illustration purposes, if we consider them to be entirely spacelike, we end up with a picture as in figure \ref{extrinsic}. The frame field $U[A](x)$ is thus defined by the system of Wilson lines and changing the congruence of paths means changing dynamical reference frame.

By restricting attention to its values on $x\in\Gamma$ we obtain a reference frame field on the boundary of the subregion. It is built from degrees of freedom entirely outside of the subregion and that is why it is an example of an extrinsic frame.

Having constructed the reference frame $U[A](x)$ we are now ready to define the gauge-invariant field via \eqref{eq:A_dr_def} to get:
\begin{equation} \label{eq:Arad_def_imfr}
    A^{\ra} = A - \dt  \left( \int _{x_0 (x)} ^x  A_{\nu} \frac{\dt x^{\nu } }{\dt \tau  } \, \dt \tau\right) =:A-\dt \Phi\,.  
\end{equation}
For later use, we note that this dressed observable can also be written as
\be
     A^\dr_\mu(x)= \int _{x_0 (x)} ^x  F_{\nu \mu } \frac{\dt x^{\nu } }{\dt \tau  } \, \dt \tau   + A_{\mu } (x_0 (x))\,.  \label{eq:alternative wilson}
 \ee
From now on, by the extrinsic edge mode field $\Phi$ we will mean the one constructed in this particular way, to contrast with the intrinsic field defined shortly. We discuss other examples of extrinsic frames in appendix~\ref{app:examples}.

What about the independence of the components of the dressed connection? First, it is clear that $A^{\rm dr}$ is generically non-vanishing as there is no reason for $A$ to be exact. Indeed, because we are in $d$-dimensional space the partial derivative $\dd$ is not the inverse of the integral over a path $\int \dt \tau $. However, what does vanish is the component of $A ^{\ra}$ \emph{tangential} to the path:
\begin{equation}\label{eq:WL_fr_dc}
    \frac{\dt x^\mu }{\dt \tau } A^{\ra} _{\mu } = \frac{\dt x^\mu }{\dt \tau } \left( A_{\mu } - \p _\mu  \int _{x_0 (x)} ^x  A_{\nu} \frac{\dt x^{\nu } }{\dt \tau  } \, \dt \tau  \right) =   A_{\mu} \frac{\dt x^{\mu } }{\dt \tau  } - \frac{\dt}{\dt \tau } \int _{x_0 (x)} ^x  A_{\nu} \frac{\dt x^{\nu } }{\dt \tau  } \, \dt \tau  = 0 \,.
\end{equation}
This is the dressing condition associated with this choice of frame -- $\mathcal{G}[A^\dr]=\frac{\dt x^\mu }{\dt \tau } A^{\ra} _{\mu }=0$.

\subsubsection{Dressing via Hodge Decomposition on the Subregion Boundary} \label{subsubsec:intrinsic}

We now describe an example of an \emph{intrinsic} frame. While in the previous case we had many choices of possible Wilson lines that lead to different frames, here there is essentially one canonical frame that we can build out of $A \big|_{\Gamma}$ alone.\footnote{In \cite{Riello:2020zbk,Riello:2021lfl} another example of intrinsic frame is provided. However, that involves the fields inside the whole region and not just on the boundary.}

We consider the pullback of the bare connection on a Cauchy slice $\p\Sigma$ of $\Gamma$: $A\big|_{\partial\Sigma}$. This is a one-form on a compact, closed 2-dimensional Riemannian manifold. This means it admits a \emph{unique} decomposition into  ``divergence-free'' and ``curl-free''/exact pieces, via the Hodge theorem. We are interested in the exact piece, which can be extracted via a non-local integral on $\partial\Sigma$ by smearing $A$ against the following kernel:
\begin{align}
\Tilde{\Phi}\big|_{\partial\Sigma}:=\int_{\partial\Sigma}\left(\Delta^{-1}_{\partial\Sigma}\star_{\partial\Sigma}\dt \star_{\partial\Sigma}\right)A\big|_{\partial\Sigma}\,,\label{eq:Hodge}
\end{align}
where $\Delta^{-1}_{\partial\Sigma}$ is the unique (up to a constant) Green's function of the Laplace operator on $\partial\Sigma$ and $\star_{\partial\Sigma}$ is the Hodge star operator on $\partial\Sigma$.

Under a small gauge transformation, $A$ changes precisely by a shift in its exact piece, so we have again the covariance property \eqref{eq:covariance}:
\begin{align}
    \Tilde{\Phi}\big|_{\partial\Sigma}\mapsto \Tilde{\Phi}\big|_{\partial\Sigma} + \lambda\big|_{\partial\Sigma}\,.
\end{align}
By extracting the exact piece, as above, on every Cauchy slice, we get a boundary reference frame phase field $\Tilde{\Phi}$ and thereby appropriately transforming $\rm{U}(1)$-valued frame field $\Tilde{U}[A](x)=e^{i\Tilde{\Phi}(x)}$ on $\Gamma$. It is defined intrinsically from degrees of freedom on the boundary of the subregion, in contrast to the previous extrinsic frame $U[A]$. In the sequel, we will gradually switch to talking about the phases $\Phi/\Tilde\Phi$ (instead of the frames $U/\Tilde U$) since these will become the edge modes.

The corresponding dressed observable is (for now only defined on $\Gamma$):
\begin{align}
    \Tilde{A}^\dr\big|_\Gamma:=A\big|_\Gamma-\dt \Tilde{\Phi}\big|_\Gamma\,.
\end{align}
This is nothing but the ``divergence-free'' part of the bare field $A$. We will refer to it as the \emph{radiative} mode of the field.\footnote{This is related to the Helmholtz decomposition, splitting transverse from longitudinal degrees of freedom \cite{Riello:2021lfl,TBH}.} As expected, it is gauge-invariant. The dressing condition for this intrinsic frame is: $\mathcal{G}[\Tilde{A}^\dr]=\star_{\partial\Sigma}\dt \star_{\partial\Sigma}\Tilde{A}^\dr\big|_{\partial\Sigma}=0$.\footnote{The intrinsic frame introduced here is, strictly speaking, not the only possible example. For instance, we can imagine adding to it any functional built from other intrinsic gauge-invariant observables, e.g.\ $\Tilde \Phi' = \Tilde \Phi + f[\Tilde A^\dr\big|_\Gamma]$.  This constitutes a distinct frame still transforming covariantly. Thus, while not unique, our argument in section~\ref{subsec:fr_reor} implies that none of them admit reorientations and they are thus qualitatively the same. In this work, we will neglect such frames and focus solely on the geometric splitting as in \eqref{eq:Hodge}. In fact, the above non-uniqueness of the frame is not specific to intrinsic frames and applies similarly to extrinsic ones. In particular, this extra freedom plays an important role in resolving long-standing ambiguities in cavity QED systems with quantum reference frames \cite{TBH}.}

\subsection{Frame Reorientations}\label{subsec:fr_reor}

We will see in section \ref{sec:BC} that the symmetries of subregional theories built with relational observables are \emph{frame reorientations} \cite{Carrozza:2021gju}. Here we describe what they are and how their existence depends on the type of reference frame.

Frame reorientations are transformations that leave the bare field on the boundary $\Gamma$ unchanged, but shift the reference frame phase (and thus the orientation of the frame): 
\begin{equation}\label{eq:fr_reor}
\begin{split}
    A\big|_\Gamma&\mapsto A\big|_\Gamma\,,\\
    \Phi&\mapsto\Phi-\rho\,,\\
    A^\dr\big|_\Gamma &\mapsto A^\dr\big|_\Gamma + \dt \rho \,,
\end{split}
\end{equation}
where $\rho$ is a field-independent function on $\Gamma$. A reorientation thus changes the relational observable $A^{\rm dr}$ and gives rise to an entire family of invariant observables labelled by $\rho$. For example, $A^{\rm dr}+\dd{\rho}$ now encodes the value of $A$ conditional on the frame phase $\Phi$ being in configuration $\Phi=-\rho$.

From here we can already see that the intrinsic frame phase $\Tilde{\Phi}$ does \emph{not} admit any frame reorientations. This is because there is no way to shift its value without shifting $A\big|_\Gamma$ as well. In other words, there is no way to shift the radiative mode $\Tilde{A}^\dr$ by an exact piece, by definition. This fact will play an important role throughout the paper.

On the contrary, extrinsic frames \emph{do} admit various kinds of frame reorientations, which we describe below. The complement degrees of freedom that we use to build the reference frame directly determine the permissible set of functions to which $\rho$ belongs. Further restrictions might come from our choice of boundary conditions on $\Gamma$, as we discuss in detail in section \ref{sec:Soft_bc}. For the extrinsic frame phase $\Phi$ defined earlier in terms of Wilson lines attached to $i^0$, and for the boundary conditions we will eventually place on the subregion theory, we have that $\rho:\partial\Sigma\to \mathbb{R}$ can be any (smooth) function on $\partial\Sigma$ which is time-independent: $\Dot{\rho}=0$. 

In the remainder of this subsection, we study the origin and physical interpretation of these symmetries, for the extrinsic frame phase $\Phi$. Such a transformation occurs if we change the field configuration in a region in the complement, spacelike separated from the subregion of interest. Essentially, we are tweaking the initial data of the global theory, which non-locally affects the initial data of the subregion. We present two distinct methods of doing this, which are:
\begin{itemize}
    \item Large gauge transformations of the global theory.
    \item Complement ``localized'' changes of initial data of the global theory.
\end{itemize}

\paragraph{Large Gauge Transformations:}
~\newline
An important class of transformations acting on the boundary that is relevant to this work consists of large gauge transformations (transformations with non-trivial support on the global boundary):
\begin{equation}
    A (x) \mapsto A(x) + \dt  \alpha (x), \quad \alpha |_{\B} \neq 0\,. 
\end{equation}
These transformations modify both the immaterial frame $U[A]$ and the dressed field $A^\dr$ as:
\begin{equation}
\begin{split}
    U (x, x_0 (x)) & \mapsto   e^{i \alpha (x ) } U (x, x_0 (x))  e^{-i\alpha (x_0 (x))}\,, \\
    A^{\dr} _\mu  (x) & \mapsto  A _\mu ^{\dr} (x)  +  \p _\nu \alpha (x_0 (x)) \f{\p x_0^\nu(x)}{\p x^\mu}\,.
\end{split}
\end{equation}
The relational field only ``sees'' the pullback of $\alpha $ on the global boundary $\B $. This is consistent with the small gauge invariance of $A^\dr$ and displays the mechanism by which reference frames bring boundary quantities in the bulk. Notice that a small gauge transformation can always undo the effect of $\alpha$ on the bare field $A$ on $\Gamma$. Therefore it is always a combination of a small and large gauge transformation that satisfies the definition of frame reorientation from \eqref{eq:fr_reor}. 
Large gauge transformations are, however, not the only thing we can do at the endpoints of the Wilson lines. In their definition \eqref{eq:Wilson} we said that the initial value $U(x_0)$ is free for us to specify. We have the freedom to change it, and this will affect the dressed field in the following way:
\begin{equation}
    \begin{split}
         U (x, x_0 (x)) & \mapsto U (x, x_0 (x))  e^{-i\rho (x_0(x))}\,, \\
          A^{\dr} _\mu  (x) & \mapsto  A _\mu ^{\dr} (x)  +  \p _\nu \rho (x_0(x) ) \f{\p x_0^\nu(x)}{\p x^\mu}\,.
    \end{split}
\end{equation} 
This transformation also acts solely on the frame and not on $A$, in line with \eqref{eq:fr_reor}.  
In order to establish equivalence between the two, we need to make sure that the set of possible functions for $\alpha$ is the same as it is for $\rho $. This is subject to the boundary conditions we impose on $\Gamma $ and $\B$. Ultimately, on the phase space presented in section \ref{sec:BC}, the two transformations will be indistinguishable.

\paragraph{Localized Changes of Complement Data:}
~\newline
Next, we present another mechanism that can generate a frame reorientation. This is achieved by  adding some electromagnetic field in the complement $\d F_{\mu \nu }$, that is non-vanishing in the directions tangential to the path. Let us label this complement variation as: $\delta\Bar{A}$. 

From \eqref{eq:Arad_def_imfr} it is clear that this changes the value of the frame phase $\Phi$. In order for this to generate successful frame reorientations, this variation must satisfy two key properties. At each moment in time, its support must lie entirely outside the subregion boundary. This ensures that inside the subregion, $ A^\dr   + \dt \rho [\d \bar{A} ] $ still satisfies the dressing condition \eqref{eq:WL_fr_dc}. Secondly, as we will see in section \ref{sec:Soft_bc} we must have $\p _t \rho [\d \bar{A}] = 0$. 
In sections \ref{subsec:GM} and \ref{sec:charge action}, while considering a special type of soft edge mode, we will see an explicit example of this, realized by the action of spacelike-separated \emph{hard photon modes}. We will further see that the addition of \emph{soft photon modes} also succeeds in generating a frame reorientation, although such an operator cannot be said to be spatially disconnected from the subregion due to the delocalized nature of zero-energy photons.

Finally, notice an important distinction between the two mechanisms for generating a frame reorientation presented here. For the first, we are using the symmetries already present in the global theory -- large gauge transformations. These act on the global boundary and therefore allow us to connect the subregion physics to the Goldstone and soft mode, both defined on $\scri^+ _-$, as we will discuss extensively in the rest of the paper. On the other hand, the second type of variation, while it must be tangential to the global space of solutions, need not necessarily be integrable and so need not necessarily be a symmetry of the global theory. 
Furthermore, the location of the perturbation can be arbitrarily far away from the global boundary and may have zero support there. A priori, it does not have a direct relationship with the aforementioned soft pair functions. The manifestation of such transformations as symmetries only makes sense from within the subregion theory, as they require a complement region to be defined. In order to survive as a symmetry of the subregion, this type of variation must be also tangential to the postselected subsector of the global solution space, determined by the boundary conditions on $\Gamma$ \cite{Carrozza:2021gju}. Large gauge transformations are therefore only a \emph{subset} of the symmetries of the subregion theory.\footnote{This distinction becomes crucial in the context of a closed universe, where large gauge transformations do not exist. Nevertheless, one can, in principle, still define subregional symmetries in that setting via the second mechanism above.}

Similarly, one can imagine a fictitious complement region of the \emph{global} theory, in which the global large gauge transformations themselves are sourced by changes of data in this artificial exterior region. Indeed, a construction like this has been used in \cite{Chen:2023tvj,Chen:2024kuq}. There, the authors use Weyl invariance of free Maxwell theory in Minkowski, in order to embed it in an Einstein universe and compute the vacuum entanglement entropy.

\subsection{Changes of Frame}\label{subsec:framechange}

It is worth considering the effect of picking a different choice of dynamical reference frame on our dressed observables \cite{Carrozza:2021gju}. For example, these might correspond to changing extrinsic frame by changing the Wilson line system, or to changing between extrinsic and intrinsic frame.

From \eqref{eq:A_dr_def} we see that, for the phases  $\Phi_1$ and $\Phi_2$ of two distinct frames, the corresponding dressed connection observables differ by:
\begin{align}\label{CFM}
    A^\dr_2(x)-A^\dr_1(x)=\dt(\Phi_1-\Phi_2)(x)=:\dt\varphi_{12}(x),\; x\in \Gamma\,.
\end{align}
By the transformation properties of frames, this difference is gauge-invariant, regardless of how the frames are constructed, be it from complement degrees of freedom or intrinsically.\footnote{As a simple example, let $\Phi_1$ and $\Phi_2$ correspond to two different sets of complement Wilson lines connecting the asymptotic boundary to $\Gamma$ with the condition that the 1:1 map $x\in\Gamma\mapsto x_0(x)\in\mathcal{B}$ is the same. Then $\rf_1 [\bar A ](x) -\rf_2 [\bar A ](x)  = \oint _{\gamma_{1\bar2}(x)} \bar A, \;x\in\Gamma $
which is a dynamical complement gauge-invariant observable. $\gamma_{1\bar2}(x)$ is the union of the two complement Wilson lines anchored at $x\in\Gamma$. }

The map $A^\dr_1\mapsto A^\dr_2$ looks formally equivalent to the frame reorientation map in \eqref{eq:fr_reor}, since $A$ is unchanged and the frame is ``shifted'' by $\Phi_1\mapsto\Phi_2=\Phi_1-\varphi_{12}$. The difference is, of course, that the shift $\varphi_{12}=\varphi_{12}[\Bar{A}]$ is itself a dynamical field, unlike $\rho$. In fact, $\varphi_{12}$ is a relational observable, giving the value of the frame phase field $\Phi_1$ when the other frame has phase $\Phi_2=0$, i.e.\ their relative orientation. Now, for a fixed frame,  we noted earlier that reorientations generate an entire family of dressed observables, labelled by the field-independent function $\rho$. The change of frame map \eqref{CFM}, instead, maps between two distinct such families and it must be \emph{field-dependent} when the frames are independent fields, in the sense that their relative orientations depend on the global solution. 

In general, the change of frame field $\varphi_{12}(x)$ involves new degrees of freedom, that are not present in the set of observables generated by either $\{A,\Phi_1\}$ or $\{A,\Phi_2\}$ alone. For example, when $\Phi_1$ and $\Phi_2$ are the phases of two extrinsic frames, then $\varphi_{12}(x)$ is some complement-supported gauge-invariant observable for each point $x\in\Gamma$.

However, there is an important situation in which no new degrees of freedom are involved and that is when we change from any extrinsic frame phase $\Phi$ to the intrinsic one $\Tilde{\Phi}$ introduced in section \ref{subsubsec:intrinsic}. We define the variable $\varphi$ to refer specifically to this situation:
\begin{align}\label{eq:Goldstonedefn1}
    \dt\varphi:=A^\dr-\Tilde{A}^\dr\,.
\end{align}
From the dressing condition of the intrinsic frame, $\mathcal{G}[\Tilde{A}^\dr]=\star_{\partial\Sigma}\dt \star_{\partial\Sigma}\Tilde{A}^\dr\big|_{\partial\Sigma}=0$, we can deduce that $\varphi$ is nothing but the exact piece of $A^\dr\big|_{\partial\Sigma}$:
\begin{align}\label{eq:Goldstonedefn2}
    \varphi:=\int_{\partial\Sigma}\left(\Delta^{-1}_{\partial\Sigma}\star_{\partial\Sigma}\dt\star_{\partial\Sigma}\right)\;A^\dr\big|_{\partial\Sigma}\,.
\end{align}
This makes it clear that $\varphi$ is different for the phases $\Phi$ of different extrinsic frames. This gauge-invariant field will play an important role in the (extended) phase space, where it will turn out to behave like a \emph{Goldstone mode} for the subregion frame reorientation symmetries. It will be a key player in our discussions in sections \ref{subsec:extensions}, \ref{sec:Soft_bc} and \ref{sec:gold_soft}. All of the information about how the subregion theory relates to the global theory into which it is embedded will be encoded in this $\varphi$ field and it will be responsible for imprinting the action of the asymptotic soft charges onto the subregion phase space.

\section{Phase Space}
\label{sec:BC}


In the previous section, we focused primarily on using dynamical frames for constructing gauge-invariant observables. We would now like to invoke these results to explore the dynamics of the theory and compare formulations using different types of frames.
To achieve a covariant formulation in spacetime, we employ the covariant phase space formalism.

In the upcoming sections, we will use the frames primarily to bridge finite-distance edge mode physics with asymptotic quantities, by embedding the region of interest into a larger global theory. Nonetheless, in this section we momentarily present the discussion in a manner that remains agnostic to the nature of the region of interest (whether it is the finite subregion or the full global spacetime), thus providing a unified basis for both cases. In particular, this will bring the ``edge'' and ``soft'' structures of the finite subregional theory and the global theory on analogous footings.

To this end, we begin by invoking the gauge-invariant formalism, where the edge frame is incorporated into the symplectic structure. If done with the intrinsic frame, this yields a novel gauge-invariant formulation of the unextended phase space, while if done with the extrinsic frame, it yields the standard gauge-invariant formulation of the extended phase space for gauge theories \cite{Donnelly:2016auv,Geiller:2017xad,Geiller:2019bti,Carrozza:2021gju}. We present an overview of the two and their relation in subsection \ref{subsec:extensions}. In this approach boundary supported gauge transformations are always unphysical. This makes its application to subregion theories particularly useful as it avoids the problem of broken gauge invariance due to the presence of a boundary. In subsection~\ref{subsec:equivalence}, we will, however, also establish the equivalence of the gauge-invariant unextended and extended formulations with their respective gauge-fixed formulations that are sometimes invoked in the literature; the latter are simply particular descriptions of the former.

\subsection{Covariant Phase Space Review}\label{subsec:constr_ph_sp}

The essence of the covariant phase space formalism \cite{Crnkovic:1986ex,Lee:1990nz,Iyer:1994ys,Wald:1999wa,Khavkine:2014kya,Harlow:2019yfa,Gieres:2021ekc} is to construct the phase space directly from on-shell field configurations on spacetime. This leads to a naturally covariant description of the theory. 

In the following, we consider a theory living on a spacetime delimited by a timelike boundary $\Gamma$, similarly to figure \ref{lab}, except that $\Gamma$ could also be asymptotic, in line with our momentary agnostic take (e.g.\ $\Gamma$ could be $i^0$ in asymptotically flat spacetimes or the global boundary of an anti-de Sitter like spacetime).
Any spacelike or null surface $\Sigma$ whose boundary lies on $\Gamma$ acts as a Cauchy surface for the theory.\footnote{When the boundary $\Gamma $ is asymptotic, we must be careful in manipulating radial divergences. More precisely, for the global theory we have to impose certain radial fall-off conditions in order to have physically sensible solutions. More on this in section \ref{sec:sol_space}.}

Let us consider a theory with bulk Lagrangian $L$. We can define a (pre)symplectic potential density through:
\be
\label{eq:var_princ}
\delta L = \cE \delta \Xi +\dt \Theta\,,
\ee
for $\Xi$ the (off-shell) field content of the theory and $\cE \approx 0$ the equations of motion. From the (pre)symplectic potential density $\Theta$, we get the (pre)symplectic current $\omega =\delta \Theta$, where $\d$ denotes the field space exterior derivative. The latter is spacetime exact on-shell, i.e.\ $\dt \omega \approx 0$. This property, combined with Stokes' theorem, allows us to construct a well-defined (pre)symplectic form on any Cauchy slice, provided we impose suitable boundary conditions. Specifically, considering a portion of spacetime between two arbitrary Cauchy slices $\Sigma_1$ and $\Sigma_2$, from  $\dt \omega \approx 0$ we get:
 \be
 \label{eq:stokes}
 \int_{\Sigma_2-\Sigma_1} \omega +\int_{\Gamma_{[12]}} \omega \approx 0\,,
 \ee
where $\Gamma_{[12]}$ is the portion of $\Gamma$ bounded by $\p \Sigma_{1}$ and $\p\Sigma_2$. We refer to the integrals of $\omega$  on the Cauchy slices as the \textit{(pre)symplectic form}, while its integral on the boundary is the \textit{symplectic flux}. If the latter vanishes, conservation of the phase space structure is ensured.

It is important to note that the definition of the (pre)symplectic potential density $\Theta$ in \eqref{eq:var_princ} is subject to a corner ambiguity \cite{Jacobson:1993vj,Harlow:2019yfa,Donnelly:2016auv,Geiller:2019bti,Carrozza:2021gju}, meaning $\Theta$ is defined up to a spacetime exact term. This is reflected in equation \eqref{eq:stokes}, where we can freely add and subtract a corner piece, shared by both $\Gamma$ and the $\Sigma_i$'s:\footnote{We choose here $\Sigma_1$ as the initial slice and $\Sigma_2$ as the final one, $\Gamma$ is oriented outward, so that $\p \cM =$‘‘$\Sigma_2 -\Sigma_1+\Gamma$'', and: \[ \emptyset = \p\p \cM = \p\Sigma_2 -\p\Sigma_1+\p\Gamma\,.\]}
\be
 \label{eq:stokes_corner}
\int_{\Sigma_2-\Sigma_1} \omega +\int_{\Gamma_{[12]}} \omega=\int_{\Sigma_2-\Sigma_1} \omega +  \int_{\p \Sigma_2-\p\Sigma_1} \omega_\p + \int_{\Gamma_{[12]}} [\omega +\dt (\omega_\p)] \approx 0 \,.
 \ee
This freedom to shift by a corner term, $\omega \mapsto \omega +\dt \omega_\p$, modifies the definition of both the symplectic form and the symplectic flux, thus altering their properties. Different choices of corner terms can lead to different physical boundary conditions because they correspond to different symplectic fluxes. 

In gauge theories, however, we can use this freedom to our advantage, by adding a corner piece $\omega_\partial$ which turns the (pre)symplectic current into a gauge-invariant quantity, which we call $\omega^\dr$ \cite{Donnelly:2016auv,Geiller:2019bti,Carrozza:2021gju}. This will then permit us to impose gauge-invariant boundary conditions that eliminate any symplectic flux \cite{Carrozza:2021gju}. It is here that the edge mode enters; usually this is done in the extended phase space construction with the extrinsic edge mode $\Phi$. However, the same construction can be performed with the intrinsic one $\Tilde\Phi$.

Let us see this explicitly for the Maxwell Lagrangian \eqref{eq:Maxwell}:
\be
    L=-\frac{1}{2}F\wedge\star F\,.
\ee
We can easily get the bare presymplectic current:
\be
\omega = \delta A \wedge \delta \star F\,.
\ee
Invoking the extrinsic edge frame, we may now dress $\omega$ on $\Gamma$, obtaining 
as per equation \eqref{eq:A_dr_def} \cite{Carrozza:2021gju}:\footnote{This is consistent despite dressing and exterior derivative $\delta$ a priori not commuting. In \cite{Carrozza:2021gju} the authors refer to the gauge-invariant dressed observables as the \textit{radiative} ($^\text{rad}$) data. Here we will use the dressed notation ($^\dr$) because we will reserve the adjective \textit{radiative} for the data that are not \textit{soft}. See section \ref{sec:gold_soft}. The radiative data is then a subset of the dressed observables.}
\be
\omega^\dr\big|_\Gamma:=U^{-1} \rhd \omega\big|_\Gamma:=\left(\delta A^{\rm dr}\wedge\delta\star F\right)\Big|_\Gamma\,.\label{omegamma}
\ee
Hence, we have 
\be\label{eq:gaugeinvariantflux}
\omega^{\rm dr}\big|_\Gamma\approx\left(\delta A\wedge\delta\star F-\dd(\delta\Phi\delta\star F)\right)\big|_\Gamma=\left(\omega+\dd(\omega_\partial)\right)\big|_\Gamma\,.
\ee
Had we used the intrinsic frame $\Tilde U$ instead, $A^{\rm dr}$ would be replaced by $\Tilde{A}^{\rm dr}$ and $\Phi$ by $\Tilde\Phi$.
As desired, on-shell the dressing acts as the addition of an exact piece to the (pre)symplectic current. 

This naturally leads to the definition of a gauge-invariant (pre)symplectic form:\footnote{We have thus chosen $\omega_\p=-\delta\Phi\delta\star F$, which is unique up to addition of a spacetime closed form $c$; if $c$ is closed but not exact, it will modify the presymplectic form $\Omega_\Sigma$. Here, we adopt the simplest option $c=0$, which is a consistent choice \cite{Carrozza:2021gju}.} 
\be
\label{eq:sympl_dress}
\Omega_\Sigma = \int_\Sigma \delta A \wedge \delta \star F -\int_{\p \Sigma} \delta \rf \,\delta \star F\,.
\ee
Indeed, let $X_{\delta_\lambda}$ denote the field space vector field implementing the gauge transformation \eqref{eq:covariance} (which could even be field-dependent). It is straightforward to check that $X_{\delta_\lambda}\cdot\Omega_\Sigma\approx0$. Using the Cartan magic formula for the field space Lie derivative, $L_{X_{\delta_\lambda}}=X_{\delta_\lambda}\cdot\delta+\delta X_{\delta_\lambda}\cdot$, this implies on-shell gauge invariance:
\be
L_{X_{\delta_\lambda}}\Omega_\Sigma\approx0\,.
\ee
Note that this also holds for gauge transformations with support on $\Gamma$. For this conclusion, we only needed to define the reference frame \emph{on} $\Gamma$, without a need to extend it into the bulk of $\Sigma$. Later, it will at times be convenient, however, to extend the frame inside the subregion.

Our conclusion is also true on the unextended phase space with the intrinsic corner term $\Tilde\omega_\p=-\delta\Tilde\Phi\, \delta\star F$ replacing the extrinsic one $\omega_\p$, which seems to have been overlooked in the literature. We thus see that, in contrast to the suggestive discussion in \cite{Donnelly:2016auv}, one does not \emph{need} to extend the phase space in order to obtain a gauge-invariant formulation (though if one does not, there will be no corner symmetries, as we will see shortly). In particular, this constitutes a manifestly gauge-invariant formulation of the unextended phase space that recovers the one in \cite{Riello:2021lfl} (see also \cite{Riello:2020zbk}). In the latter works, the authors did not add an intrinsic corner term like us; instead they took the standard bulk piece of the presymplectic form and via a Helmholtz decomposition in the bulk split the gauge-invariant part from a non-invariant boundary one. This non-invariant piece was set to zero on `superselection' sectors corresponding to fixed configurations of $\star F$ on $\partial\Sigma$. Our intrinsic corner term simply cancels the non-invariant one from \cite{Riello:2021lfl} so the end results agree.  

Furthermore, from \eqref{eq:stokes_corner}, we infer that:
\be
\Omega_{\Sigma_2}-\Omega_{\Sigma_1} + \int_{\Gamma_{[12]}} \omega^\dr \approx 0\,,
\ee
which entails that conservation of the presymplectic form can be achieved by (necessarily gauge-invariant) boundary conditions that eliminate the symplectic flux $\omega^\dr\big|_\Gamma \, \hat \approx\, 0$,\footnote{It is important to note that while the vanishing of the symplectic flux $\omega^\dr|_\Gamma\,\hat{\approx}\,0$ is a sufficient condition for a well-defined theory, it is not necessary. A weaker condition, $\omega^\dr|_\Gamma\,\hat{\approx}\,\dt \beta$, can be imposed, with the extra term included in the corner part of the symplectic form. For example, this kind of boundary condition turns out necessary in gravity  \cite{Carrozza:2022xut}. 
In addition, coming back to the corner term non-uniqueness of the previous footnote,
modifying the definition by $\omega^\dr \mapsto \omega^\dr +\dt c$, with $\dt c\,\hat \approx\,0 \big |_\Gamma$ but $c \neq 0 \big |_{\p \Sigma}$, would not alter the boundary conditions, but change $\Omega$.  Here, we adopt the simplest option, $c=0$, for each choice of frame.}
  so that $\Omega_{\Sigma_1}\,\hat{\approx}\,\Omega_{\Sigma_2}$ \cite{Carrozza:2021gju}. For this reason, we will henceforth drop the Cauchy slice label $\Sigma$ from the presymplectic form. We will explore different boundary conditions achieving conservation later in section \ref{sec:Soft_bc}.

In addition, $\Omega$ \emph{may}  admit non-degenerate directions corresponding to the frame reorientations, when they exist. We recall from subsection \ref{subsec:fr_reor} that these act only on the frame phase $\rf$ leaving invariant the field configuration within $\Sigma$:%
\be
\label{eq:reorient}
\delta_\rho A =0\,,\q
\delta_\rho \rf = -\rho\,.
\ee
The corresponding generators, for \textit{field-independent} reorientations, are the Hamiltonian integrable charges: 
\be
\label{eq:charge}
\delta Q_\rho = X_{\delta_\rho} \cdot \Omega\,\q\Rightarrow\q Q_\rho =\int_{\p \Sigma} \rho \star F\,.
\ee
We already noted in section \ref{subsec:fr_reor} that such reorientations -- and thus charges -- only exist for extrinsic frames. However, we will see that \emph{also} for extrinsic frames they are not guaranteed and depend on boundary conditions.

\begin{figure}[h!]
    \centering

\tikzset{every picture/.style={line width=0.75pt}} 

\begin{tikzpicture}[x=0.75pt,y=0.75pt,yscale=-1,xscale=1]

\draw    (274.1,250.94) .. controls (296.14,214.75) and (236.18,157.37) .. (265.28,125.59) .. controls (294.38,93.81) and (285.56,77.04) .. (282.04,45.26) ;
\draw    (389.62,250.94) .. controls (394.03,213.87) and (418.72,173.26) .. (402.85,138.83) .. controls (386.98,104.4) and (385.21,85.86) .. (404.61,46.14) ;
\draw   (282.04,45.26) .. controls (282.04,35.5) and (309.48,27.6) .. (343.32,27.6) .. controls (377.17,27.6) and (404.61,35.5) .. (404.61,45.26) .. controls (404.61,55.01) and (377.17,62.91) .. (343.32,62.91) .. controls (309.48,62.91) and (282.04,55.01) .. (282.04,45.26) -- cycle ;
\draw   (274.1,250.94) .. controls (274.1,241.19) and (299.96,233.29) .. (331.86,233.29) .. controls (363.76,233.29) and (389.62,241.19) .. (389.62,250.94) .. controls (389.62,260.7) and (363.76,268.6) .. (331.86,268.6) .. controls (299.96,268.6) and (274.1,260.7) .. (274.1,250.94) -- cycle ;
\draw [color={rgb, 255:red, 74; green, 144; blue, 226 }  ,draw opacity=1 ]   (338.3,148.19) .. controls (373.57,121.7) and (367.58,165.31) .. (402.85,138.83) ;
\draw [color={rgb, 255:red, 155; green, 155; blue, 155 }  ,draw opacity=1 ]   (335.83,111.46) .. controls (371.1,84.98) and (355.23,129.12) .. (390.5,102.64) ;
\draw [color={rgb, 255:red, 155; green, 155; blue, 155 }  ,draw opacity=1 ]   (343.59,173.79) .. controls (382.57,154.72) and (371.99,175.02) .. (408.14,164.43) ;
\draw [color={rgb, 255:red, 155; green, 155; blue, 155 }  ,draw opacity=1 ]   (335.83,198.68) .. controls (369.52,205.22) and (379.92,193.56) .. (404.61,190.03) ;
\draw [color={rgb, 255:red, 155; green, 155; blue, 155 }  ,draw opacity=1 ]   (325.95,96.99) .. controls (361.23,70.5) and (355.23,114.11) .. (390.5,87.63) ;
\draw [color={rgb, 255:red, 208; green, 2; blue, 27 }  ,draw opacity=1 ]   (402.85,138.83) ;
\draw [shift={(402.85,138.83)}, rotate = 0] [color={rgb, 255:red, 208; green, 2; blue, 27 }  ,draw opacity=1 ][fill={rgb, 255:red, 208; green, 2; blue, 27 }  ,fill opacity=1 ][line width=0.75]      (0, 0) circle [x radius= 3.35, y radius= 3.35]   ;
\draw    (338.3,148.19) ;
\draw [shift={(338.3,148.19)}, rotate = 0] [color={rgb, 255:red, 0; green, 0; blue, 0 }  ][fill={rgb, 255:red, 0; green, 0; blue, 0 }  ][line width=0.75]      (0, 0) circle [x radius= 3.35, y radius= 3.35]   ;

\draw (410.25,127.05) node [anchor=north west][inner sep=0.75pt]  [color={rgb, 255:red, 208; green, 2; blue, 27 }  ,opacity=1 ]  {$\Phi ( x_{0})$};
\draw (323.48,242.7) node [anchor=north west][inner sep=0.75pt]    {$\Sigma _{i}$};
\draw (332.12,36.13) node [anchor=north west][inner sep=0.75pt]    {$\Sigma _{f}$};
\draw (238.29,137.76) node [anchor=north west][inner sep=0.75pt]    {$\Gamma $};
\draw (300.49,145.71) node [anchor=north west][inner sep=0.75pt]    {$\Phi ( x)$};
\draw (337.83,115.86) node [anchor=north west][inner sep=0.75pt]    {$\textcolor[rgb]{0.29,0.56,0.89}{W}\textcolor[rgb]{0.29,0.56,0.89}{_{\gamma _{x}}}\textcolor[rgb]{0.29,0.56,0.89}{[}\textcolor[rgb]{0.29,0.56,0.89}{A}\textcolor[rgb]{0.29,0.56,0.89}{]}$};

\end{tikzpicture}

    \caption{ Extension of the edge mode into the bulk of the subregion. It is defined via a congruence of Wilson lines anchored on $\Gamma$ via: $\Phi(x):=-i\ln W_{\gamma_x}[A]+\Phi(x_0)$, where $W_{\gamma_x}[A]$ is a Wilson line along the curve $\gamma_x$ mapping the point $x$ to $x_0$ on $\Gamma$.}
    \label{edge extension}
\end{figure}
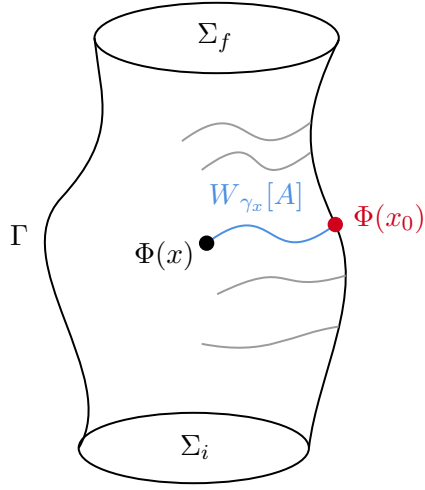

In order to turn the presymplectic form \eqref{eq:sympl_dress} into a non-degenerate symplectic one, resulting in the physical phase space, we have to mod out its degenerate directions $X_{\delta_\lambda}$ from the space of solutions \cite{Lee:1990nz}. In practice, this can be done either by applying a suitable gauge-fixing prescription inside the region, or, equivalently, extending the frame into the bulk to write $\Omega$ in a manifestly gauge-invariant expression that can be evaluated on the space of orbits.  
To illustrate the latter point, we can, for example, ``shoot in'' Wilson lines from $\Gamma$ inwards, as shown in figure \ref{edge extension}, and invoke the obvious extension of \eqref{eq:A_dr_def}. This can be done for \emph{both} extrinsic and intrinsic edge mode frames. The symplectic form then reads:
\be\label{physomega}
    \Omega\,\approx\int_{\Sigma}\delta A^\dr\wedge\delta\star F\,.
\ee

However, it is important to recognize that such an extension merely provides a particular description of the subregion theory, by picking a particular label, $A^\dr\big|_\Sigma$, for the gauge orbits. It is the boundary edge mode living on $\Gamma$ that is integral to the definition of the subregion theory.

\subsection{Phase Space (Un)Extensions}\label{subsec:extensions}

The discussion above about the freedom to add a corner piece to the symplectic form was agnostic about the nature of the edge mode $\Phi/\Tilde\Phi$. Irrespective of this, the outcome of manifest boundary-anchored gauge invariance is achieved. However, extrinsic and intrinsic edge modes lead to quite different phase spaces. 

Here, we consider only the kinematical setting of a spatial subregion $\Sigma$ with boundary $\p\Sigma$ on a global Cauchy slice, as done in much of the literature on edge modes and as appropriate for example for exploring the phase space associated with a causal diamond. This means that here it will not be necessary to impose any boundary conditions. We will transition to the dynamical setting of a timelike tube with boundary $\Gamma$, where boundary conditions are needed, from the next section onward.

\paragraph{Intrinsic Edge Mode:}

The edge mode $\Tilde{\Phi}$ is built from degrees of freedom intrinsic to $\p\Sigma$. The addition of its corresponding corner piece to the presymplectic form (call it $\Omega_\text{int}$) does not involve an actual phase space extension, but rather a convenient re-definition of variables which allows to work directly with gauge-invariant quantities.

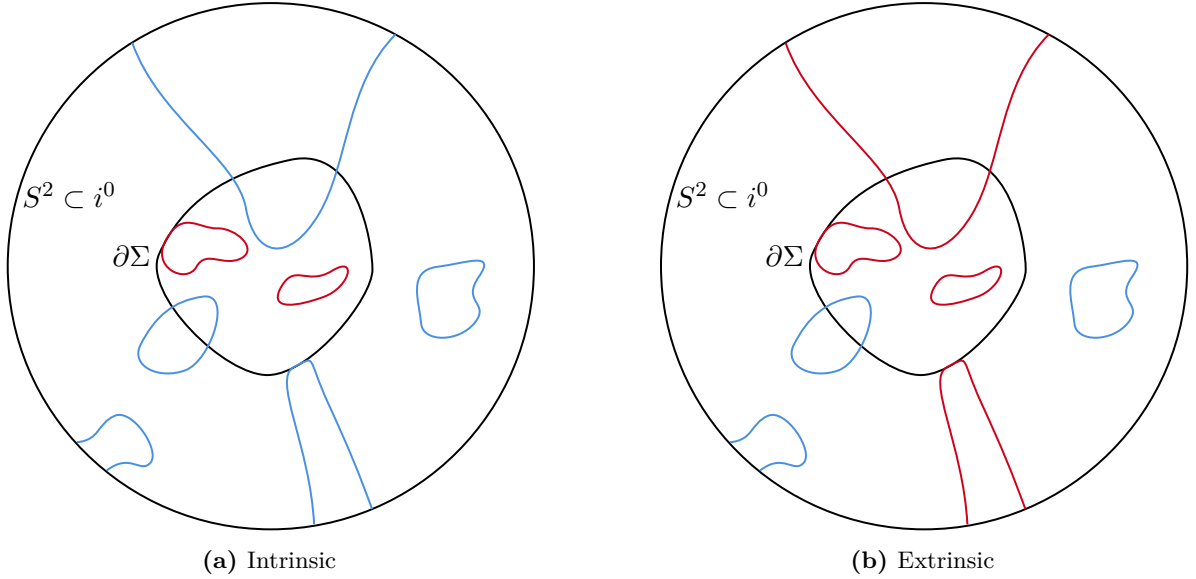
\begin{figure}[h!]
    \centering
    \begin{subfigure}[t]{0.5\textwidth} 
        \centering

\tikzset{every picture/.style={line width=0.75pt}} 

\begin{tikzpicture}[x=0.75pt,y=0.75pt,yscale=-1,xscale=1]

\draw   (209,156) .. controls (209,83.1) and (268.1,24) .. (341,24) .. controls (413.9,24) and (473,83.1) .. (473,156) .. controls (473,228.9) and (413.9,288) .. (341,288) .. controls (268.1,288) and (209,228.9) .. (209,156) -- cycle ;
\draw    (284.97,149.81) .. controls (296.95,123.62) and (318.07,109.59) .. (351.07,102.59) .. controls (384.07,95.59) and (391.97,143.81) .. (391.97,157.81) .. controls (391.97,171.81) and (362.07,209.59) .. (340.07,210.59) .. controls (318.07,211.59) and (275.95,169.62) .. (284.97,149.81) -- cycle ;
\draw [color={rgb, 255:red, 74; green, 144; blue, 226 }  ,draw opacity=1 ]   (271.49,43.87) .. controls (284.49,80.87) and (324.16,102.29) .. (328.49,126.87) .. controls (332.81,151.44) and (350,155) .. (365,130) .. controls (380,105) and (378.62,64.67) .. (403.62,39.67) ;
\draw [color={rgb, 255:red, 208; green, 2; blue, 27 }  ,draw opacity=1 ]   (345.43,167.85) .. controls (352.67,153.33) and (358.67,166.33) .. (371.67,158.33) .. controls (384.67,150.33) and (380.71,166.54) .. (367.67,171.33) .. controls (354.63,176.13) and (340.67,178.33) .. (345.43,167.85) -- cycle ;
\draw [color={rgb, 255:red, 74; green, 144; blue, 226 }  ,draw opacity=1 ]   (275.43,195.85) .. controls (282.67,181.33) and (292.05,173.23) .. (307.05,171.23) .. controls (322.05,169.23) and (311.09,203.44) .. (298.05,208.23) .. controls (285.01,213.02) and (270.67,206.33) .. (275.43,195.85) -- cycle ;
\draw [color={rgb, 255:red, 74; green, 144; blue, 226 }  ,draw opacity=1 ]   (362.7,285.49) .. controls (360.7,245.49) and (341.16,212.3) .. (352.7,206.49) .. controls (364.24,200.69) and (358.82,200.87) .. (369.82,224.87) .. controls (380.82,248.87) and (384.82,258.87) .. (391.82,277.87) ;
\draw [color={rgb, 255:red, 208; green, 2; blue, 27 }  ,draw opacity=1 ]   (288.67,142.33) .. controls (297,127.67) and (305,137.33) .. (314.17,137.01) .. controls (323.34,136.69) and (335,147.33) .. (326.17,152.01) .. controls (317.34,156.69) and (308.67,147.33) .. (303.67,156.33) .. controls (298.67,165.33) and (280,156.67) .. (288.67,142.33) -- cycle ;
\draw [color={rgb, 255:red, 74; green, 144; blue, 226 }  ,draw opacity=1 ]   (416.43,184.85) .. controls (415.33,173.33) and (409.33,158.33) .. (424.33,156.33) .. controls (439.33,154.33) and (454.33,148.33) .. (445.33,161.33) .. controls (436.33,174.33) and (451.33,174.33) .. (443.33,184.33) .. controls (435.33,194.33) and (417.33,194.33) .. (416.43,184.85) -- cycle ;
\draw [color={rgb, 255:red, 74; green, 144; blue, 226 }  ,draw opacity=1 ]   (243.67,244.33) .. controls (255.62,243.07) and (253.62,235.07) .. (262.62,231.07) .. controls (271.62,227.07) and (285.69,248.48) .. (280.69,254.48) .. controls (275.69,260.48) and (270.65,249.2) .. (258.69,258.48) ;

\draw (260,144.4) node [anchor=north west][inner sep=0.75pt]    {$\p \Sigma $};
\draw (215,112.4) node [anchor=north west][inner sep=0.75pt]    {$S^2 \subset i^0$};

\end{tikzpicture}
        \caption{Intrinsic}\label{fig:intrinsic}
    \end{subfigure}%
    ~ 
    \begin{subfigure}[t]{0.5\textwidth} 
        \centering

\tikzset{every picture/.style={line width=0.75pt}} 

\begin{tikzpicture}[x=0.75pt,y=0.75pt,yscale=-1,xscale=1]

\draw   (209,153) .. controls (209,80.1) and (268.1,21) .. (341,21) .. controls (413.9,21) and (473,80.1) .. (473,153) .. controls (473,225.9) and (413.9,285) .. (341,285) .. controls (268.1,285) and (209,225.9) .. (209,153) -- cycle ;
\draw    (284.97,146.81) .. controls (296.95,120.62) and (318.07,106.59) .. (351.07,99.59) .. controls (384.07,92.59) and (391.97,140.81) .. (391.97,154.81) .. controls (391.97,168.81) and (362.07,206.59) .. (340.07,207.59) .. controls (318.07,208.59) and (275.95,166.62) .. (284.97,146.81) -- cycle ;
\draw [color={rgb, 255:red, 208; green, 2; blue, 27 }  ,draw opacity=1 ]   (271.49,40.87) .. controls (284.49,77.87) and (324.16,99.29) .. (328.49,123.87) .. controls (332.81,148.44) and (350,152) .. (365,127) .. controls (380,102) and (378.62,61.67) .. (403.62,36.67) ;
\draw [color={rgb, 255:red, 208; green, 2; blue, 27 }  ,draw opacity=1 ]   (345.43,164.85) .. controls (352.67,150.33) and (358.67,163.33) .. (371.67,155.33) .. controls (384.67,147.33) and (380.71,163.54) .. (367.67,168.33) .. controls (354.63,173.13) and (340.67,175.33) .. (345.43,164.85) -- cycle ;
\draw [color={rgb, 255:red, 74; green, 144; blue, 226 }  ,draw opacity=1 ]   (275.43,192.85) .. controls (282.67,178.33) and (292.05,170.23) .. (307.05,168.23) .. controls (322.05,166.23) and (311.09,200.44) .. (298.05,205.23) .. controls (285.01,210.02) and (270.67,203.33) .. (275.43,192.85) -- cycle ;
\draw [color={rgb, 255:red, 208; green, 2; blue, 27 }  ,draw opacity=1 ]   (362.7,282.49) .. controls (360.7,242.49) and (341.16,209.3) .. (352.7,203.49) .. controls (364.24,197.69) and (358.82,197.87) .. (369.82,221.87) .. controls (380.82,245.87) and (384.82,255.87) .. (391.82,274.87) ;
\draw [color={rgb, 255:red, 208; green, 2; blue, 27 }  ,draw opacity=1 ]   (288.67,139.33) .. controls (297,124.67) and (305,134.33) .. (314.17,134.01) .. controls (323.34,133.69) and (335,144.33) .. (326.17,149.01) .. controls (317.34,153.69) and (308.67,144.33) .. (303.67,153.33) .. controls (298.67,162.33) and (280,153.67) .. (288.67,139.33) -- cycle ;
\draw [color={rgb, 255:red, 74; green, 144; blue, 226 }  ,draw opacity=1 ]   (416.43,181.85) .. controls (415.33,170.33) and (409.33,155.33) .. (424.33,153.33) .. controls (439.33,151.33) and (454.33,145.33) .. (445.33,158.33) .. controls (436.33,171.33) and (451.33,171.33) .. (443.33,181.33) .. controls (435.33,191.33) and (417.33,191.33) .. (416.43,181.85) -- cycle ;
\draw [color={rgb, 255:red, 74; green, 144; blue, 226 }  ,draw opacity=1 ]   (243.67,241.33) .. controls (255.62,240.07) and (253.62,232.07) .. (262.62,228.07) .. controls (271.62,224.07) and (285.69,245.48) .. (280.69,251.48) .. controls (275.69,257.48) and (270.65,246.2) .. (258.69,255.48) ;

\draw (260,141.4) node [anchor=north west][inner sep=0.75pt]    {$\p \Sigma $};
\draw (215,109.4) node [anchor=north west][inner sep=0.75pt]    {$S^2 \subset i^0$};

\end{tikzpicture}
        \caption{Extrinsic}\label{fig:extrinsic}
    \end{subfigure}
    \caption{Examples of observables for the two classes of edge modes. In red we show observables which are present in the respective phase spaces. In blue we show observables which are not present in either. The extrinsic edge mode adds some complement-supported observables to the intrinsic phase space. However, it itself misses many of the observables of the global theory. }\label{fig:observables}
\end{figure}

The gauge-invariant observables present in this \emph{unextended} phase space include, in addition to the usual local observables, subregion-supported non-local observables of the type shown in figure \ref{fig:intrinsic}. This phase space knows \emph{nothing} about the complement region.

A consequence of this fact is the lack of corner symmetries in this subregion theory. In particular, the lack of frame reorientations (as explained in section \ref{subsec:fr_reor}) means that there are no  boundary-supported, integrable and non-degenerate directions of $\Omega_\text{int}$. Physically, this comes about because this theory has no information about its relationship to the global theory into which it is embedded. 

\paragraph{Extrinsic Edge Mode:}

In the case of the edge mode $\Phi$ built from Wilson lines in the complement, the addition of the corner piece to the presymplectic form (call it $\Omega_\text{ext}$) signifies an actual extension of the phase space. 

This \emph{extended} phase space includes, in addition to the ones above, \emph{some} gauge-invariant non-local observables with support \emph{across}  $\p\Sigma$, in both the subregion and the complement, as shown in figure \ref{fig:extrinsic}. This phase space knows \emph{something} about the complement region.

Now there \emph{will} be physical symmetries for this kinematical subregion theory, namely the frame reorientations of  section \ref{subsec:fr_reor}, and these will carry over to the dynamical setting with a timelike boundary $\Gamma$ later, provided they survive the imposition of boundary conditions, with which they have to be compatible, cf.\ section \ref{sec:Soft_bc}. The corresponding charges are those of \eqref{eq:charge}. This contrast with the intrinsic case above highlights the relational character of symmetries themselves. The symmetry transformations of this subregion theory are precisely those that act on the variables that encode the relationship between it and the global theory into which it is embedded. Corner symmetries thus correspond to different relations between the region and its complement.

The crucial point is this: in  works on subregions in gauge theories such as \cite{Donnelly:2016auv,Geiller:2017xad,Geiller:2019bti,Freidel:2023bnj}, the edge mode is added to the phase space in order to have manifest gauge-invariance at the boundary. But without knowledge of how the edge mode is constructed we cannot tell what symmetries we have for the subregion. In the above, we go from no corner symmetries to possibly infinitely many simply by changing from intrinsic to extrinsic edge modes. This highlights the importance of understanding the edge mode physically, as described in \cite{Carrozza:2021gju, Carrozza:2022xut} and here, and not as a mere mathematical convenience.

\subsubsection{Extrinsic $=$ Intrinsic $\cross$ Goldstone}\label{subsubsec:splitting}

So far, we have discussed the \emph{pre}symplectic structures. It is instructive to analyze what happens once we quotient by degenerate directions. A neat gauge-invariant factorization structure emerges.\footnote{This is similar in spirit to the recent proposal in \cite{Ball:2024hqe}, obtained in a gauge-fixed description. However, through our manifestly gauge-invariant approach, the physical interpretation advocated for here is new. We will explain the similarities and differences in detail in section \ref{subsec:equivalence}.} 

To eliminate all of the degenerate directions, we extend the boundary reference frames into the bulk of the subregion, e.g. via Wilson lines as shown in figure \ref{edge extension}. Different choices of extension correspond to different choices of equivalence class representatives. The theory is the same for every choice: the resulting physical phase space encodes the same set of gauge-invariant observables. For both the extended and unextended phase space, i.e.\ both for extending the intrinsic or extrinsic edge frame into the bulk, the symplectic form is then of the form \eqref{physomega} (for the intrinsic case we simply replace $A^\dr$ by $\Tilde{A} ^\dr$). Elimination of the degenerate directions corresponding to small gauge transformations is then straightforward, as they leave all variables above invariant.

In section \ref{subsec:framechange} we discussed the field-dependent transformation that maps between the extrinsic frame $\Phi$ and the intrinsic frame $\Tilde{\Phi}$, via equations \eqref{eq:Goldstonedefn1} and \eqref{eq:Goldstonedefn2}, which led us to define a new variable $\varphi$: the Goldstone mode. This leads to the following on-shell relationship between extrinsic and intrinsic phase spaces:\footnote{Leaving aside for now the conceptual differences, explained in section \ref{subsec:equivalence}, this is the same splitting of the symplectic form as in \cite{Riello:2021lfl,Ball:2024hqe}, with two differences. Firstly, their gauge-fixing condition is related to a ``Coulomb-like'' dressing condition for $A^\dr$ inside the subregion, rather than the ``radial-like'' choice we made here.  The other difference is that there the authors perform a splitting of $\star F\big|_\Sigma$ as well. This is because, in actuality, only a piece of $\star F\big|_\Sigma$ contributes to the bulk piece, with the other yielding a manifestly zero contribution. Here, we do not address this technicality, as it does not change the main message, but an interested reader can always refer to \cite{Riello:2021lfl,Ball:2024hqe}.}
\bsub
\be
     \Omega_\text{ext}&\approx\Omega_\text{int}+\Omega_\text{Gold}\,,\\
     \Omega_\text{Gold}&:=\int_{\partial\Sigma}\delta\varphi\;\delta\star F\,,\\
     \Omega_\text{ext}&\approx \int_\Sigma\delta\Tilde{A}^\dr\wedge\delta\star F+ \int_{\partial\Sigma}\delta\varphi\;\delta\star F\,.
\ee
\esub
In other words, the extended phase space \emph{factorizes} into a bulk (intrinsic) and a corner (Goldstone) piece. This is the equation underpinning figure \ref{fig:observables}. The unextended phase space ($\Omega_\text{int}$) only knows about intrinsic observables. The extension of the phase space (to $\Omega_\text{ext}$) via certain complement observables \emph{always} corresponds to the addition of a gauge-invariant \emph{Goldstone} symplectic pair ($\varphi,\star F\big|_{\partial\Sigma}$), which lives entirely on the corner. The particular Goldstone mode $\varphi$ depends on the extrinsic frame $\Phi$ in question. 

We noted before that, in order to have corner symmetries, we must extend the phase space by ``sufficiently many'' cross-boundary observables to record information about the relation between the subregion and its complement. A single extrinsic frame only extends the phase space by a subset of all the observables of the global theory. We could, of course, include more shared observables by extending the phase space further via the addition of more gauge-invariant corner terms with new Goldstone modes corresponding to the further extrinsic edge modes. We do not take this route here because our choice of extrinsic frame will already imprint the full asymptotic symmetry algebra into the subregion theory.

We note that the (gauge-invariant) unextended, intrinsic phase space does \emph{not} have a corner piece.  Of course, we are always allowed to write the presymplectic structure as in \eqref{eq:sympl_dress} with the explicit corner term involving $\Tilde{\Phi}$ and $\star F\big|_{\partial\Sigma}$. However, the point is that, once we quotient by the gauge orbits, there is no gauge-invariant corner sector in the symplectic structure and $\star F\big|_{\partial\Sigma}$ drops out of the phase space.\footnote{A similar conclusion is reached in \cite{Riello:2020zbk,Riello:2021lfl}, where it is argued that $\star F\big|_{\p\Sigma}$ must be  ``superselected''. There, the starting point is the non-invariant presymplectic form \emph{prior} to the addition of any corner edge piece. It thus corresponds to an unextended phase space. The authors go on to argue that in order to preserve boundary gauge invariance, one must \emph{by hand} eliminate $\star F\big|_{\p\Sigma}$ from the phase space via superselection. The end result is a symplectic form equivalent to our $\Omega_\text{int}$ above, which we directly obtain from a manifestly gauge-invariant approach.} Even though there is some ambiguity in the symplectic conjugate to $\star F\big|_{\partial\Sigma}$ at the level of the pre-phase space, this ambiguity completely disappears at the level of the phase space: $\star F\big|_{\partial\Sigma}$ only appears in the extended phase space and it is always conjugate to the Goldstone mode $\varphi$.

\subsection{Equivalence of Gauge-fixed and Gauge-invariant Phase Spaces}\label{subsec:equivalence}

Suppose for a moment that $\Gamma$ is a finite subregion boundary. Gauge transformations supported on $\Gamma$, but without support on the asymptotic boundary, are unphysical. To implement this fact there are two standard approaches:
\begin{enumerate}
    \item \underline{Gauge-fixing}: We take a representative of the gauge orbit on $\Gamma$ by imposing a gauge-fixing condition on the boundary. This leads to a gauge-fixed phase space in which the degenerate directions associated with boundary gauge transformations have been eliminated. This is implicitly used in e.g.\ \cite{Harlow:2019yfa,  Ball:2024hqe,Blommaert:2018oue,Bub:2024nan}, as we will explain shortly.

    \item \underline{Dressing}: We dress the bare connection to an edge mode field on $\Gamma$ and work with manifestly boundary gauge-invariant quantities, like $A^\dr\big|_\Gamma$, as we have done in this paper. The phase space thus obtained is defined on the quotient space of orbits of boundary gauge transformations. This is used in e.g.\ \cite{Donnelly:2016auv,Geiller:2019bti,Carrozza:2021gju,Carrozza:2022xut}.
\end{enumerate}
The point we want to emphasize here is that the first approach is simply a choice of description of the second. They are, therefore, equivalent. This is true for both unextended and extended phase spaces. While this should not be surprising, this observation brings clarity into the relation of different approaches pursued in the literature and their respective treatment of boundary symmetries. In fact, the same conclusion could be applied when $\Gamma$ is an asymptotic boundary, in which case the standard formulation, in which large gauge transformations are physical, would arise as the gauge-fixed description of an asymptotically extended phase space where they are not.

The equivalence notwithstanding, a lot of insight is gained by working at the gauge-invariant/dressing level, since the understanding of which transformations are physical or not is  more transparent. Section \ref{subsec:extensions} also serves as a clear example of the power of this framework, as it makes relational structures between the subregion and the complement explicit. We now see how these are manifested in a gauge-fixed description, where the physical interpretation is somewhat more obscure, unless we inherit it from the gauge-invariant level.

We emphasize that, here, we are focused on the equivalence with respect to \emph{boundary} gauge transformations. Showing the equivalence with respect to bulk gauge transformations proceeds in an obvious analogous manner. For the unextended case, the construction of a gauge-invariant phase space by directly quotienting out \emph{all} gauge transformations, including those on $\Gamma$, is discussed in \cite{Gomes:2016mwl,Gomes:2018dxs,Riello:2021lfl,Riello:2020zbk} without a specific reference to edge modes. As in the previous subsection, we focus momentarily on the phase spaces associated with a \emph{spatial} subregion, e.g.\ a causal diamond, for which we do not need to impose any boundary conditions unlike in the dynamical setting of the timelike tube which we explore from the next subsection onward.

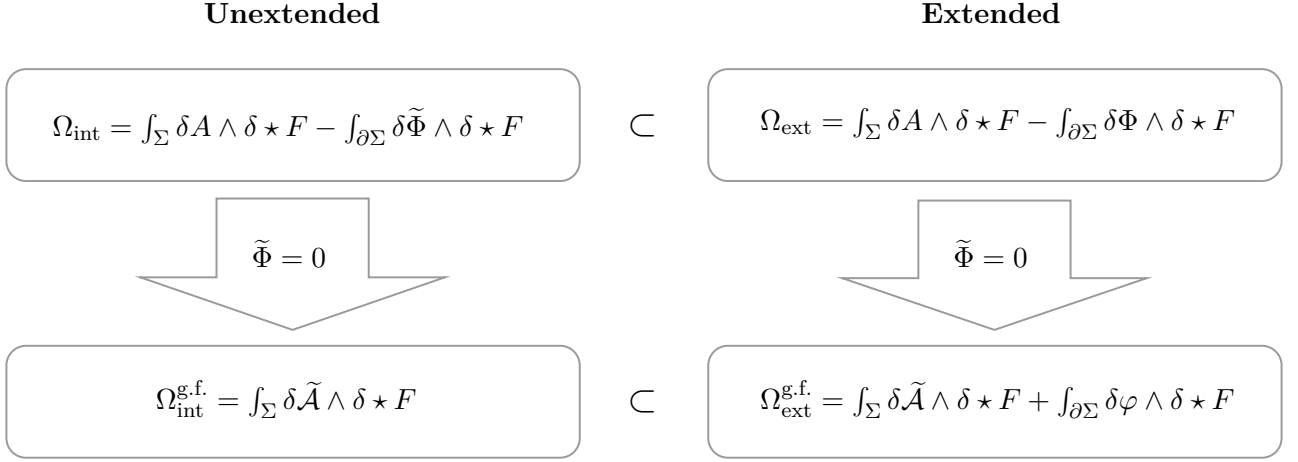
\begin{figure}[h!]
    \centering

\tikzset{every picture/.style={line width=0.75pt}} 

\begin{tikzpicture}[x=0.75pt,y=0.75pt,yscale=-1,xscale=1]

\draw  [color={rgb, 255:red, 155; green, 155; blue, 155 }  ,draw opacity=1 ] (5.78,68.78) .. controls (5.78,63.36) and (10.18,58.96) .. (15.6,58.96) -- (283.58,58.96) .. controls (289,58.96) and (293.4,63.36) .. (293.4,68.78) -- (293.4,105.27) .. controls (293.4,110.69) and (289,115.09) .. (283.58,115.09) -- (15.6,115.09) .. controls (10.18,115.09) and (5.78,110.69) .. (5.78,105.27) -- cycle ;

\draw  [color={rgb, 255:red, 155; green, 155; blue, 155 }  ,draw opacity=1 ] (5.78,207.58) .. controls (5.78,202.15) and (10.18,197.75) .. (15.6,197.75) -- (283.58,197.75) .. controls (289,197.75) and (293.4,202.15) .. (293.4,207.58) -- (293.4,244.06) .. controls (293.4,249.49) and (289,253.89) .. (283.58,253.89) -- (15.6,253.89) .. controls (10.18,253.89) and (5.78,249.49) .. (5.78,244.06) -- cycle ;

\draw  [color={rgb, 255:red, 155; green, 155; blue, 155 }  ,draw opacity=1 ] (357.88,207.58) .. controls (357.88,202.15) and (362.28,197.75) .. (367.7,197.75) -- (635.68,197.75) .. controls (641.1,197.75) and (645.5,202.15) .. (645.5,207.58) -- (645.5,244.06) .. controls (645.5,249.49) and (641.1,253.89) .. (635.68,253.89) -- (367.7,253.89) .. controls (362.28,253.89) and (357.88,249.49) .. (357.88,244.06) -- cycle ;

\draw  [color={rgb, 255:red, 155; green, 155; blue, 155 }  ,draw opacity=1 ] (73,163.41) -- (111.18,163.41) -- (111.18,123.87) -- (187.54,123.87) -- (187.54,163.41) -- (225.73,163.41) -- (149.36,189.77) -- cycle ;

\draw  [color={rgb, 255:red, 155; green, 155; blue, 155 }  ,draw opacity=1 ] (357.88,68.78) .. controls (357.88,63.36) and (362.28,58.96) .. (367.7,58.96) -- (635.68,58.96) .. controls (641.1,58.96) and (645.5,63.36) .. (645.5,68.78) -- (645.5,105.27) .. controls (645.5,110.69) and (641.1,115.09) .. (635.68,115.09) -- (367.7,115.09) .. controls (362.28,115.09) and (357.88,110.69) .. (357.88,105.27) -- cycle ;

\draw  [color={rgb, 255:red, 155; green, 155; blue, 155 }  ,draw opacity=1 ] (425.1,163.82) -- (463.28,163.82) -- (463.28,124.89) -- (539.64,124.89) -- (539.64,163.82) -- (577.82,163.82) -- (501.46,189.77) -- cycle ;

\draw (479.42,143.01) node [anchor=north west][inner sep=0.75pt]    {$\Tilde{\Phi } =0$};
\draw (382.5,213.67) node [anchor=north west][inner sep=0.75pt]    {$\Omega _\text{ext}^\text{g.f.} =\int _{\Sigma } \delta \Tilde{\mathcal{A}} \wedge \delta \star F+\int _{\partial \Sigma } \delta \varphi \wedge \delta \star F$};
\draw (382.5,76.98) node [anchor=north west][inner sep=0.75pt]    {$\Omega _\text{ext} =\int _{\Sigma } \delta A \wedge \delta \star F-\int _{\partial \Sigma } \delta \Phi \wedge \delta \star F$};
\draw (127.32,143.01) node [anchor=north west][inner sep=0.75pt]    {$\Tilde{\Phi } =0$};
\draw (79.28,213.67) node [anchor=north west][inner sep=0.75pt]    {$\Omega _\text{int}^\text{g.f.} =\int _{\Sigma } \delta \Tilde{\mathcal{A}} \wedge \delta \star F$};
\draw (27.4,76.98) node [anchor=north west][inner sep=0.75pt]    {$\Omega _\text{int} =\int _{\Sigma } \delta A \wedge \delta \star F-\int _{\partial \Sigma } \delta \Tilde{\Phi } \wedge \delta \star F$};
\draw (315.91,218.67) node [anchor=north west][inner sep=0.75pt]  [font=\Large]  {$\subset $};
\draw (103.32,24.18) node [anchor=north west][inner sep=0.75pt]   [align=left] {\textbf{Unextended}};
\draw (463.42,24.18) node [anchor=north west][inner sep=0.75pt]   [align=left] {\textbf{Extended}};
\draw (315.91,79.98) node [anchor=north west][inner sep=0.75pt]  [font=\Large]  {$\subset $};

\end{tikzpicture}
    \caption{Schematic diagram illustrating the equivalence between the two approaches. The top row corresponds to the \emph{boundary gauge-invariant} (un)extended phase spaces. The bottom row corresponds to the \emph{boundary gauge-fixed} phase spaces. We use the same boundary gauge-fixing condition $\Tilde{\Phi}=0$ for both unextended and extended phase spaces. Here, $\Tilde{\mathcal{A}}$ refers to the boundary gauge-fixed bare connection, for this choice of gauge-fixing condition. We see that the factorization structure survives in the gauge-fixed description as well. Any gauge-fixed phase space is just one possible description of the gauge-invariant phase space.}
    \label{fig:g.f. dressed}
\end{figure}

Figure \ref{fig:g.f. dressed} summarizes how the equivalence works.  We start at the gauge-invariant level in the top row. We then gauge-fix on  $\p\Sigma$. To do so, we have to choose an appropriate gauge-fixing functional on the boundary fields. Since both the intrinsic ($\tilde \Phi$) and extrinsic ($\Phi$) frame are complete and parameterize the gauge orbit faithfully, any such boundary gauge-fixing can be interpreted as a condition on one of them.  
A  particularly simple example for both the extended and unextended phase spaces is the gauge-fixing condition on the exact piece of $A|_{\p\Sigma}$: $\Tilde{\Phi}=0$. 
We use the symbol $\Tilde{\mathcal{A}}$ to denote the bare connection restricted to this (boundary) gauge-fixing surface.

Upon pullback onto the \emph{same} gauge-fixing surface, the symplectic forms are shown in the bottom row of figure \ref{fig:g.f. dressed}. The fixing of the exact piece of $\tilde{\mathcal{A}}$, along with the lack of a corner term in the gauge-fixed intrinsic phase space is consistent with there being no symmetries.

In contrast, we see that the gauge-fixed extrinsic phase space includes an additional corner Goldstone piece. This follows because on the gauge-fixing surface we have that $\Phi^\text{g.f.}=\Tilde{\Phi}^\text{g.f.}-\varphi=-\varphi$, where we used that $\varphi$ is a gauge-invariant field. We thus \emph{do have symmetries}, corresponding to shifts of the Goldstone field: $\varphi\mapsto\varphi+\rho$. This is inherited from the frame reorientations acting as symmetries at the gauge-invariant level, and the charges take the same form as in \eqref{eq:charge}.

It is also clear from figure \ref{fig:g.f. dressed} that the inclusion of the unextended phase space inside the extended phase space survives cleanly at the gauge-fixed level. This continues to hold even if we use different gauge-fixing conditions for the two phase spaces.

As an example, we could use a different, equally good, gauge-fixing condition for the extended phase space, namely $\Phi=0$. This also fixes the boundary gauge completely, because $\Phi$ is also the phase of a complete reference frame. On this gauge-fixing surface we have that $\Tilde{\Phi}^\text{g.f.}=\varphi$. We use the symbol $\mathcal{A}$ for the bare connection restricted to this (boundary) gauge-fixing surface. The symplectic form then reads:
\be
\Omega^\text{g.f.}_\text{ext}=\int_\Sigma\delta\mathcal{A}\wedge\delta\star F\, . \label{eq:g.f. ext}
\ee
which looks similar to $\Omega^\text{g.f.}_\text{int}$ in the gauge $\Tilde{\Phi}=0$ in figure~\ref{fig:g.f. dressed}. The difference here is that $\mathcal{A}$, including its exact piece, is now \emph{not} fixed in this gauge because $\Phi$ is the extrinsic frame phase built from complementary data. Thus, there exists a non-degenerate direction in phase space which performs the following ``gauge-looking'' transformation on the  connection in this gauge:
\be
    \mathcal{A}\mapsto \mathcal{A} + \dt \rho \, .\label{eq:g.f. frame reorientations}
\ee
This is again achieved by a shift of the Goldstone field: $\varphi\mapsto\varphi+\rho$. It is thus orthogonal to the gauge orbit and it corresponds to a physical symmetry. Of course, the conclusion about the symmetry structure of the extended phase space is therefore consistent for both (equally good) choices of boundary gauge-fixing conditions: $\Tilde{\Phi}=0$ and $\Phi=0$.

This discussion should help to clarify  misleading statements in the literature pertaining to the question of whether or not boundary-supported gauge transformations are to be thought of as ``physical''. We see that  confusion in this regard may arise at the gauge-fixed level if not treated carefully; from the gauge-invariant level it is clear that the answer is  no.

Let us briefly explain why the approach based on the widespread view that ``boundaries break gauge invariance'' nevertheless yields technically correct results that are equivalent with our manifestly gauge-invariant approach, yet misses some of the latter's physical insights. 

On the one hand, we have the extended phase space, given by $\Omega_\text{ext}$ after quotienting by degenerate directions, or, equivalently, by \eqref{eq:g.f. ext} in the gauge-fixed language, where boundary gauge transformations are treated as unphysical from the start. On the other hand, in parts of the literature, e.g.  \cite{Harlow:2019yfa,Ball:2024hqe,Chen:2024kuq,Bub:2024nan,Canfora:2024awy}, one finds the phase space obtained by treating boundary-supported gauge transformations as physical, \emph{without} extension via complement-supported degrees of freedom. Such a symplectic form would look like:\footnote{The subscript $\text{gauge}\to \text{``phys''}$ is supposed to denote the ``upgrade'' of boundary gauge transformations to physical transformations.}
\be
    \Omega_{\text{gauge}\to \text{``phys''}} = \int_\Sigma \delta A\wedge \delta\star F\,, \label{eq:gauge turned phys}
\ee
where we use the symbol $A$ for the un-gauge-fixed bare connection to remind the reader that no boundary gauge-fixing is being performed in this approach. Although this phase space looks intrinsic, since no extra degrees of freedom have been added, it is, in fact, strictly larger than the one corresponding to $\Omega_\text{int}$ in figure \ref{fig:g.f. dressed}, because transformations that were degenerate there (and are hence to be quotiented out) are here treated as physical.

Inspecting \eqref{eq:g.f. ext} and \eqref{eq:gauge turned phys} one sees that actually:
\be
    \Omega_{\text{gauge}\to \text{``phys''}}=\Omega_\text{ext}^\text{g.f.}\,,\,\label{eq:confusion}
\ee
meaning that this phase space should be understood rather as an \emph{extended} phase space. The equality is made manifest in the gauge $\Phi=0$, but it holds at the level of the gauge-invariant extended phase space. Then, we can conclude that transformations of the form $A\mapsto A + \dt \lambda$, $\lambda\big|_\Gamma\neq0$, which are the non-degenerate directions of $\eqref{eq:gauge turned phys}$, should be thought of as gauge-fixed incarnations of frame reorientations of cross-boundary data, as explained around \eqref{eq:g.f. frame reorientations} above, and \emph{not} as boundary-supported gauge transformations.

Taking this latter approach does, however, miss out on some important physics. In particular, the knowledge of which cross-boundary data we are extending the subregional phase space by is completely lost. This is manifested in the fact that \eqref{eq:confusion} is a \emph{many-to-one} equality. Every extended phase-space reduces to the phase space of \eqref{eq:gauge turned phys} once we gauge-fix on the corresponding extrinsic edge mode taking value $\Phi=0$. The physics of subregion/complement relationality is only seen at the level of the manifestly gauge-invariant description of the extended phase space.

We, thus, advocate that the use of the phase space \eqref{eq:gauge turned phys} in, for instance, computations of edge mode contributions to entanglement entropy, as done recently in \cite{Ball:2024hqe,Ball:2024xhf}, can be understood in the language of algebra extensions via subregion dressing to complement degrees of freedom. We believe this can shed light on the meaning of said contribution and it is part of ongoing work.

On the other hand, a somewhat different reading of the equivalence in \eqref{eq:confusion}  has been invoked in \cite{Riello:2020zbk,Riello:2021lfl} as an argument against the physical significance of the phase space extension itself, originally introduced in \cite{Donnelly:2016auv}, leading to statements that ``edge modes break gauge invariance'', or that the extended phase space has no genuinely physical symmetries. Our discussion clarifies that this is, however, not the case. 

Part of our paper can be seen as advocating for the utility of the manifestly gauge-invariant description for understanding correctly the physics at hand. This shows clearly that, in contrast to widespread claims in the literature, boundaries do \emph{not} break gauge invariance. 

In our conclusion that the approach of \eqref{eq:gauge turned phys} is equivalent to an extended phase space with an extrinsic edge mode frame, it was crucial that we were in the kinematical setting of a spatial subregion, on which no boundary conditions were imposed. As we will see in section~\ref{sec:Soft_bc}, this discussion may change somewhat in the dynamical setting of a timelike tube, where we must impose boundary conditions. In particular, the boundary conditions may be strong enough to land us back on the intrinsic phase space.

\subsection{Initial Data Evolution and the Physicality of $A^\dr_t|_\Gamma$} \label{subsec:in_dat_ev}

Before classifying the types of boundary conditions, it is crucial to understand what information is needed in order to evolve the initial data. In this subsection, we will thus consider the full timelike boundary $\Gamma$ and not only its cut $\p\Sigma$, unlike in the kinematical setting of the previous subsections.

In gauge theories, the answer is not as straightforward as a mere Dirichlet/Neumann choice at $\Gamma$ for the fundamental fields. This is because not every term in the symplectic flux corresponds to a variable in phase space. In particular, we have that (cf.\ \eqref{omegamma}):
\begin{align}
    \delta A^\dr_t \dt t \wedge \delta \l \star F \r \big|_{\Gamma} \subset \omega^\dr\big|_\Gamma \, , \label{a_t flux}
\end{align}
whereas $A^\dr_t$ does \emph{not} appear in the symplectic form $\Omega$. What are we then to make of a condition on $A^\dr_t\big|_\Gamma$? The key point is that this cannot be put on the same footing as boundary conditions coming from the other terms in the symplectic flux.

As we will see shortly, it turns out that the value $A^\dr_t\big|_\Gamma$ must \emph{always} be specified in the extended phase space.  It is simply not consistent to fix the normal electric field along $\Gamma$ to annihilate the symplectic flux \eqref{a_t flux}, while leaving $A^\dr_t\big|_\Gamma$ free. This is not enough data to solve the equations of motion. Essentially, $A^\dr_t\big|_\Gamma$ keeps track of the complement information necessary to evolve the Goldstone mode. Ignoring this subtle point may lead one to over-constrain the phase space when imposing boundary conditions, as we will discuss in detail in section \ref{sec:Soft_bc}.

We emphasize that this fixing is distinct from gauge-fixing the primary constraint via $A_t$ (which is not invariant in contrast to $A_t^{\rm dr}$) on a Cauchy slice in canonical formulations of Maxwell theory. $A_t^{\rm dr}$ (or $A_t$ in a gauge-fixed formulation) on $\Gamma$ is rather analogous to the role of boundary lapse and shift in general relativity.

Now, let us show this explicitly by attempting to evolve our initial data. For the sake of illustration, we will momentarily consider the case in which $\Gamma$ is a tube of constant radius in Minkowski spacetime, and we take our Cauchy slices to be constant in Minkowski time. The conclusion is general, but in this way the equations are particularly transparent.

At this point, it is convenient to extend our edge mode frame inside the bulk as well, so that we work with gauge-invariant variables throughout, as explained around equation \eqref{physomega}. Without loss of generality, we take the bulk Wilson lines to be purely radial.\footnote{
Other examples of dressing conditions are given in Appendix \ref{app:examples}, leading to the same results presented here.} This leads to the dressing condition: $\mathcal{G}[A^\dr\big|_\Sigma]=A^\dr_r=0$, as shown in \eqref{eq:WL_fr_dc}. 

The symplectic form \eqref{physomega} now looks like:
\begin{align}
    \Omega&\approx\int_\Sigma \epsilon_\Sigma \left(\delta A^\dr_a\, \curlywedge \, \delta F^{ta}\right) \\
    &\approx\int_\Sigma \epsilon_\Sigma \left(\delta \Tilde{A}^\dr_a\; \curlywedge \;\delta F^{ta}\right) +\int_{\partial\Sigma} \epsilon_{\p \Sigma} \left(\delta\varphi\; \curlywedge \; \delta  F^{rt}\right) \,,\label{eq:sympl_bulkdressed}
\end{align}
where ($t,r,x^a$) are the standard Minkowski polar coordinates.

We define the notation: $\star F_{ra}=:\sqrt{-g}\epsilon_{ab}E^b $ and $\star F_{ab}=:\sqrt{-g}\epsilon_{ab}E^r$, where $g$ is the determinant of the full spacetime metric and $\epsilon_{ab}$ is the fully antisymmetric tensor with $\epsilon_{12}=+1$ and $\epsilon^{12}=-1$. 
Also notice that, on $\Sigma$, we have that: $ F\big|_\Sigma=\dt A^\dr\big|_\Sigma=\dt \Tilde{A}^\dr\big|_\Sigma$, since $A^\dr\big|_\Sigma=\Tilde{A}^\dr\big|_\Sigma+\dt\varphi$. We recall that the extension of $\varphi$ into $\Sigma$ is done such that both dressed observables satisfy the same dressing condition: $\mathcal{G}[A^\dr\big|_\Sigma]=\mathcal{G}[\Tilde{A}^\dr\big|_\Sigma]$. This convenient choice  means having $\partial_r\varphi=0$.

We now want to evolve the initial data $(\Tilde{A}^\dr_a\big|_\Sigma, E^a\big|_\Sigma, \varphi, E^r\big|_{\p\Sigma})$. The evolution equations for the momenta, obtained from the restrictions of the Maxwell equations $\dd\star F\approx0$ to $\Sigma$ and $\p\Sigma$, respectively, involve knowledge of the initial data only:\footnote{We refer the reader to the notations introduced in section \ref{sec:intro}. Here $|_{\Sigma}$ refers to the pullback of the fields to the initial slice, not of the Maxwell equations, otherwise we will simply get the Gauss law.}\bsub\be
\p _t E^a \big|_\Sigma & = - \frac{1}{r^2} \gamma ^{ab} \;\p _r ^2 \Tilde{A} _b ^{\dr}\big|_\Sigma  - \frac{2}{r^4} \gamma ^{ab} \gamma^{cd}\; D_c D_{[d} \Tilde{A}^\dr _{b]}\big|_\Sigma\,,\\
\p_t E^r\big|_{\p\Sigma}&=\frac{1}{r^2}D^a\left(\p_r \Tilde{A}^\dr_a\right)\Big|_{\p \Sigma}\,. \label{eq:Er evolution}
\ee\esub
This is not the case for the evolution equations for the configuration fields:
\bsub\be
\p_t\Tilde{A}^\dr_a\big|_\Sigma &=r^2\gamma_{ab}\;E^b\big|_\Sigma +\p_a \Tilde{A}^\dr_t\big|_\Sigma \label{eq:tilde evolution}\,,\\
\p_t\varphi&=A^\dr_t\big|_{\p\Sigma}-\Tilde{A}^\dr_t\big|_{\p\Sigma} \,,\label{eq:gold evolution}
\ee\label{eq:config_evol}\esub
the first of which is just the definition of $F_{ta}$, while the second follows from the definition of the Goldstone mode. We see that $A^\dr_t$ and $\Tilde{A}^\dr_t$ appear in the evolution equations, even though they are not part of the phase space.  
They are determined by the initial conditions and the Gauss constraint, $\dd\star F\big|_\Sigma\approx0$, \emph{up to} their values on $\Gamma$, in the following manner. 
From the initial values of the momenta ($E^a\big|_\Sigma, E^r\big|_{\p\Sigma}$) we solve the constraint to get $E^r\big|_\Sigma$. But from $E^r\big|_\Sigma=\p_r A^\dr_t=\p_r \Tilde{A}^\dr_t$ we can reconstruct the whole profiles of $A^\dr_t\big|_\Sigma$ and $\Tilde{A}^\dr_t\big|_\Sigma$, only if we know their boundary values. 

Here comes a key difference between the intrinsic and extrinsic dressed observables. The dynamics of the former depends only on the initial data in the subregion, unlike the dynamics of the latter, because the intrinsic frame $\tilde\Phi$ only has support in the subregion in contrast to the extrinsic one $\Phi$.  This fixes $\Tilde{A}^\dr_t \big| _{\p \Sigma }$ in terms of (local) initial data: 
\be
\begin{split}
\Tilde{A}^\dr_t\big|_{\p\Sigma}&=A_t\big|_{\p\Sigma}
- \p_t \Tilde \Phi = \int_{\p\Sigma} \Delta^{-1}_{\p\Sigma}  D^a \l \p_a A_t -  \p_t A_a\r\\
&=r^2\int_{\p\Sigma}\Delta^{-1}_{\p\Sigma}\;\left(D_a E^a\right)\Big|_{\p\Sigma}\approx -\int_{\p\Sigma}\Delta^{-1}_{\p\Sigma}\;\p_r\left(r^2 E^r\right)\Big|_{\p\Sigma}\,,
\end{split}\label{eq:tilde fixed}
\ee
where we used the Gauss constraint in the last line. This means we can solve \eqref{eq:tilde evolution} without the need for additional information. 

On the other hand, the whole boundary profile $A^\dr_t|_\Gamma$ is completely undetermined from the subregion data and must be specified by hand. It encodes the relevant complement information necessary to evolve the extrinsic edge mode. 
In conclusion, we \emph{must} provide the value of $A^\dr_t\big|_\Gamma$ in order to be able to solve \eqref{eq:gold evolution}.
This specification \emph{already} eliminates the piece of the symplectic flux in \eqref{a_t flux}. Sticking to this necessary and sufficient condition will prove crucial in writing down the boundary conditions of section \ref{sec:Soft_bc}.

\paragraph{Manifestation in the Gauge-fixed Approach:}

We now illustrate what the necessary gauge-invariant boundary condition $\delta A^\dr_t\big|_\Gamma=0$  looks like in a gauge-fixed description. For this, it will suffice to discuss the partial gauge-fixing associated to the time profile of the gauge parameter.\footnote{A complete gauge-fixing would then further involve time-independent gauge transformations.} Also, here we are only analyzing the boundary gauge-fixing, to parallel the discussion of the previous subsection.

\begin{figure}[h!]
    \centering

\tikzset{every picture/.style={line width=0.75pt}} 

\begin{tikzpicture}[x=0.75pt,y=0.75pt,yscale=-1,xscale=1]

\draw   (259,32.2) .. controls (259,24.91) and (264.91,19) .. (272.2,19) -- (387.8,19) .. controls (395.09,19) and (401,24.91) .. (401,32.2) -- (401,71.8) .. controls (401,79.09) and (395.09,85) .. (387.8,85) -- (272.2,85) .. controls (264.91,85) and (259,79.09) .. (259,71.8) -- cycle ;

\draw   (59,188.2) .. controls (59,180.91) and (64.91,175) .. (72.2,175) -- (187.8,175) .. controls (195.09,175) and (201,180.91) .. (201,188.2) -- (201,227.8) .. controls (201,235.09) and (195.09,241) .. (187.8,241) -- (72.2,241) .. controls (64.91,241) and (59,235.09) .. (59,227.8) -- cycle ;

\draw   (259,188.2) .. controls (259,180.91) and (264.91,175) .. (272.2,175) -- (387.8,175) .. controls (395.09,175) and (401,180.91) .. (401,188.2) -- (401,227.8) .. controls (401,235.09) and (395.09,241) .. (387.8,241) -- (272.2,241) .. controls (264.91,241) and (259,235.09) .. (259,227.8) -- cycle ;

\draw   (459,188.2) .. controls (459,180.91) and (464.91,175) .. (472.2,175) -- (587.8,175) .. controls (595.09,175) and (601,180.91) .. (601,188.2) -- (601,227.8) .. controls (601,235.09) and (595.09,241) .. (587.8,241) -- (472.2,241) .. controls (464.91,241) and (459,235.09) .. (459,227.8) -- cycle ;

\draw    (258.95,52) -- (129.95,52) -- (129.95,175) ;
\draw    (400.95,52) -- (529.95,52) -- (529.95,175) ;
\draw    (329.67,84.67) -- (329.67,174.84) ;

\draw (257,118.4) node [anchor=north west][inner sep=0.75pt]    {$A_{t}\big|_{\Gamma } =0$};
\draw (75,114.4) node [anchor=north west][inner sep=0.75pt]    {$\dot{\Tilde{\Phi }} =0$};
\draw (476,118.4) node [anchor=north west][inner sep=0.75pt]    {$\dot{\Phi } =0$};
\draw (296,40.4) node [anchor=north west][inner sep=0.75pt]    {$\delta A_{t}^{\dr}\big|_{\Gamma } =0$};
\draw (94,197.4) node [anchor=north west][inner sep=0.75pt]    {$\delta \dot{\Phi } =\delta \Tilde{A}_{t}^{\dr}\big|_{\Gamma }$};
\draw (305,197.4) node [anchor=north west][inner sep=0.75pt]    {$\delta \dot{\Phi } =0$};
\draw (500,197.4) node [anchor=north west][inner sep=0.75pt]    {$\delta A_{t}\big|_{\Gamma } =0$};

\end{tikzpicture}
    \caption{Figure showing various useful/common gauge-fixing conditions (on the edges) applied to the gauge-invariant boundary condition at the top. The bottom entries represent three different manifestations of \emph{the same} physical boundary condition, in different gauges. }
    \label{fig:a_t g.f.}
\end{figure}
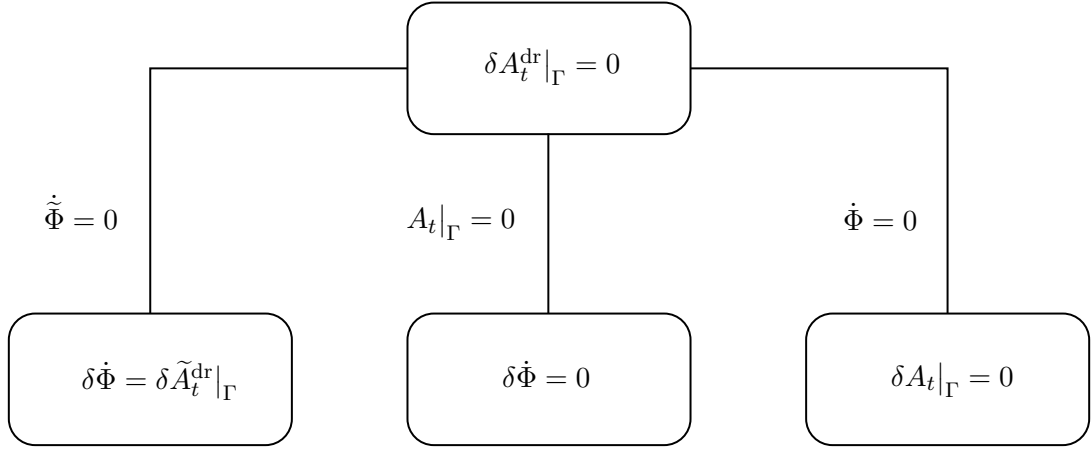

We explore three natural (partial) gauge-fixing conditions: $\dot{\Tilde \Phi}=0$, $A_t\big|_\Gamma=0$ and $\dot{\Phi}=0$, as summarized in figure \ref{fig:a_t g.f.}.

If we gauge-fix on $\p \Sigma $ to the cut $\dot{\Tilde \Phi} =0$, and using the same notation as in the last subsection, 
we get that the gauge-fixed connection satisfies: $\Tilde \cA_t\big|_\Gamma = \Tilde A_t^\dr\big|_\Gamma$, which is thus fixed via equation \eqref{eq:tilde fixed}.
Hence, in this gauge:
\be
\delta A^\dr_t\big|_\Gamma=0 \iff \delta \p_t \Phi\big|_{\dot{\Tilde \Phi}=0} =\delta \Tilde{A}^\dr_t\big|_\Gamma\,.
\ee
We see that the extrinsic edge mode dynamics is fixed by the regional initial data up to a phase space constant profile. It is this profile that encodes the complement information and that implements a postselection on the global solution space. 

A different gauge condition, which is commonly used, is the \emph{temporal gauge}: $A_t=0$.  In this gauge (here indicated by the superscript $^0$), the dynamics of the intrinsic edge mode becomes fixed by the initial data: $\p_t\Tilde \Phi^0=-\Tilde A^\dr_t\big|_\Gamma$, via \eqref{eq:tilde fixed}. The gauge-invariant boundary condition now fixes the dynamics of the extrinsic edge mode to be a constant on phase space:
\be
\delta A^\dr_t\big|_\Gamma=0\iff \delta\p_t\Phi^0=0\,.
\ee
Here, it becomes even more transparent that the extra data provided by $A^\dr_t\big|_\Gamma$ fixes the dynamics of the complement-supported observables, represented by the extrinsic edge mode $\Phi^0$.

Lastly, we can gauge-fix on the cross-section $\dot{\Phi}=0$. This is a non-local gauge condition. Again, using the notation of the last subsection, we get that the boundary condition becomes:
\be
    \delta A^\dr_t\big|_\Gamma=0\iff \delta \mathcal{A}_t\big|_\Gamma=0\,.
\ee
So, here the Lagrange multiplier is fixed by the boundary condition itself, rather than by the requirement that the constraint is satisfied. This has the implication that $\p_t \Tilde \Phi\big|_{\dot{\Phi}=0}=A^\dr_t\big|_\Gamma-\Tilde A^\dr_t\big|_\Gamma$, meaning that the dynamics of the intrinsic edge mode is determined by the initial data, up to a complement-supported free function. We see that this gauge condition has the effect of transferring the non-locality from $\Phi$ onto $A_t\big|_\Gamma$, and consequently onto the intrinsic edge mode $\Tilde \Phi$.\footnote{We point out that the discussion in \cite{Ball:2024hqe, Ball:2024xhf,Canfora:2024awy} most closely parallels this gauge-fixing condition. There, they take $ A_t\big|_\Gamma=0$ as one of their entries in the ``dynamical edge mode'' boundary conditions. Because they are working with an extended phase space (for otherwise they would not have a Goldstone sector), we see that it is only in this gauge that the gauge-invariant boundary condition matches with their form. This connection is somewhat obscured by the fact that those authors do not introduce an extrinsic edge mode explicitly in their construction. In accordance with our discussion in the previous subsection, it is only in the gauge $\Phi\big|_\Gamma=0$ (which includes the partial gauge condition $\dot{\Phi}=0$) that the corner symmetries ``look like'' boundary-supported gauge transformations. Bringing all this evidence together, we infer that the physical conclusions drawn in those papers are consistent gauge-fixed versions of our gauge-invariant statements, in the implicit boundary gauge $\Phi=0$. We disagree, however, with the physical interpretation advocated for in those papers, that boundary-supported gauge transformations at finite distance are physical. We refer the reader to the discussion in section \ref{subsec:equivalence}.} 

Thus, the precise manifestation of the boundary condition fixing $A^\dr_t\big|_\Gamma$ varies for different choices of gauge-fixing cuts.  Our previous observation at the gauge-invariant level thereby becomes less transparent at the gauge-fixed level; namely that, on the extended phase space,  $\Tilde A_t^\dr\big|_\Gamma$ is a redundant observable, determined by the initial conditions, while $A_t^\dr\big|_\Gamma$ is an independent \emph{physical} quantity, with its physicality acquired through determining the time evolution of the Goldstone mode $\varphi$. Once again, this discussion underscores the clarity and utility of the gauge-invariant approach.

\section{Soft Boundary Conditions}\label{sec:Soft_bc}

As discussed in the previous section, to ensure the well-posedness of the variational principle and the conservation of the symplectic form between two Cauchy surfaces, a sufficient condition is the vanishing of the gauge-invariant  symplectic flux:
\be\label{eq:sympl_flux2}
\omega^\dr|_\Gamma = \delta A^{\ra}\wedge\delta \star F |_{\Gamma}\,\hat \approx \,0\,.
\ee

Throughout this subsection, we work with a general asymptotically flat metric, with the notation presented in section \ref{sec:notations}. In particular, we take a time foliation whose slices hit the boundary $\Gamma$ normally. The symplectic flux density written in these coordinates is:
\begin{equation}
\label{eq:sympl_flux_expand}
\begin{split}
    \omega^\dr |_\Gamma   &= \epsilon_\Gamma\, m _\mu \l  \d A  ^{\ra} _t \curlywedge \d  F^{\mu t} + \d A  ^{\ra} _a \curlywedge \d F^{\mu a } \r\\
    &= \epsilon_\Gamma\, \l  \d A  ^{\ra} _t \curlywedge \d  F^{r t} + \d A  ^{\ra} _a \curlywedge \d F^{r a } \r\,.
\end{split}
\end{equation}

\subsection{Traditional Boundary Conditions and 
Their Problems} \label{sec:traditional bcs}

The traditional approach to having a vanishing symplectic flux is fixing \textit{half} of the symplectically conjugated data on the boundary flux \eqref{eq:sympl_flux_expand}. While this approach gives a perfectly well-defined theory, it does not permit the existence of frame reorientations as a symmetry of the theory, as we will show in this section. We recall that frame reorientations act as $ A ^{\dr } \mapsto A^{\dr } + \dt \rho $.

Traditional Dirichlet boundary conditions are defined as:
\be
\d A^\dr  |_{\Gamma } = 0\,.
\ee
Frame reorientations spoil these, unless $\rho=const$. But in the absence of matter inside the subregion, the corresponding total charge vanishes on-shell:
\be
Q_\text{Dir}=\int_{\p \Sigma} \rho \star F = \rho \int_{\Sigma}  \dt \star F \approx 0\,.
\ee
This constraint is associated with the standard Gauss law, hence it does not add any physically relevant information to the theory.
All the other frame reorientations spoil the boundary conditions and are thus \textit{meta-symmetries}: symplectomorphisms between different regional phase spaces (i.e.\ different theories) \cite{Carrozza:2021gju}, as they map one set of boundary conditions to another. Standard Dirichlet boundary conditions eliminate the Goldstone sector of the theory, by fixing the value of $\varphi$.

In the complementary case, or Neumann boundary conditions, we have:
\be \label{eq:Neumann constraints}
\d \star F|_\Gamma=0 , \quad \Rightarrow \quad \d Q_\text{Neu} = 0 \, .
\ee 

We see that these boundary conditions also constrain degrees of freedom from the phase space, namely $\star F\big|_{\p\Sigma}$, thus also eliminating the Goldstone sector. This is because \eqref{eq:Neumann constraints} introduces additional first-class edge constraints. They require further \textit{gauge-fixing} or dressing of the frame reorientation directions in addition to the usual gauge transformations \cite{Carrozza:2021gju}. In other words, the charges that generate frame reorientations are constrained, and therefore we are forced to remove the directions of frame reorientations from the phase space, in order to have an invertible symplectic form.  

Before moving on, we point out another shortcoming of the Neumann boundary conditions. Our analysis of initial data evolution from section \ref{subsec:in_dat_ev} shows that $A_t ^\dr |_\Gamma$ necessarily has to be fixed in order to solve the dynamics. Therefore, Neumann boundary conditions are over-constraining as they fix more functions on the boundary than necessary to make the symplectic flux vanish. Relatedly, by Maxwell's equations on $\Gamma$, $\dd\star F|_\Gamma=0$, the independent degrees of freedom of $\star F\big|_\Gamma$ are the two components of $F^{ra} |_{\Gamma }$ and the value of $F^{rt}  |_{\p\Sigma }$ on a single slice, so the Neumann boundary conditions involve data which is dependent on each other, and are therefore not only over-constraining but also redundant. The ‘‘redundancy'' arises from the fact that we do not need to specify the time evolution $\p_t F^{rt}|_\Gamma$, as it is already determined by the other components of $\star F|_\Gamma$. The ‘‘over-constraining'' occurs because we do not need to specify $F_{rt}$ at all, because of $A^\dr_t$.

In conclusion, standard boundary conditions (including mixed and Robin boundary conditions \cite{Carrozza:2021gju}) are overly restrictive and do not allow for an infinite-dimensional symmetry algebra for the subregion theory, which is at the heart of a lot of interesting physics. We refer to Appendix \ref{app:BC_2} for a more detailed description of these ‘‘traditional'' boundary conditions. 

\subsection{Soft Boundary Conditions}
\label{subsec:soft_bc}

In this subsection, we present a new type of boundary conditions on the extended phase space that allows for arbitrary \emph{time-independent} frame reorientations, along with the charges that generate them. Essentially, they leave the Goldstone symplectic pair free (or in other words dynamical), justifying the qualification of ``soft'', while in addition imposing Dirichlet, Neumann or Robin conditions on the remaining intrinsic (radiative) data. Crucial to this is the split of $A^\dr$ in terms of $\Tilde{A}^\dr$ and $\varphi$, as introduced in \eqref{eq:Goldstonedefn1}:
\begin{equation}\label{eq:tA _def}
    A^\dr  = \Tilde{A} ^\dr + \dt \varphi\, ,
\end{equation}
as well as the factorization structure of the extended phase space into intrinsic and Goldstone pieces, as explained in section \ref{subsubsec:splitting}.

As discussed in section \ref{subsec:extr_intr_ref_fr}, both $A^\dr $ and $\Tilde{A} ^\dr $ are gauge-invariant dressed observables dressed with $\Phi $ and $\Tilde{\Phi }$ respectively. We extend both frames in the bulk of the subregion, such that both $A^\dr$ and $\Tilde{A}^\dr$ satisfy the same dressing condition -- $m^{\mu } A^\dr _\mu  =0 $ -- with a spacelike vector $m_\mu$, tangential to the Wilson lines, that coincides with the normal vector on $\Gamma$.

The Wilson lines within the subregion induce a foliation of $\Sigma$ into compact codimension-2 surfaces, diffeomorphic to both $\p \Sigma $ and $S^2$. Therefore, we can use Hodge's theorem on every such surface and express $A ^\dr _a \big|_\Sigma $ as:\footnote{Both $\rho$ and $h$ are only defined up to a radial function. That piece turns out to be irrelevant for what follows because there are always $\p_a$ in front. The same holds for $\mathfrak{p}$ and $\mathfrak{h}$ below.}
\begin{equation}\label{eq:HD_bulk}
    A _a ^\dr = \l \dt _{\p \Sigma } \varrho  + \star _{\p \Sigma } \dt _{\p \Sigma } h \r _a  = \p _a \varrho   + g _{ab} \epsilon _{\ \p \Sigma } ^{bc} \p _c h \, ,
\end{equation}
where $\varrho |_{\p \Sigma } = \varphi$. Note that both $\rho$ and $h$ are functions on spacetime, not just the coordinates $x^a$. We can do the same decomposition for $g_{ab}F^{rb}\big|_\Sigma$:
\begin{equation}
    g_{ab}F^{rb}  = \l \dt _{\p \Sigma } \mathfrak{p}   + \star _{\p \Sigma } \dt _{\p \Sigma } \mathfrak{h} \r _a = \p _a \mathfrak{p}    + g _{ab} \epsilon _{\ \p \Sigma } ^{bc} \p _c \mathfrak{h} \, . 
\end{equation}
Because we have chosen the coordinate $r$ such that $\dt r$ is normal to $\Gamma $, and 
inside the subregion we extend the frame with Wilson lines along $r$, at the boundary we can identify $\mathfrak{p} \big|_\Gamma   = \p _r \varrho \big|_\Gamma $ and $\mathfrak{h} \big|_\Gamma = \p _r h \big|_\Gamma$. With other coordinates or bulk frame extensions this will not necessarily be true. Expressed in these new variables, the symplectic flux is:
\begin{equation}
\begin{split} \label{eq:symp_flux}
  \Omega _{\text{flux}}  \approx &  \int _{\Gamma _{[1,2]} }  \epsilon_\Gamma  \Big(   \d \p _a  h \curlywedge  \d \Tilde{\nabla} ^a \mathfrak{h}     + \d  \underbrace{\l \Tilde{A}^\dr _t  +  \p _t \varphi \r }_{A_t ^\dr }\curlywedge \d F^{rt}  + \d  \varphi \curlywedge \sqrt{g} ^{-1} \d \p _t \l \sqrt{g}  F^{tr} \r \Big). 
\end{split}
\end{equation}
Here comes the crucial bit: we want to keep the corner piece of the symplectic form $ \d \varphi \curlywedge \d F^{tr} $ from \eqref{eq:sympl_bulkdressed}, but also ensure a vanishing symplectic flux. The only way to achieve this is by demanding:
\begin{equation}\label{eq:sbc_prelim}
\begin{split}
  \d A_t ^\dr  \big|_{\Gamma} =   \d \l \Tilde{A}^\dr _t\big|_\Gamma  + \p _t \varphi  \r &= 0\,, \\
   \d \p _t \l \sqrt{g} F^{tr} \r  \big|_{\Gamma} \approx \d \p _a  \l \sqrt{g} F^{ra} \r  \big|_{\Gamma} = \sqrt{g} \delta \tilde{\nabla} ^2 \mathfrak{p}    \big|_{\Gamma} &= 0\,. 
\end{split}
\end{equation}
The first condition is necessary to evolve the Goldstone mode, as shown in subsection \ref{subsec:in_dat_ev}, while the second condition, which is new, dictates the time evolution profile of the boundary charges. This new class of boundary conditions is thus identified by fixing the time derivative the evolution isn't fixed when just fixing the time derivative, but not the initial data of the corner variables while their initial values remain free, thereby allowing a non-trivial corner symmetry group. What remains from the symplectic flux \eqref{eq:symp_flux} is the first term, which depends on the boundary divergence-free components of $\Tilde{A}^\dr_a$ and $F^{ra}$. The way we fix these leads to different types of what we call \emph{soft} boundary conditions, summarized in table \ref{table:sbc_types} below.  The name ‘‘soft'' originates from the connection between the regional symmetries (enhanced by these boundary conditions), asymptotic symmetries and memory effect (cf.\ section \ref{subsec:memory}).

\begingroup
\renewcommand{\arraystretch}{1.7}
\begin{table}[h!]
    \centering
        \begin{tabularx}{0.8\textwidth}{|>{\centering\arraybackslash}X|>{\centering\arraybackslash}X|>{\centering\arraybackslash}X|}
            \hline
            \textbf{(Soft) Dirichlet} & \textbf{(Soft) Neumann} & \textbf{(Soft) Robin}\\
            \hline
            \multicolumn{3}{|c|}{$\d A^\dr _t  \big |_{\Gamma } = 0$} \\
            \hline
            \multicolumn{3}{|c|}{$\sqrt{g} ^{-1} \d \p _t \l \sqrt{g} F^{rt} \r \big |_{\Gamma }  =\d \tilde{\nabla} ^2  \mathfrak{p}   \big|_{\Gamma} =  0$} \\
            \hline
            $\delta \p_a h\big|_\Gamma=0$& $\delta \p_a \mathfrak{h}\big|_\Gamma=0$ & $\d \l \alpha \p_a h   + \beta   \p_a \mathfrak{h}  \r \big|_{\Gamma }    = 0$\\
            \hline
        \end{tabularx}
        \caption{Classification of boundary conditions on $\Gamma$ allowing for an infinite-dimensional symmetry algebra for the subregion theory. We call them \emph{soft boundary conditions}. Crucial to this classification is the different treatment of the intrinsic and Goldstone sectors. }
        \label{table:sbc_types}
\end{table}
        
\endgroup

In the last column of table \ref{table:sbc_types}, the variable $ \l \alpha \p_a h  + \beta  \p_a \mathfrak{h}  \r |_{\Gamma } $ is part of a canonical symplectic pair, defined as a linear (canonical) transformation on $\p _a h |_{\Gamma }$ and $\p _a \mathfrak{h} |_{\Gamma }$: 
\begin{equation}
    \{ \l \alpha \p_a h  + \beta \p _a \mathfrak{h}  \r, \, \l \gamma \p_a h  + \zeta \p _a \mathfrak{h} \r \} |_{\Gamma }, \quad \alpha \zeta -\beta\gamma =1 \, ,
\end{equation}
where $\alpha, \beta ,\gamma ,\zeta $ are field-independent coefficient functions on $\Gamma$. We can imagine defining other variables that are more general functions of $\p_a h |_{\Gamma }$ and $ \p_a \mathfrak{h} |_{\Gamma }$ (these would be non-linear canonical transformations), which would lead to yet more general classes of soft boundary conditions. 
For an illustration of the boundary Lagrangian that implements the soft boundary conditions, interested readers are directed to Appendix \ref{app:post_DBC}, which follows the algorithm presented in \cite{Carrozza:2021gju}. Finally the asymptotic boundary conditions at $i^0$ for the global solution space are discussed in section \ref{subsec:review_gss} with comparison to other literature. 

To provide physical intuition for the boundary conditions in the first two columns, recall that the electromagnetic tensor depends solely on the radiative degrees of freedom, $\Tilde{A}^\dr$. In the case of soft Dirichlet, the condition $\delta \p_a h=0$ effectively fixes the orthogonal component of the magnetic field at the boundary, while for the soft Neumann boundary conditions, we instead fix its tangential components, via $\delta\p_a \mathfrak{h}=0$ (as $\mathfrak{p}$ is already fixed, $\delta\p_a \mathfrak{h}=0$ fixes completely $F^{ra}\big|_{\Gamma}$). However, unlike traditional boundary condition, the soft Dirichlet case does not  fix the tangential electric field.
Doing so would imply that we are fixing $\Tilde{A}^\dr_t$, which depends on the initial conditions rather than the boundary conditions. This distinction is essential: while soft boundary conditions act like standard Dirichlet/Neumann/Robin b.c. for the magnetic field components at the corner, they do not for the electric field on $\Gamma$. That is only partially constrained by boundary conditions (e.g.\ $\d \p_t \l \sqrt{g} F^{rt} \r=0$) and remains generally dependent on the initial data, making it a dynamical quantity.

We note that the boundary condition, $\delta A^\dr_t\big|_\Gamma=0$, is the only one which is not local to the subregion. It is a condition on non-local observables shared between the subregion and the complement. It thus imposes a non-local post-selection on the global phase space, by means of constraining the initial data in the complement. For example, in the case of the extrinsic frame $\Phi$ being built out of Wilson lines to the asymptotic boundary we get, from \eqref{eq:alternative wilson}, that:
\be
 \delta A^\dr_t\big|_\Gamma= \d \l \int E_i \dt x^i + A^\dr _t \big| _{i^0}  \r  = 0 \iff \delta \Delta V\big|_\Gamma := \d \int E_i \dt x^i = 0\,,
\ee
where the two equalities are to be understood as being imposed on the subregion and global phase spaces respectively, and we have used the fact $A_t ^\dr \big| _{i^0} $ must be fixed as well for a well defined global theory. We see that, from the global perspective, we are only considering those solutions with a given profile for the ``potential difference'' $\Delta V$ between the asymptotic and finite boundaries. Constraining the complement data in this way gives us the necessary external information to be able to evolve dynamically the shared observables we added to our intrinsic phase space. This is encoded in the dynamics of the Goldstone mode $\varphi$, which, via $\p_t\varphi=A^\dr_t\big|_\Gamma-\Tilde A^\dr_t\big|_\Gamma$, depends on both this external potential difference and the intrinsic initial data.

Another key aspect of the soft boundary conditions is the separate treatment of spacelike and timelike directions, as they are explicitly dependent on the choice of a time direction. Unlike the traditional ones, they are thus \emph{a priori} not covariant. One might conceptually associate this choice with the properties of a particular observer or experimental setup, as suggested by figure \ref{lab}. 

Having discussed both soft and traditional boundary conditions, we now briefly conclude the discussion about the phase space equivalence of \eqref{eq:confusion}, in the dynamical setting of a timelike subregion tube. This is best exemplified by comparing \cite{Harlow:2019yfa} and \cite{Ball:2024hqe}. While both start from the symplectic form $\Omega_{\text{gauge}\to \text{``phys''}}$, the former imposes traditional Dirichlet boundary conditions, thereby also fixing the exact piece of $A$ and thus the Goldstone symplectic pair, effectively constraining $\Omega_\text{ext}$  to $\Omega_\text{int}$ and recovering our intrinsic phase space. Conversely, the latter authors impose soft Neumann (or in their language ``dynamical edge mode'') boundary conditions, thereby preserving the full $\Omega_\text{ext}$. The boundary conditions of both papers can be reached by imposing the gauge $\Phi=0$ on our manifestly gauge-invariant boundary conditions, in line with the discussion around \eqref{eq:confusion}. This shows the crucial role of boundary conditions in constraining, or not, the subregion initial data.

As a final comment on \cite{Ball:2024hqe}, while the authors presented the soft Neumann boundary conditions,  here we have generalized this analysis, and in doing so also arrived at the soft Dirichlet boundary conditions, explicitly realising, in a gauge-invariant language, what in the context of \eqref{eq:confusion} can be referred to as ``Dirichlet up to large gauge'' boundary conditions. For this, it was crucial to differentiate between conditions on the curl-free part of $g_{ab}F^{rb}$ ($\mathfrak{p}$), which must always be imposed, and conditions on the divergence-free part ($\mathfrak{h}$), which are not strictly necessary.

\subsection{Symplectic Form}\label{subsec:sym_form}

Now that we have ensured the vanishing of the symplectic flux, we can analyse the final structure of the symplectic form. We need its inverse in order to define the charges and Hamiltonian, and their respective action on phase space variables. This task is simplified by the fact that $\Omega$ is already ``block anti-diagonal'', as we now show. Crucially, the soft boundary conditions guarantee the independence of the Goldstone and intrinsic phase space factors.

Just like for $F^{ra}$ and $A^\dr_a$ before, we write $F^{ta}$ and $\tilde{A} ^\dr _a$ in terms of their curl-free and divergence-free pieces, via a Hodge decomposition on every codimension-2 cut of $\Sigma$:
\begin{equation}
\begin{split}\label{eq:intr_hodge_split}
 &\tilde{A} ^\dr _a = \p _a\tilde{\varrho} + g _{ab} \epsilon _{\ \p \Sigma } ^{bc} \p _c h\,, \quad \tilde{\varrho}= \varrho - \varphi\,, \quad \tilde{\varrho} \big|_{\p \Sigma } =0 \, ,\\
    &g_{ab}F^{tb}   = \p _a  \Pi _{\varrho }  + g _{ab} \epsilon _{\ \p \Sigma } ^{bc} \p _c \Pi _h   \, .  \\
\end{split}
\end{equation}
Introducing the extra notation -- $ \mathcal{Q} := F^{tr} |_{\p \Sigma,} \, $, the symplectic form reads: 
\begin{equation}\label{eq79}
\begin{split}
    \Omega = &  \int _{\Sigma} \,  \epsilon_{\Sigma } \, \l \d \Tilde{A}^\dr _a \curlywedge \d F^{ta} \r  + \int _{\p \Sigma } \, \epsilon_{\p \Sigma } \, \d \varphi \curlywedge \d \mathcal{Q}  \\ 
   =&  \int _\Sigma \, \epsilon_\Sigma  \, \l   \d \p _a \tilde{\varrho} \curlywedge \d \Tilde{\nabla }^a \Pi _\varrho   + \d \p _a h \curlywedge \d  \Tilde{\nabla } ^a  \Pi _h    \r  + \int _{\p \Sigma } \, \epsilon_{\p \Sigma } \, \d \varphi \curlywedge \d \mathcal{Q}  \  \\
    \approx & \int _\Sigma \, \epsilon_\Sigma  \, \l  \d \tilde{\varrho}  \curlywedge  \sqrt{g} ^{-1} \d \p_r \l \sqrt{g} F^{tr} \r + \d \p _a h \curlywedge \d  \Tilde{\nabla } ^a  \Pi _h    \r  + \int _{\p \Sigma } \, \epsilon_{\p \Sigma } \, \d \varphi \curlywedge \d \mathcal{Q}  \ \\
    =& \int _\Sigma \, \epsilon_\Sigma  \, \l -  \d  \p _r \tilde{\varrho}  \curlywedge \d  F^{tr}  + \d \p _a h \curlywedge \d \Tilde{\nabla } ^a  \Pi _h    \r +  \int _{\p \Sigma } \, \epsilon _{\p \Sigma }  \,  \d \varphi  \curlywedge \d \mathcal{Q}\, ,
\end{split}
\end{equation}
where to get to the third line we used the equations of motion, and for the fourth, we simply integrated by parts along the radial direction. By definition, $\p _a h $ and $\p _a \Pi _h  $ are independent degrees of freedom from all the rest. Furthermore, from its definition, $\varphi$ is independent of $\tilde{A}^\dr$  (including $\p _r \tilde{\varrho} $) and $F^{tr}$. Next, as $\d \mathcal{Q} = \d F^{tr} |_{\p \Sigma } $, at first glance it seems like this variable is symplectically conjugated to $\d \varphi$ \emph{and} $- \d \p _r \tilde{\varrho} |_{\p \Sigma} $. However, as pointed out previously, thanks to our choice of coordinates from section \ref{sec:notations} and dressing $ \d \p _r \tilde{\varrho} |_{\p \Sigma}$ is equal to $\d \mathfrak{p} |_{\p \Sigma }$, and the soft boundary conditions from table \ref{table:sbc_types} say that it is always zero!  This is all we need to conclude that $\Omega$ is indeed block-diagonal and we can write its inverse formally:\footnote{Even though our proof relied on the boundary conditions, choice of coordinates and bulk frame extension, we expect it to be true in general as well, although different techniques might be necessary to show it.}
\begin{equation}
    \begin{split} \label{eq:sym_form_inv}
         \Omega  ^{-1}  &=  \int _{\Sigma} \,  \epsilon_{\Sigma } \, \l \frac{\d  }{\d F^{ta}}  \frac{\d }{\d \Tilde{A}^\dr _a  }-\frac{\d }{\d \Tilde{A}^\dr _a } \frac{\d }{\d F^{ta} }    \r  +   \int _{\p \Sigma } \, \epsilon _{\p \Sigma }\, \l \frac{\d }{\d \mathcal{Q}  }  \frac{\d }{\d \varphi }  -\frac{\d }{\d \varphi} \frac{\d }{\d \mathcal{Q} } \r \, .
    \end{split}
\end{equation}

\paragraph{Charges:} 
~\newline 
A crucial property of the soft boundary conditions is that they allow for frame reorientations $\varphi \mapsto \varphi  + \rho $. The function $\rho$ is not completely general - the boundary conditions from table \ref{table:sbc_types} and the dressing condition constrain it in the following way: 
\begin{equation}
\begin{split}
    &\mathcal{G} [\dt \rho] = 0\,,\\
    &\d A_t ^\dr \big|_\Gamma  = 0 \ \Rightarrow \ \p _t \rho \big|_\Gamma  = 0\,.
\end{split}
\end{equation}
The dressing condition is a differential equation which determines the extension of $\rho$ inside $\Sigma$ in terms of a free function on $\p\Sigma$.

Frame reorientations are symmetries of the post-selected theory and their associated charges are: 
\begin{equation}
    X_{\d _\rho } \cdot  \Omega = \d Q_\rho =  \int _{\p \Sigma } \, \volC   \, \rho \,  \d \mathcal{Q}  \, ,
\end{equation}
where $X_{\d _\rho}$ is a vector on phase space that induces the reorientation by $\rho$. Notice that $\rho $ is a constant function on phase space, and therefore the $\d Q_\rho$ is integrable. This means $\d _\rho$ fulfils the symmetry transformation requirements, described in section \ref{sec:out_res}.  As a reminder, these do \emph{not} involve time independence of $Q_\rho$. That would be true for only a small subset of the soft boundary conditions for which $\p _t \l \sqrt{g} \mathcal{Q}\r$ is not only fixed, but also 0. In this work we allow it to be an arbitrary (well-behaved) function on $\Gamma$, in order to describe as general as possible interactions between the subregion and its complement.  Finally, as $\rho$ is an arbitrary function on the sphere (that can be labelled by spherical harmonics) there are infinitely many charges for each on-shell field configuration, which fixes $\mathcal{Q}$!

The action of these charges on the other phase space variables is computed with the Poisson brackets defined by the inverse symplectic form from \eqref{eq:sym_form_inv}:
\begin{equation}
    \{ Q_\rho , O(x)\} = \int _{\p \Sigma }\sqrt{g} \, \dt y ^2 \,  \rho (y) \, \frac{\d O (x)}{\d \varphi  (y)} \implies \{ Q_\rho  , A^\dr (x)\} =  \dt \rho (x)\,.
\end{equation}
From here we can also conclude that the set of all these charges forms an Abelian algebra with no central extension:
\begin{equation}
    \{ Q_\rho , Q_{\rho '}  \} = 0\,.
\end{equation}

\paragraph{Hamiltonian:}
~\newline
Using the covariant phase space methods (contracting the symplectic form with the phase space vector field $X_{\p _t}$, associated to time translations) and some algebra we can also compute the Hamiltonian. The result is: 
\begin{equation}
    \begin{split}
        \d H  & = X_{\p _t} \cdot \Omega \\
        & =\int _{\Sigma } \, \epsilon  _\Sigma  \,  \l \p _t \Tilde{A} _a ^\dr \d F^{ta} - \d \Tilde{A} _a ^\dr \sqrt{g} ^{-1} \p _t \l \sqrt{g} F^{ta} \r  \r  + \int _{\p \Sigma }\, \epsilon _{\p \Sigma } \, \l \p _t  \varphi \d F^{tr}    -\d \varphi \sqrt{g} ^{-1}\p _t \l \sqrt{g} F^{tr} \r \r \\
        &=\int _{\Sigma }\, \epsilon _\Sigma \, \l   F_{ta} \d F^{ta} +  F_{tr} \d  F^{tr}  - \frac{1}{2} \d F_{ba} F^{ba} -\d F_{ra}  F^{ra} \r  \\
         & \qquad \qquad \qquad \qquad  + \int _{\p \Sigma } \, \epsilon _{\p \Sigma } \, \l  \d \Tilde{A} ^\dr _a F^{ra}  +  \d \p _a \varphi F^{ra} +  A_t ^\dr  \d F^{tr} \r\,.
    \end{split}
\end{equation}

Next, we need to discuss integrability. The Hamiltonian for this theory is always integrable. To see this, we can momentarily use a different coordinate choice, in which the timelike vector of the Hamiltonian takes a simple form $n_\mu  = \dt t$, $n ^\mu  = \p _t$. Then we have:
\begin{equation}
\begin{split}
    \d H = -& \int _{\Sigma }\, \epsilon _\Sigma \, \l -\frac{1}{2} F_{\mu \nu } \d F^{\mu \nu } + n^\rho F_{\mu \rho} \d \l n^{\sigma} F^{\mu} _{ \  \sigma } \r  \r   \\ 
      &  -  \int _{\p \Sigma } \, \epsilon _{\p \Sigma } \, \l  \d \Tilde{A} ^\dr _a F^{ra} - \d \varphi \sqrt{g} ^{-1} \p _a \l \sqrt{g} F^{ra} \r  + \d \l  A_t ^\dr   F^{tr} \r \r\,.
\end{split}
\end{equation}
We can immediately see that the bulk term is phase space exact. To show it for the boundary piece, we need to use the boundary conditions. While for any of type of soft boundary conditions we have $\d \varphi \sqrt{g} ^{-1} \p _a \l \sqrt{g} F^{ra} \r  = \d \l \varphi \sqrt{g} ^{-1} \p _a \l \sqrt{g} F^{ra} \r \r$, the term $\d \Tilde{A} ^\dr _a F^{ra}$ is either zero or not, depending on the type we choose. Therefore, the expression for the Hamiltonian depends on the boundary conditions and indirectly on the boundary action $\ell _{\text{soft}}$ (see appendix \ref{app:post_DBC}), even though the symplectic form does not. This is consistent with a more standard definition of the Hamiltonian as the Legendre transform of the Lagrangian. After some manipulations of the expression above we find:
\begin{equation}
\begin{split}
    H  = \int _\Sigma  \, \epsilon _{\Sigma } \, \f{ E ^2 + B^2} {2}   - \int _{\p \Sigma } \, \epsilon _{\p \Sigma } \, \l  A_t ^\dr F^{tr} + \varphi \Tilde{\nabla }^2 \mathfrak{p}  \r  \quad 
\end{split}
- \quad \left\{
    \begin{split}
        & \text{Soft Dirichlet} \quad 0 \, .\\
       &  \text{Soft Neumann} \quad  \p _a h \Tilde{\nabla }^a \mathfrak{h} \, .\\
       & \text{Soft Robin} \quad  -\frac{b}{2a} \p _a \mathfrak{h}  \Tilde{\nabla} ^a \mathfrak{h}\,. 
    \end{split} \right.
\end{equation}

The form of this Hamiltonian tells us that shifting the value of $\varphi$ does not change the energy of the field, when it is in its vacuum state ($F=0$). Therefore, frame reorientations are symmetries that take us from one vacuum to another. From the point of view of the subregion, classically these vacua are indistinguishable.\footnote{However, in the quantum theory one expects that they will correspond to orthogonal Hilbert space vectors which is what happens for the theory defined in Minkowski spacetime. This has been explored in various papers, \cite{Strominger:2013lka, He:2014cra, Kapec:2015ena, Campiglia:2015qka, Kapec:2017tkm}, and while it is an important insight, a more recent line of work  \cite{Prabhu:2022zcr, Prabhu:2024lmg} has shown that these vacua states do not belong to the standard Fock space (because their norm diverges), and advocates for an algebraic formulation of QED scattering.}

\subsection{Intrinsic Boundary Conditions}
 
This analysis of the symplectic form is also helpful in understanding the boundary conditions on the unextended phase space. This question is relevant if we are interested in keeping track of observables local to the subregion only and would like a well-defined variational principle for those. 

The intrinsic symplectic flux is:
\be
\begin{split}
    \omega^\dr_\text{int}\big|_\Gamma  =  \epsilon_\Gamma\, \l  \d \p _a  h \curlywedge  \d \Tilde{\nabla} ^a \mathfrak{h} 
 + \delta\Tilde A^\dr_t\curlywedge \delta  F^{rt} \r  \, .
\end{split}
\ee

For the first term, we once again have the choice of fixing half of the degrees of freedom of $\l h, \mathfrak{h}\r$.  For the second we do not. The boundary value of $F^{rt}$ must be fixed in order to have an invertible symplectic form, as we have removed its symplectic conjugate $\varphi$ from the phase space. 

However, for a similar reason, we must also impose $\d \Tilde{A}^\dr _t \big|_{\Gamma } = 0$. To understand why, we can look at the intrinsic symplectic form in terms of the quantities defined in \eqref{eq:intr_hodge_split}:
\begin{equation}
\begin{split}
    \Omega _{\text{int}} =  \int _\Sigma \, \epsilon_\Sigma  \, \l   \d \p _a \Tilde{\varrho} \curlywedge \d \Tilde{\nabla }^a \Pi _{\varrho  }   + \d \p _a h \curlywedge \d  \Tilde{\nabla } ^a  \Pi _h    \r  \, . 
\end{split}
\end{equation}
 The restriction of the first term on the boundary is: 
\begin{equation}
  \l   \d \p_a\Tilde{\varrho} \curlywedge \d \Tilde{\nabla }^a \Pi _{\Tilde{\rho } }  \r \big|_{\p \Sigma } =  \d \p_a \Tilde{\varrho} \big|_{\p \Sigma }\curlywedge \d \Tilde{\nabla }^a \Tilde{A} _t ^\dr \big|_{\p \Sigma }  \, ,\label{eq:int sympl}
\end{equation}
where we have used $\Tilde{\nabla }^2 \Pi _{\Tilde{\rho } } : = \p _a \l \sqrt{g} F^{ta} \r $, which on $\p \Sigma $ is equal to $D^2 \tA ^\dr _t \big|_{\p \Sigma } $, by its definition \eqref{eq:tilde fixed}.
But on the intrinsic phase space $ \Tilde{\varrho} \big|_{\p \Sigma }  = 0 $. This means that the phase space direction corresponding to shifts in $\Tilde{A} _t ^\dr \big|_{\p \Sigma } $ is degenerate (in contrast to the extended phase space in section~\ref{subsec:in_dat_ev}). So its value must also be must be quotiented, leaving us to pick a particular profile for $\tA_t ^\dr \big|_{\Gamma}: =  \Pi _{\Tilde{\rho } }  \big|_{\Gamma } $ that would be a constant on the intrinsic phase space. 

\begingroup
\renewcommand{\arraystretch}{1.7}
\begin{table}[h!]
    \centering
        \begin{tabularx}{0.6\textwidth}{|>{\centering\arraybackslash}X|}
            \hline
            \textbf{Intrinsic Boundary Conditions} \\
            \hline
            $\delta \tilde A^\dr_t\big|_\Gamma=0$ \\
            \hline
            $\delta F^{rt}\big|_\Gamma=0$ \\
            \hline
            $\delta \p_a h\big|_\Gamma=0$ \big/ $\delta \p_a \mathfrak{h}\big|_\Gamma=0$ \big/ $\d \l \alpha \p_a h   + \beta   \p_a \mathfrak{h}  \r \big|_{\Gamma }    = 0$\\
            \hline
        \end{tabularx}
        \caption{Summary of boundary conditions on the \emph{unextended} phase space. The second line is a stronger version of the second line in table \ref{table:sbc_types} because $F^{rt}\big|_{\p\Sigma}$ is now no longer part of the phase space. All the conditions are (quasi)local to the subregion. They also come in Dirichlet/Neumann/Robin types, respectively. There are no corner charges.}
        \label{table:intrinsic bcs}
\end{table}

We see that to have a well-defined theory on the unextended phase space, we must not only do a post-selection via boundary conditions, but also a super-selection, as we are fixing more boundary degrees of freedom than necessary for a vanishing symplectic flux. The fixing of $F^{rt}\big|_\Gamma$ was advocated for also in \cite{Riello:2021lfl},\footnote{If the reader goes back to section \ref{subsec:in_dat_ev}, they can convince themselves that one needs to know the value of $F^{rt}\big|_{\p\Sigma}$ on \emph{every} slice in order to be able to evolve the intrinsic initial data associated with $\tilde A^\dr$. This point was also explained in \cite{Riello:2021lfl}, in the context of a causal diamond. On the unextended phase space, we have that $F^{rt}\big|_{\p\Sigma}$ commutes with every observable (possibly up to a phase space constant), as can be seen from inspecting equation \eqref{eq:int sympl}. These properties grant the observable $F^{rt}\big|_{\p\Sigma}$ the status of a \emph{superselection} variable. This discussion justifies the treatment in \cite{Donnelly:2015hxa,Donnelly:2014fua,Ball:2024hqe,Ball:2024xhf},  where sums over superselection sectors are carried out.} however the fixing of $\tilde A^\dr_t\big|_\Gamma$ is new. Table \ref{table:intrinsic bcs} summarizes the types of \emph{intrinsic boundary conditions} for the unextended phase space.

\section{Explicit Setup in Minkowski}\label{sec:sol_space}

So far we have considered a general asymptotically flat setup in developing the formalism of intrinsic/extrinsic frames, unextended/extended phase spaces, Goldstone modes and soft boundary conditions. Now we would like to consider an explicit scenario which allows one to write down formulae that directly connect the finite region edge physics to the asymptotic soft physics. In doing so, we will see the explicit manifestation of several general observations made so far in this article, but we will also be able to connect the asymptotic memory effect with a finite version of it, as we will see in section \ref{sec:gold_soft}.

The specification we adopt from now on involves the following three choices:
\begin{itemize}
    \item we work in Minkowski spacetime,
    \item our subregion is bounded by a tube at constant radius, and
    \item our extrinsic frame $\Phi$ is built from \emph{null} Wilson lines.
\end{itemize}
This setup is depicted in figure \ref{fig:subregion_rad_Wl}. From now on, the coordinates $\l t, r, x^a\r$ defined in section \ref{sec:notations} are the standard flat space spherical coordinates $\l t, r, \theta, \phi \r$. In the following sections we will also use retarded null coordinates $\l u, \tilde{r}, \theta, \phi \r$, where even though $\tilde{r}=r$, the distinction between them is important as $V_{\tilde{r}} = V_r + V_t$ for a general covariant vector $V_\mu$. The subregion is defined by $r =r_s$ and it has a finite lifetime - from $t^-$ to $t^+$. In subsection \ref{subsec:big_edges}, we will also discuss what happens in the limit where $t ^\pm  \to \pm \infty$.  Charged massive matter fields may exist outside the subregion of interest. This matter can be anything from charged point particles to massive Dirac spinors, and we denote these fields generically by $\psi $.

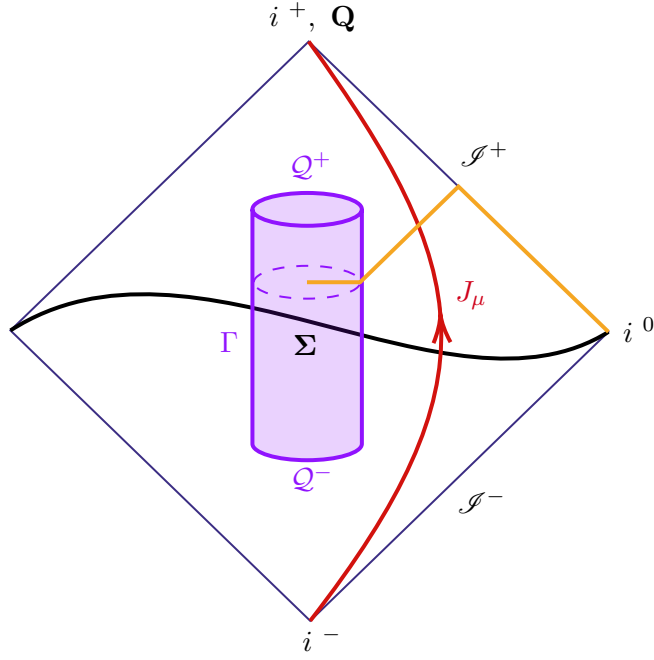
\begin{figure}[h!]
    \centering

\tikzset{every picture/.style={line width=0.75pt}} 

\begin{tikzpicture}[x=0.75pt,y=0.75pt,yscale=-1,xscale=1]

\draw  [color={rgb, 255:red, 59; green, 45; blue, 131 }  ,draw opacity=1 ][line width=0.75]  (326.54,46.58) -- (476.91,192.24) -- (327.29,337.19) -- (176.92,191.52) -- cycle ;
\draw [line width=1.5]    (176.91,191.52) .. controls (269.72,132) and (396.62,242.18) .. (476.92,192.24) ;
\draw [color={rgb, 255:red, 210; green, 18; blue, 14 }  ,draw opacity=1 ][line width=1.5]    (326.54,46.58) .. controls (409.67,161.33) and (420.33,217.33) .. (327.29,337.19) ;
\draw [shift={(392.6,183.47)}, rotate = 86.66] [color={rgb, 255:red, 210; green, 18; blue, 14 }  ,draw opacity=1 ][line width=1.5]    (14.21,-4.28) .. controls (9.04,-1.82) and (4.3,-0.39) .. (0,0) .. controls (4.3,0.39) and (9.04,1.82) .. (14.21,4.28)   ;
\draw  [color={rgb, 255:red, 144; green, 19; blue, 254 }  ,draw opacity=1 ][fill={rgb, 255:red, 144; green, 19; blue, 254 }  ,fill opacity=0.19 ][line width=1.5]  (353.21,130.77) -- (353.21,248.39) .. controls (353.21,252.93) and (340.92,256.62) .. (325.76,256.62) .. controls (310.6,256.62) and (298.31,252.93) .. (298.31,248.39) -- (298.31,130.77) .. controls (298.31,126.23) and (310.6,122.54) .. (325.76,122.54) .. controls (340.92,122.54) and (353.21,126.23) .. (353.21,130.77) .. controls (353.21,135.32) and (340.92,139.01) .. (325.76,139.01) .. controls (310.6,139.01) and (298.31,135.32) .. (298.31,130.77) ;
\draw  [color={rgb, 255:red, 144; green, 19; blue, 254 }  ,draw opacity=1 ][dash pattern={on 4.5pt off 4.5pt}] (299,167.33) .. controls (299,162.79) and (310.94,159.1) .. (325.67,159.1) .. controls (340.39,159.1) and (352.33,162.79) .. (352.33,167.33) .. controls (352.33,171.88) and (340.39,175.57) .. (325.67,175.57) .. controls (310.94,175.57) and (299,171.88) .. (299,167.33) -- cycle ;
\draw [color={rgb, 255:red, 245; green, 166; blue, 35 }  ,draw opacity=1 ][line width=1.5]    (325.67,167.33) -- (352.33,167.33) -- (401.67,119.33) -- (476.91,192.24) ;

\draw (401.11,95.49) node [anchor=north west][inner sep=0.75pt]  [font=\normalsize]  {$\scri^{+}$};
\draw (399.72,269.88) node [anchor=north west][inner sep=0.75pt]  [font=\normalsize]  {$\scri^{-}$};
\draw (482.82,183.45) node [anchor=north west][inner sep=0.75pt]  [font=\normalsize]  {$i\ ^{0}$};
\draw (317.18,192.23) node [anchor=north west][inner sep=0.75pt]  [font=\normalsize]  {$\mathbf{\Sigma }$};
\draw (315.24,100.7) node [anchor=north west][inner sep=0.75pt]  [font=\normalsize,color={rgb, 255:red, 144; green, 19; blue, 254 }  ,opacity=1 ]  {$\mathcal{Q}^{+}$};
\draw (315.84,257.5) node [anchor=north west][inner sep=0.75pt]  [font=\normalsize,color={rgb, 255:red, 144; green, 19; blue, 254 }  ,opacity=1 ]  {$\mathcal{Q}^{-}$};
\draw (305.25,25.38) node [anchor=north west][inner sep=0.75pt]  [font=\normalsize]  {$i\ ^{+} ,\ \mathbf{Q}$};
\draw (322.01,339.01) node [anchor=north west][inner sep=0.75pt]  [font=\normalsize]  {$i\ ^{-}$};
\draw (280.67,190.73) node [anchor=north west][inner sep=0.75pt]  [color={rgb, 255:red, 144; green, 19; blue, 254 }  ,opacity=1 ]  {$\Gamma $};
\draw (398.67,165.07) node [anchor=north west][inner sep=0.75pt]  [color={rgb, 255:red, 208; green, 2; blue, 27 }  ,opacity=1 ]  {$J_{\mu }$};

\end{tikzpicture}
    \caption{ The explicit physical setup of interest, which is studied from this section onwards. We have a subregion, finitely extended in both space and time, embedded in Minkowski spacetime. Its timelike boundary is denoted by $\Gamma $, and the initial and final electric charge density by $\mathcal{Q} ^\pm $, respectively. The reference frame used for its description is constructed via Wilson lines going in a null direction form $\Gamma$ to $\scri ^+$, and then to $\scri ^+ _-$. There may be massive matter fields outside the subregion, whose current is depicted as $J_\mu$. This sources a global charge density $\mathbf{Q}$ at $\scri^+_+$. }
    \label{fig:subregion_rad_Wl}
\end{figure}

To build the subregion relational observables, we use an extrinsic reference frame field built from null Wilson lines. Outside the subregion, the lines follow null paths defined by $u=t-r=const$, with the angular coordinates also held fixed, until they reach $\scri ^+$. Afterwards, they follow light rays on $\scri ^+$ (so $\theta, \phi $ are still constant) until they reach $\scri ^+ _-$. These lines are illustrated in figure \ref{fig:subregion_rad_Wl}. With these Wilson lines, we can define the dressed field $
\bar{A} ^\dr$ everywhere in the complement region $\bar{\mathcal{R}}$ and \emph{on} the boundary $\Gamma$ in the following way:
\begin{equation}\label{eq:adr_ru}
\begin{split}
    \bar{A}^\dr _\mu (u',\tilde{r}',\theta, \phi)  &= \bar{A} _\mu (u',\tilde{r}',\theta, \phi)   -  \p _\mu \l \int _{\infty} ^{\tilde{r}' } \bar{A} _{\tilde{r}} (u',\tilde{r},\theta, \phi) \dt \tilde{r}  + \int _{u= - \infty} ^{u'} \bar{A} _u (u, \infty , \theta, \phi) \dt u \big|_{\scri ^+} \r \\
    &= \int _{\infty} ^{\tilde{r}' }  \bar{F} _{r \mu } \dt \tilde{r}  + \int _{u= - \infty} ^{u'}  \bar{F}_{u \mu } \dt u +  \bar{A}_\mu (-\infty, \infty , \theta , \phi ) \, .
\end{split}
\end{equation}

The main advantage of this construction is that the dressing condition outside the subregion:
\begin{equation} \label{eq:dr_cond_ru_out}
\bar{A} _{\tilde{r} }^{ \dr } \big|_{\bar{\mathcal{R}}}=0=\bar{A}^\dr  _u \big|_{\scri ^+} 
\end{equation}
takes the same form as the retarded radial gauge condition $\bar{\mathcal{A}} _{\tilde{r}} =0=\bar{\mathcal{A}} _u \big|_{\scri ^+}$, used in the literature of asymptotic symmetries (where a gauge-fixed route is usually taken). Because of the equivalence of gauge-fixed and gauge-invariant approaches, shown in subsection \ref{subsec:equivalence}, any results presented in this section are straightforwardly translated to existing ones, derived with gauge-fixed quantities.\footnote{Note that whatever matter we have in the complement, it must also be dressed. This is achieved by acting with the operator representation of the frame on the matter fields, as outlined in \cite{Carrozza:2021gju}:
\begin{equation}
    U[A, \psi ] \rhd \psi  = e^{-i \Phi [A, \psi ]} \psi  =: \psi ^\dr\,. 
\end{equation}}

This was the construction of the reference frame $\Phi$ outside and \emph{on} $\Gamma$, and now we must extend it into the interior of the subregion. Inside, the Wilson lines are extended radially inward in a \emph{space}-like direction to follow the standard radial coordinate. The dressing condition inside is:
\begin{equation}\label{eq:dr_cond_ins}
    A^\dr _r \big|_\mathcal{R} = 0\,.
\end{equation}

Finally, we point out that here we do \emph{not} introduce a new global edge field $\Phi^\text{gl} $ on the asymptotic boundary, which would make the global theory small \emph{and} large gauge-invariant, with the symmetries being promoted to (global) frame reorientations, even though this could in principle be done according to the agnostic discussion in sections~\ref{sec:ref_frame} and~\ref{sec:BC}. There are two reasons for this. Firstly, even though we can add an edge mode to the global boundary and recast large gauge transformations as frame reorientations, we do not need to do so. This is because here we are not interested in embedding our spacetime into a larger one, and a global frame would not bring additional physical insights for our purposes. However, it is worth noting that embedding the global theory in an even larger universe has shown to be useful in some circumstances, for example when computing the vacuum entanglement entropy at a cut of $\scri ^+$ \cite{Chen:2023tvj, Chen:2024kuq}. Such a construction fits into the framework developed in this paper. Secondly, in this way we keep the term ``large gauge transformations'' (not vanishing at $\scri ^+ _-$) solely for symmetries of the global solution space, differentiating them clearly from the much larger class of frame reorientations, which are the symmetries of the subregion solution space.

\subsection{Review of Global Solution Space}\label{subsec:review_gss}

Now we briefly review the global solution space of free Maxwell theory, in terms of an analytic expansion in $r$ of the dressed field $\bar{A}^\dr$ around $\scri^+$. This follows closely the works of \cite{He:2014cra, Kapec:2014zla, Strominger:2017zoo}. 

Finiteness of the global energy and momentum: 
\begin{equation}
    P ^{\mu} = \int _{\scri ^+ } \epsilon_\scri  \, T^{\tilde{r} \mu} , \quad T^{\mu \nu } = F^{\mu \rho } F_\rho ^{ \ \nu } - \frac{1}{4} g^{\mu \nu } F_{\rho \sigma } F^{\rho \sigma }\,,
\end{equation}
together with the dressing condition \eqref{eq:dr_cond_ru_out} imply the following radial fall-off for the dressed potential, as $\tilde{r} \to\infty$: 
\begin{equation}
    \Bar{A} _u ^\dr \sim O(\tilde{r}^{-1}) , \quad \Bar{A}_a ^\dr \sim O(1)\, , \quad \Bar{A} _{\tilde{r}} ^\dr = 0 \, .
\end{equation}
Furthermore, finiteness of the energy density integral on $\scri ^+$: 
\begin{equation}
     P^u = \int _{\scri ^+ } \sin \theta \, \dt u \dt \theta \dt \phi   \,  \l  \p _u \bar{A}_{\theta } ^{ \dr} \p _u \bar{A}^{ \dr} _\theta  + \csc ^2 \theta   \p _u \bar{A}_{\phi } ^{ \dr} \p _u \bar{A} _\phi  ^{\dr} \r < \infty \, 
\end{equation}
requires the following behaviour on $\scri^+$, as $u\to-\infty$:
\be
\p _u \bar{A}_a ^\dr\big|_{\scri ^+ _-}  = 0.
\ee
From these, a generic solution to the leading order in $\tilde{r}$ expansion of the Maxwell equations, $\tilde{r} \p_u \bar{A}^\dr_u=D^a\bar{A}^\dr_a+O(\tilde{r}^{-1})$, is of the form:
\begin{equation}\label{eq:large_r_sol}
\begin{split}
    &\bar{A}_\theta ^\dr  = \p _\theta \Tilde{\phi} + \csc \theta \p _\phi b +O(\tilde{r}^{-1})\,, \quad \Tilde{\phi}  = \Tilde{\phi } (u, \theta, \phi )\,,\quad b = b (u, \theta, \phi )  \\
    &\bar{A}_\phi ^\dr  = \p _\phi \Tilde{\phi}- \sin \theta \p _\theta  b  + O(\tilde{r}^{-1}) \,,\\
    &\bar{A}_u  ^\dr = \frac{1}{\tilde{r}} \l \Tilde{\mathbf{Q}}  + D^2 \Tilde{\phi}\r + O(\tilde{r}^{-2})\,,  \quad  \Tilde{\mathbf{Q} } = \Tilde{\mathbf{Q} } \l \theta, \phi \r \,,
\end{split}
\end{equation}
where usually it is taken that $ \lim _{u \to \pm \infty } b = 0 $ (``no magnetic charges condition'' or other words no radial magnetic field), a choice we therefore also make here. We have once again split the angular part of $\bar{A}^\dr$ in terms of curl-free ($\tilde\phi$) and divergence-free ($b$) parts. As of now, $\Tilde{\mathbf{Q}}$ is a free function on the celestial sphere. It will be fixed shortly.

Next, we define the following key (small) gauge-invariant, global quantities:
\begin{equation}
  \begin{split}\label{eq:gl_GM_SM_def}
  & \quad \lim _{u \to \pm \infty } \Tilde{\phi} = \phi ^\pm\,, \\
                \textbf{Goldstone mode (GM):}& \quad   \phi  =\frac{1}{2} \l \phi ^+ + \phi ^- \r\,, \\
        \textbf{Soft mode:}& \quad \p _a N  =\p _a \l  \phi ^+ - \phi ^- \r\,.
  \end{split}
\end{equation}
Both  the Goldstone and soft modes $\phi,N$ can be arbitrary functions on the celestial sphere and we will discuss their meaning in detail in the next section. Briefly, $N$ is called ``soft'' mode as it is the zero energy mode of the radiation at $\scri ^+$. The crucial distinction between them is that the GM transforms under large gauge transformations as $\phi \mapsto \phi + \alpha\big|_{\scri^+_-} $, while the soft mode is completely large gauge-invariant. This can be seen by remembering that the action of a large gauge transformation on the dressed observable $\bar{A}^\dr$ respects the dressing condition (here \eqref{eq:dr_cond_ru_out}), meaning that $\bar{A}^\dr_a(u)\mapsto \bar{A}^\dr_a(u)+\p_a \alpha\big|_{\scri^+_-}$. This means that $\tilde \phi(u)\mapsto \tilde\phi(u)+\alpha\big|_{\scri^+_-}$, from which the result follows. 

Next, we introduce the following convenient, (large) gauge-invariant variable: 
\begin{equation}\label{eq:hphi_def}
    \hat{\phi } := \Tilde{\phi } - \phi , \qquad \lim _{u \to \pm \infty } \hat{\phi } = \pm \frac{1}{2}N\,.
\end{equation}
With this variable, we can re-express the solution in \eqref{eq:large_r_sol} as:
\begin{equation}\label{eq:large_r_sol_2}
\begin{split}
    \bar{A}_\theta ^\dr  &= \p _\theta \phi  + \p _\theta \hat{\phi} + \csc \theta \p _\phi b  +O(\tilde{r}^{-1}) \,, \\
     \bar{A}_\phi ^\dr  &= \p _\phi \phi + \p _\phi \hat{\phi}- \sin \theta \p _\theta  b  + O(\tilde{r}^{-1}) \,,\\
     \bar{A}_u  ^\dr &= \frac{1}{\tilde{r}} \l \Tilde{\mathbf{Q}}+ D^2 \phi  +  D^2 \hat{ \phi }\r+ O(\tilde{r}^{-2}) \, .
\end{split}
\end{equation}
The free function $\Tilde{\mathbf{Q}}$ is fixed by the requirement that at $\scri ^+ _+$, the radial electric field $\tilde{r}^2F_{\tilde{r} u} \big | _{\scri ^+ _+}$ is equal to the total charge density $\mathbf{Q}$ sourced by massive fields at $i^+$, which ensures $F_{\tilde{r} u}$ is continuous across $\scri ^+_+$. Using the fact that, for our dressing condition, $F_{\tilde{r}u}=\p_{\tilde{r}} \bar{A}^\dr_u$, we get that:
\begin{equation}
   \Tilde{\mathbf{Q}} = \mathbf{Q} -  D^2 \phi  - \frac{1}{2} D^2 N  \implies \lim _{u \to +\infty }\tilde{r}\bar{A}_u ^{\dr }\big|_{\scri^+}  = \mathbf{Q}, \quad \lim _{u \to - \infty }\tilde{r} \bar{A}_u ^{\dr}\big|_{\scri^+} = \mathbf{Q} - D^2 N  \, . \label{eq:continuity}
\end{equation}
This in turn implies that $\star F \big| _{\scri _- ^+ }= D^2 N - \mathbf{Q}$. So the global charge is written in terms of the soft mode and the matter charge at $i^+$.

This form of the solution is particularly nice, as it clearly separates the variable that transforms under large gauge transformations -- namely $\phi $ -- from the rest of the degrees of freedom that do not. 

Solving Maxwell's equations order by order in  $\tilde{r}$, one finds that the subleading components in the large $\tilde{r}$ expansion of $\bar{A}^\dr$ (that we can call $A^{\dr, n }$) are determined by the free data at leading order ($\tilde{\phi} $, $b$ and $\mathbf{Q} - D^2 N$), up to the boundary values of the angular part of the dressed potential: $\bar{A}^{ \dr\, (n) }_a \big|_{\scri ^+_-}$. However, these extra pieces of free data will not appear in the symplectic form evaluated \emph{on} $\scri^+$, as they are subleading in $\tilde{r}$. This appears surprising, yet in \cite{Campiglia:2018dyi} the authors show the sub-leading charges are related to the sub-subleading soft photon theorem, described in \cite{Hamada:2018vrw}, and give evidence that such a relationship exists at all orders.

Overall, the global solution space is specified by:\footnote{Notice that because we are using a null Cauchy slice, we seem to require only \emph{half} of the degrees of freedom in order to fully specify a solution. This is what is called a \emph{characteristic initial value problem}. In contrast to the situation on a spacelike Cauchy slice, the fields do \emph{not} commute with themselves along $\scri^+$, as can be seen from the fact that they are symplectically conjugate to their $u$-, i.e.\ tangential derivative in \eqref{eq:omega scri}. This fact will be important in section \ref{sec:charge action}.} 
\begin{equation}
    \mathcal{S}_{\text{global}} = \{ \hat{\phi}(u,x^a),\,  b(u,x^a) , \,  \phi (x^a),\,  N(x^a) 
    ,\, \mathbf{Q}(x^a)   \} \cup \{ \text{Initial data for matter at $i^+$}  \}\,,
\end{equation}
with the symplectic form evaluated on $\scri^+\cup i^+$ taking the form (cf.\ \eqref{eq79}): 
\be
    \mathbf{\Omega} & = \int _{\scri^+}\,\epsilon _{\scri^+}  \, \l \d \p _a \hat{\phi } \curlywedge \d D^a  \p _u \hat{\phi } + \d \p _a b \curlywedge \d D^a \p _u  b \r + \int _{\scri ^+ _-} \, \epsilon _{\scri ^+ _-} \,   \d \phi \curlywedge \d \l  D^2 N  - \mathbf{Q}  \r + \mathbf{\Omega}_\text{matter} \, . \label{eq:omega scri}
\ee

The large $\tilde{r}$ analytic properties of the solutions above are with respect to the null $u=t-r $ and radial $\tilde{r}$ coordinates and unfortunately do not necessarily translate to an asymptotic analytic expansion in the more standard Minkowski $\{t, r\}$ coordinate system. Therefore, we do not present a large $r$ expansion around $i^0$ of the solution space developed here. Instead, we give more general arguments regarding relevant properties of the electromagnetic field around $i^0$. 

\paragraph{Global Boundary Conditions:}

Because $i^0$ is a global timelike boundary we must impose some boundary conditions there.  If we wish to preserve the infinite set of large gauge transformations as physical symmetries of the global theory, we need to impose one of the soft boundary conditions presented in table \ref{table:sbc_types}. To decide which one, we once again demand a finite energy and momentum on a spacelike Cauchy surface $\Sigma $ ending at $i^0$. The asymptotic nature of $\Sigma $ requires (at least)  that $\sqrt{g}  T^{\mu t}  \big |_{i^0}= 0 $ for this to be true. This condition is satisfied when almost all components of $F_{\mu \nu }\big|_{i_0}$ vanish. This can be straightforwardly seen by writing $\sqrt{g}  T^{\mu t}  \big |_{i^0}= 0 $ in terms of $F_{\mu \nu }$.  Therefore the relevant boundary conditions are the large $r$ limit of soft Neumann of the following type:\footnote{We remind the reader that, at the global boundary, the dressed and bare observables become the same, since we are not including a global edge mode ($\Phi^\text{gl}$) in this analysis: $\bar{A}^\dr\big|_{i^0}=\bar{A}\big|_{i^0}$.}
\begin{equation}
   \sqrt{g} F^{ra} \big|_{i^0}  = 0 , \qquad \bar{A}_t   \big|_{i^0}  = 0. \label{eq:global bcs}
\end{equation}
The first condition above ensures finite and conserved electric charge density: 
\begin{equation}
 \p _t \l \sqrt{g} F ^{tr} \r \big|_{i^0} = 0 \, , \qquad \sqrt{g} F^{rt} \big|  _{i^0} < \infty. \label{eq:global bcs 2}
 \end{equation}
The additional requirement coming from the fall-off conditions are:
\begin{equation}\label{eq:bcsio+}
    F_{ta} \big|_{i^0}  = F_{tr} \big|_{i^0}  = 0 .
\end{equation}
Notice that this partially constrains the non-exact part of $A_a ^\dr |_{i^0}$, or the radial magnetic field, to be time-independent $\p _t F_{\theta \phi}\|_{i ^0}=0.$ However, since at $\scri ^+ _-$ we have previously imposed vanishing radial magnetic field, we must also impose $F_{\theta \phi } \big|_{i^0} = 0.$
In conjunction, this amounts to imposing the large $r$ limit of both soft Dirichlet and Neumann boundary conditions:
\begin{equation}
    \p _t \l \sqrt{g} F ^{tr} \r \big|_{i^0} =  \sqrt{g} F^{ra} \big|_{i^0}  = 0 , \quad  A ^\dr \big|_{i^0} = \dt \phi ^- , \quad  \p _t \phi ^- = 0 \, .
\end{equation}

In \cite{He:2020ifr,He:2023bvv} the authors study the asymptotic structure and covariant phase space of non-Abelian gauge theory. Using similar considerations of finite energy flux trough $\scri ^+$, vanishing asymptotic radial magnetic field, and vanishing symplectic flux at $i^0$, the authors impose boundary conditions that reduce to the above in the Abelian case.

An important  distinction between finite and infinite distance is that different fixed profiles of $A^\dr _t $ at finite $r$ embed the subregions into different global solution spaces, whereas different choices of $A^\dr _t \big|_{i^0}$ do not have physical consequences. For this reason, one can freely impose $A^\dr _t \big|_{i^0}=0$, as we do here and as is usually done.

\paragraph{Subregion solution space}

Finally, we briefly remind the reader of the subregion solution space. It has a timelike boundary and spacelike Cauchy surface, so we cannot do the same type of expansion as we did for the global theory. All we can say from the analysis of the previous sections (cf.\ section~\ref{subsec:in_dat_ev}) is that its solution space is specified by:
\begin{equation}
    \mathcal{S}_{\text{subregion}} = \{ \Tilde{A}_a ^\dr \big|_\Sigma , \, F^{ta} \big|_\Sigma , \,  \mathcal{Q}  \big|_{\p \Sigma }, \,  \varphi \big|_{\p \Sigma } \}\,,
\end{equation}
where we have chosen some type of soft boundary conditions from table \ref{table:sbc_types} (all of which are here equally good).  For future reference we also define here $\mathcal{Q} ^ \pm  :=  F^{rt} \big |_{t=t^\pm , \, r=r_s} $ .

\section{Asymptotic and Finite-Distance Goldstone and Soft Modes}
\label{sec:gold_soft}

In this section, we will revisit the standard notions of asymptotic Goldstone and soft modes and their connection with the asymptotic memory effect, before proposing their quasi-local counterparts at finite distance using our timelike tube setup and exploring the link between the two settings.

\subsection{Goldstone Mode}\label{subsec:GM}

As already mentioned in the introduction, a line of seminal work (\cite{Kapec:2014zla, Pasterski:2015zua, Pasterski:2017kqt} and many others) has established an equivalence between different phenomena in the infrared physics of gauge theories and gravity. One side of this triangle relates soft theorems of scattering amplitudes with the Ward identities of the asymptotic charges generating large gauge transformations. The existence and validity of both require accepting that the electromagnetic vacuum is infinitely degenerate:
\begin{equation}
    F = 0   \implies A^\dr  =  \dt \phi \,.
\end{equation}

For the global theory, this is the same $\phi$  as defined in equation \eqref{eq:gl_GM_SM_def}. Two functions $\phi $ with different profiles on $\scri ^+_-$ specify distinct solutions and we can interpolate between the two with a large gauge transformation. At the classical level, it seems unnatural to accept $\phi $ as a physical degree of freedom as it cannot be measured by any local experiment on radiation. However, $\phi$ is symplectically conjugate to a measurable quantity, namely the memory or soft mode (as we discuss in the next subsection). Furthermore, quantum mechanically, vacuum states labelled by a different $\phi $ need to be orthogonal states in the Hilbert space in order to avoid IR divergences of the scattering matrix.\footnote{It turns out that even that approach has issues, as it leads to non-normalizable states. For a recent detailed discussion on this see \cite{Prabhu:2022zcr}.}

For the subregion theory, the equivalent variable is $\varphi$, which carries over the concept of the Goldstone mode to the quasi-local context. While the quasi-local vacuum $F|_\Sigma = 0$ in a subregion does not always correspond to the global vacuum, they are indistinguishable, at least classically, from the subregion's perspective, when we restrict ourselves to probing the solution with local observables only. The subregion Goldstone mode corresponds to the (relational) spontaneous breaking of \emph{reorientation symmetry} by the quasi-local vacuum, as discussed in section \ref{subsec:fr_reor}. We can move between different such vacua via a complement transformation amounting to a frame reorientation. As discussed in section \ref{subsec:fr_reor}, these are more general than large gauge transformations.\footnote{It also is worth noting that in some works on asymptotic symmetries \cite{Chen:2023tvj,Chen:2024kuq}, the Minkowski space is embedded within a larger universe with sources outside the original spacetime. Nevertheless, even in these settings, the Goldstone mode is identified with the degeneracy parameter of the ‘‘vacuum'' in the internal patch.}

We now examine how $\varphi $ is related to degrees of freedom at the global boundary, including the global Goldstone mode $\phi$. From the definition of the dressed field in \eqref{eq:alternative wilson}, we have that, for spacelike/null Wilson lines, at any cut $\p \Sigma $ the exact piece of $A^\dr _a  \big| _{\p \Sigma} $ is related to the exact piece  of $\bar{A} _a \big| _{\p \mathbf{\Sigma } }  $ at the global boundary in the following way: 
\begin{equation}
\begin{split}
   \underbrace{\Tilde{\nabla }^a  A^\dr _a  \big| _{\p \Sigma} }_{:= \Tilde{\nabla}^2 \varphi}  &=  \  \Tilde{\nabla } ^a  \int _{x_0 (x)} ^x   F _{\mu a }  \frac{\dt x ^\mu }{\dt \tau }  \,  \dt \tau  +  \underbrace{  \Tilde{\nabla } ^a  \bar{A} _a \big| _{\p \mathbf{\Sigma } }  }_{:= \Tilde{\nabla}^2 \tilde \phi } \, .
\end{split}
\end{equation}
The middle term is non-trivial to deal with. If the metric is sufficiently independent of $\tau$  along the Wilson paths,\footnote{In four dimensional spacetime $\mathbb{R}^{(3,1)} \simeq \mathbb{R}^{(1,1)} \times S^2$ this sufficiency condition translates to $\tau $ independence of the inner product of two linearly independent null vectors normal to $S^2$.} then one can push the operator $\Tilde{\nabla }^a$ inside the integral and then use the equations of motion. On top of that, if the Wilson path is null, the result simplifies even further. Both of these conditions are satisfied by our choice of frame \eqref{eq:adr_ru}, justifying our explicit setup, and the resulting relationship between $\varphi $ and $\phi $ is:
\begin{equation}
\begin{split} \label{eq:v_scri+}
  \varphi (t) 
     =& \int _{S^2} \Delta ^{-1} _{S^2} \l   r^2 F_{ru} \big| _{\p \Sigma _t }  - r^2 F_{ru} \big| _{\p S^2 _u  }   +\int _{r_s} ^{\infty }  j_r (u ) \,  \dt r  \r +  \hat{\phi } (u  )   +\phi \,, \\
      =& \int _{S^2} \Delta ^{-1} _{S^2} \l   r^2 F_{ru} \big| _{\p \Sigma _t }  + \mathbf{Q}   +\int _{r_s} ^{\infty }  j_r (u) \, \dt r   \r +  2 \hat{\phi } (u  )  - \frac{1}{2}N  +\phi \,, 
\end{split}
\end{equation}
where $j_\mu$ is the matter current in the complement,  $\p \Sigma _t  $ is a cut of $\Gamma $   at constant time $t$ and $S^{2}_u $ is a cut of $\scri ^+ $  at constant $u = t-r_s$, where $r_s$ is the location of $\Gamma$. Note that, to go to the second line, we used the form of $\bar{A}^\dr $ around $\scri ^+$ presented in \eqref{eq:large_r_sol_2}. We will also employ the simplifying notation $j(u):=\int_{r_s}^\infty j_r(u)\dt r$ from now on. 

This formula is important as it explicitly encodes the inherent non-locality of the Goldstone sector of the subregional theory, which has some information about the complement and asymptotic data. It exhibits the regional observable $\varphi$ in terms of observables defined on the global phase space. Particularly interesting is the isolated appearance of certain asymptotic observables: $\phi$ -- asymptotic Goldstone mode, $N$ -- asymptotic soft mode, $\hat{\phi}$ -- hard mode localized on asymptotic cut. This clean separation justifies the choice of soft edge mode employed here:
\be
    \varphi \sim \,... \,+ \text{Hard Photon Mode} + \text{Soft Photon Mode} + \text{Asymptotic GM},
\ee
where ... signifies other complement observables not supported solely on $\scri^+\cup\scri^+_-$. For general soft edge modes (built from, for instance, spacelike Wilson lines), this dependence on null asymptotic data will still be present, although it will generically be more difficult to extract.

Here we remind the reader of the discussion on frame reorientations from section \ref{subsec:fr_reor}. There, it is explained that the subregion frame reorientations can be achieved via either large gauge transformations, or by changing the initial data of the complement in a particular way.  The equations above clearly illustrate this. Large gauge transformations $\phi \mapsto \phi +\alpha $, indeed change the value of $\varphi $, but so can changes in all the other terms in the expression above. To decide whether the latter are genuine frame reorientations, however, we have to make sure that those directions in the global phase space preserve all other subregional variables, as well as the boundary conditions. In section \ref{sec:charge action} we provide explicit examples of global (asymptotic) observables, different from the asymptotic charge, which generate a subregional frame reorientation.

\subsection{Memory and Soft Modes}\label{subsec:memory}

The memory effect in Maxwell theory refers to the overall change in momentum of a charged (inertial) particle accumulated during the passage of an electromagnetic wave. At initial ($s _i$) and final ($s_f$) proper times, respectively, the particle is moving at a constant velocity, but in between it experiences acceleration caused by the electromagnetic wave. The exact formula for this momentum change is:
\begin{equation}
\begin{split} \label{eq:momentum_kick}
\frac{m}{q}  \Delta \, \l  \frac{\Dot{\Tilde{x}}_\mu}{\sqrt{-\Dot{\Tilde{x}}^2}} \r:=
 \frac{m}{q}\l  \frac{\Dot{\Tilde{x}}_\mu}{\sqrt{-\Dot{\Tilde{x}}^2}}  \bigg | _{s_f} -   \frac{\Dot{\Tilde{x}}_\mu}{\sqrt{-\Dot{\Tilde{x}}^2}}  \bigg | _{s_i} \r =  \int _{s_i} ^{s_f} \,  F_{\nu \mu } \frac{\dt \Tilde x ^\nu }{\dt s } \dt s \,,
\end{split}
\end{equation}
where the particle trajectory is given by the curve $\Tilde x:\mathbb{R}\to \mathcal{M}$ and we take the particle to have mass $m$ and charge $q$.  The right-hand side is proportional to the integral of $F_{\mu \nu }$ along the particle's worldline. Notice that it is completely gauge-invariant. 

Finding the solution $\Tilde{x} _{\mu} (s)$ to this equation for a general curved spacetime and electromagnetic tensor is quite difficult. But we can use the \emph{test particle approximation}, in which $q/m\ll1$, and  do a perturbative expansion of the worldline -- $\Tilde{x}(s) ^\mu  = \Tilde{x}(s) ^\mu _0  + \frac{q}{m} \Tilde{x}(s) ^\mu _1 + \mathcal{O}\l  \frac{q}{m}\r ^2 $. At zeroth order, $\Tilde{x} ^\mu  _0 (s)$ is simply the unperturbed timelike geodesic. The effect of $F_{\mu \nu }$ appears at first order of the expansion. One can easily show that the leading order momentum kick $\Delta \dot{\Tilde{x}} _{\mu \, 1 }$ (on a general time-dependent curved background) is equal to the integral of  $ F_{\nu \mu } \frac{\dt \Tilde x _0 ^\nu }{\dt s }$ along the \emph{unperturbed geodesic}. This integral is what we call the \emph{memory mode}.

Instead of the test particle approximation, one can also use the large $\tilde{r}$ limit of this affect -- when the particle is very far away from the source. This has been studied first in \cite{Bieri:2013hqa} and later connected to the asymptotic memory mode in \cite{Pasterski:2015zua}. 

Thus, when $\Gamma $ is traced by timelike geodesics (of, say, some neutral detectors), we have a more operational and quasi-local definition of a memory effect, that depends on the value of the electric field \emph{on the boundary} of our subregion. The tangential components of the electric field are \emph{not} fixed by the soft boundary conditions and, just like in the asymptotic case, we find that the memory is a function on the subregional phase space. This highlights an advantage of working with timelike tubes, as opposed to causal diamonds as already discussed in section \ref{sec:out_res}.

Closely related to the concept of particle memory is the zero-energy, or \emph{soft mode} of electromagnetic radiation. This is defined simply as: 
\begin{equation}\label{eq:soft_mode}
    \lim _{\omega \to  0 } \int _{-\infty } ^{+\infty } e^{i \omega t }F_{ t a} \, \dt t\,,
\end{equation}
where $t$ is some preferred timelike/null direction, and we have purposely integrated the tangential electric field, as it is the one carrying the information about radiation. If we pick the timelike direction to be a geodesic, we recover the leading momentum kick formula for a test particle obtained from the \emph{infinite} time limit of \eqref{eq:momentum_kick}.

With these definitions at hand, we will now move on to compare the asymptotic memory effect defined at $\scri ^+ $ to the subregion memory effect defined at $\Gamma $.

\paragraph{Asymptotic Memory Effect:}
~\newline
Let us start with the case of $\scri^+$. Its null and asymptotic nature makes it a special surface.
In particular, the memory mode is directly part of the symplectic structure, and the charges are also directly related to the memory. The soft mode at $\scri ^+ $ is defined as:
\begin{equation}
    \int_{\scri^+}  \dt u F_{ua}  = \p _a \l \phi^+ - \phi ^- \r = \p _a N\,. \label{eq:global memory = transition}
\end{equation}
The equality follows from \eqref{eq:large_r_sol_2}, the dressing condition $\bar{A}^\dr_u\big|_{\scri^+}=0$ and the ``no magnetic charges condition''. This is also the global memory mode, which tests particles hovering asymptotically around $\scri ^+$ experience.

It follows from the symplectic form in \eqref{eq:omega scri} and the (large) gauge invariance of all variables except $\phi$ that for a large gauge transformation parametrized by $\alpha$: 
\be
\label{eq:Qglobal}
X_{\delta_\alpha} \cdot \mathbf{\Omega}  = \delta \mathbf{Q} _{\alpha } =  \d \int_{\scri ^+ _-} \, \epsilon_{\scri ^+ _-} \, \alpha \, \l  D^2  N - \mathbf{Q}\r\,.
\ee
This means that \emph{asymptotically} the memory/soft mode is the \emph{generator} of large gauge transformations, along with the global charge at $i^+$. This is the manifestation of one of the edges of the infrared triangle. Such a simple relationship will no longer hold at the subregion level, as we now see.

\paragraph{Subregion Memory Effect:}
~\newline
Let us now adapt the construction above to a finite region with \emph{timelike} boundary $\Gamma$. Unlike its global version, we define the subregion memory mode as an integral over a \emph{finite} time interval (along some timelike geodesic in $\Gamma$) and it therefore is \emph{not} equal to the soft mode.  In particular:
\be
\textbf{Subregion memory:} \, \quad \mathcal{N}_a := \int _{t^-} ^{t^+ } \, F_{ta}\big|_{\Gamma } \dt t     = A_a ^{\dr} \big|_{t=t^+} - A_a ^{\dr }  \big|_{t=t^-}  - \p _a \int _{t^-} ^{t^+} A_t ^\dr\big|_{\Gamma } \dt t  \,. \label{eq:subregion memory = transition}
\ee
Notice that, because of the non-vanishing radial magnetic field at $t ^\pm $, the formula above is not a total derivative, unlike \eqref{eq:global memory = transition}. In addition, there is an extra contribution coming from the subregion boundary field $A^{\rm dr}_t|_\Gamma$ (which is a subregion phase space constant).

Using the split $A^\dr_a=\tilde A^\dr_a + \p_a\varphi$, we see that it is still true that the subregional memory mode is related to a vacuum transition $\Delta\varphi=\varphi(t^+)-\varphi(t^-)$. However, this relationship is now not as simple as the global version in \eqref{eq:global memory = transition}, further involving the boundary conditions ($A^\dr_t\big|_\Gamma$) and the dynamical intrinsic data ($\tilde A^\dr\big|_\Gamma$). Crucially, $\Delta\varphi$ is \emph{frame-reorientation invariant}, since frame reorientations parameterized by $\rho$ must satisfy $\p_t\rho=0$ to preserve the soft boundary conditions. This is the subregional analogue of the fact that the asymptotic vacuum transition $\Delta\phi=\phi^+-\phi^-$ is large gauge-invariant. 

In contrast to the global case, here it is \emph{not} true that the memory mode $\mathcal{N}_a$ is symplectically conjugate to the Goldstone mode $\varphi$. It is instead a considerably more complicated phase space function, given by \eqref{eq:subregion memory = transition}. This, in particular, means that the subregional memory mode is \emph{not} the generator of the frame reorientation symmetries. A finite distance version of the edge of the infrared triangle described above is thus not available.

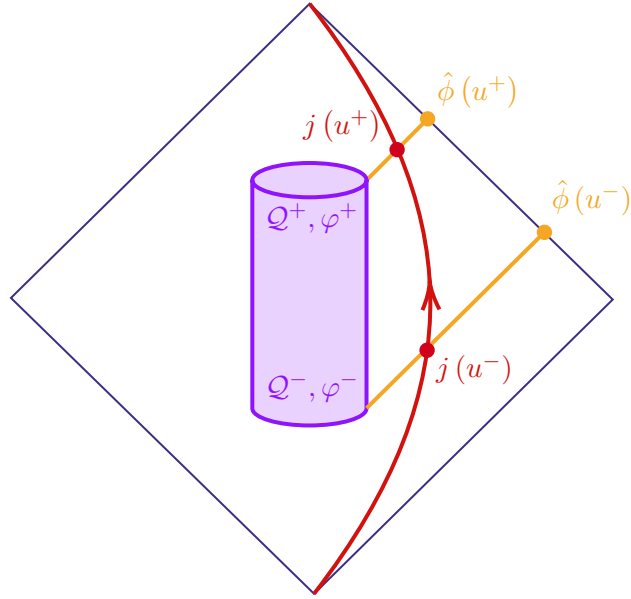
\begin{figure}[h!]
    \centering

\tikzset{every picture/.style={line width=0.75pt}} 

\begin{tikzpicture}[x=0.75pt,y=0.75pt,yscale=-1,xscale=1]

\draw  [color={rgb, 255:red, 59; green, 45; blue, 131 }  ,draw opacity=1 ][line width=0.75]  (338.64,20) -- (490.7,168.4) -- (340.87,316) -- (188.8,167.6) -- cycle ;
\draw [color={rgb, 255:red, 245; green, 166; blue, 35 }  ,draw opacity=1 ][line width=1.5]    (367.19,108.56) -- (397.85,77.77) ;
\draw  [color={rgb, 255:red, 144; green, 19; blue, 254 }  ,draw opacity=1 ][fill={rgb, 255:red, 144; green, 19; blue, 254 }  ,fill opacity=0.19 ][line width=1.5]  (367.19,108.51) -- (367.19,222.97) .. controls (367.19,227.72) and (354.36,231.57) .. (338.54,231.57) .. controls (322.71,231.57) and (309.89,227.72) .. (309.89,222.97) -- (309.89,108.51) .. controls (309.89,103.76) and (322.71,99.91) .. (338.54,99.91) .. controls (354.36,99.91) and (367.19,103.76) .. (367.19,108.51) .. controls (367.19,113.26) and (354.36,117.11) .. (338.54,117.11) .. controls (322.71,117.11) and (309.89,113.26) .. (309.89,108.51) ;
\draw [color={rgb, 255:red, 245; green, 166; blue, 35 }  ,draw opacity=1 ][line width=1.5]    (367.19,222.92) -- (456.43,134.68) ;
\draw [color={rgb, 255:red, 210; green, 18; blue, 14 }  ,draw opacity=1 ][line width=1.5]    (338.64,20) .. controls (416.48,120.37) and (421.48,211.91) .. (340.87,316) ;
\draw [shift={(398.89,159.34)}, rotate = 87.53] [color={rgb, 255:red, 210; green, 18; blue, 14 }  ,draw opacity=1 ][line width=1.5]    (14.21,-4.28) .. controls (9.04,-1.82) and (4.3,-0.39) .. (0,0) .. controls (4.3,0.39) and (9.04,1.82) .. (14.21,4.28)   ;
\draw [color={rgb, 255:red, 208; green, 2; blue, 27 }  ,draw opacity=1 ]   (397.56,193.83) ;
\draw [shift={(397.56,193.83)}, rotate = 0] [color={rgb, 255:red, 208; green, 2; blue, 27 }  ,draw opacity=1 ][fill={rgb, 255:red, 208; green, 2; blue, 27 }  ,fill opacity=1 ][line width=0.75]      (0, 0) circle [x radius= 3.35, y radius= 3.35]   ;
\draw [color={rgb, 255:red, 208; green, 2; blue, 27 }  ,draw opacity=1 ]   (382.52,93.21) ;
\draw [shift={(382.52,93.21)}, rotate = 0] [color={rgb, 255:red, 208; green, 2; blue, 27 }  ,draw opacity=1 ][fill={rgb, 255:red, 208; green, 2; blue, 27 }  ,fill opacity=1 ][line width=0.75]      (0, 0) circle [x radius= 3.35, y radius= 3.35]   ;
\draw [color={rgb, 255:red, 245; green, 166; blue, 35 }  ,draw opacity=1 ]   (397.85,77.77) ;
\draw [shift={(397.85,77.77)}, rotate = 0] [color={rgb, 255:red, 245; green, 166; blue, 35 }  ,draw opacity=1 ][fill={rgb, 255:red, 245; green, 166; blue, 35 }  ,fill opacity=1 ][line width=0.75]      (0, 0) circle [x radius= 3.35, y radius= 3.35]   ;
\draw [color={rgb, 255:red, 245; green, 166; blue, 35 }  ,draw opacity=1 ]   (456.43,134.68) ;
\draw [shift={(456.43,134.68)}, rotate = 0] [color={rgb, 255:red, 245; green, 166; blue, 35 }  ,draw opacity=1 ][fill={rgb, 255:red, 245; green, 166; blue, 35 }  ,fill opacity=1 ][line width=0.75]      (0, 0) circle [x radius= 3.35, y radius= 3.35]   ;

\draw (314.75,118.58) node [anchor=north west][inner sep=0.75pt]  [font=\normalsize]  {$\textcolor[rgb]{0.56,0.07,1}{\mathcal{Q}}\textcolor[rgb]{0.56,0.07,1}{^{+}}\textcolor[rgb]{0.56,0.07,1}{,\varphi }\textcolor[rgb]{0.56,0.07,1}{^{+}}$};
\draw (315.04,203.42) node [anchor=north west][inner sep=0.75pt]  [font=\normalsize]  {$\textcolor[rgb]{0.56,0.07,1}{\mathcal{Q}}\textcolor[rgb]{0.56,0.07,1}{^{-}}\textcolor[rgb]{0.56,0.07,1}{,\varphi }\textcolor[rgb]{0.56,0.07,1}{^{-}}$};
\draw (333.87,71.81) node [anchor=north west][inner sep=0.75pt]  [font=\normalsize,color={rgb, 255:red, 210; green, 18; blue, 14 }  ,opacity=1 ]  {$j\left( u^{+}\right)$};
\draw (400.64,194.2) node [anchor=north west][inner sep=0.75pt]  [font=\normalsize,color={rgb, 255:red, 210; green, 18; blue, 14 }  ,opacity=1 ]  {$j\left( u^{-}\right)$};
\draw (401.17,52.71) node [anchor=north west][inner sep=0.75pt]  [color={rgb, 255:red, 245; green, 166; blue, 35 }  ,opacity=1 ]  {$\hat{\phi }\left( u^{+}\right)$};
\draw (458.57,107.08) node [anchor=north west][inner sep=0.75pt]  [color={rgb, 255:red, 245; green, 166; blue, 35 }  ,opacity=1 ]  {$\hat{\phi }\left( u^{-}\right)$};

\end{tikzpicture}
    \caption{Depiction of relevant quantities involved in \eqref{eq:memory_relation_1} -- the initial and final subregional electric charge densities  ($\mathcal{Q}^\pm$), the exact piece of $A^\dr |_{\p \Sigma ^\pm }$  ($\varphi ^\pm$), as well as the exact piece of $\bar{A}^\dr _a$ on two different cuts of $\scri ^+ $ ($\hat{\phi}(u^\pm)$).}
    \label{fig:GM_formula}
\end{figure}

Despite their differences, subregion and global memories are, in certain circumstances, mathematically related.  

To see this, we begin by noting that from the relationship between the finite and asymptotic Goldstone modes derived in \eqref{eq:v_scri+}, we can write down the following equality for the subregional vacuum transition:
\begin{equation}
\label{eq:memory_relation_1}
   \begin{split}
       &  \varphi (t^+) - \varphi (t^-) = 2 \l \hat{\phi} (u^+) - \hat{\phi} (u^-)
        \r +\int _{S^2} \Delta _{S^2} ^{-1}  \l r_s ^2 \mathcal{Q}^+  - r_s ^2 \mathcal{Q} ^-    + j(u^+) - j(u^-) \r  \, ,
   \end{split}
\end{equation}
where $u^\pm=t^\pm-r_s$ and the relevant quantities are depicted in figure \ref{fig:GM_formula}. 

However, looking at \eqref{eq:subregion memory = transition}, we see that we would also need to be able to relate the divergence-free piece of $A^\dr _a \big |_{t=t^\pm}$ (related to the radial magnetic field) to the asymptotic data, if we would want to relate the subregion memory to the data on $\scri ^+$. Unfortunately, such a straightforward connection does not exist in general. This motivates us to consider the limit in which we take our tube to be infinitely extended in time. In that case, we are justified to set the initial and final subregional magnetic radial fields to zero as well, allowing for an explicit mathematical relation between the two memories. This and other types of asymptotic limits are studied in the next subsection.

\subsection{Asymptotic Limits}
\label{subsec:big_edges}

In the previous sections, we have compared features of finite-distance boundaries with those at infinite distance. This has been done by assuming that in the second case, the corresponding theory is the global one.  There are, however, several different ways of studying the asymptotic limit of a subregion, each suitable for exploring different aspects of the \textit{soft edges}. 
Let us classify them as follows, for $r_s$ the radius of the boundary, and $t^-$, $t^+$ the initial and final slices, respectively:
 \begin{itemize}
        \item[a)] Send $r_s\to\infty$, keeping $t$ constant. Then the $\Sigma$'s become complete Cauchy slices and $\Gamma$ becomes a portion of $i^0$.
        \item[b)] Send $t^+ \to \infty$, $t^- \to -\infty$ . The two slices then evolve into portions of $i^+$ and $i^-$, and $\Gamma$ forms an infinite tube extending from $i^-$ to $i^+$.
        \item[c)] Keep $r_s/t^\pm$ constant, while sending $r_s\to\infty$. In this case, the two edges lie on the future and past null asymptotes, with the tube boundary appearing as a vertical line in the compactified Penrose diagram.
 \end{itemize}
We can schematically represent these limits as in the figure \ref{fig:Soft_edges}. 

\begin{minipage}{.95\textwidth}
    \centering
    \includegraphics[width=0.8\linewidth]{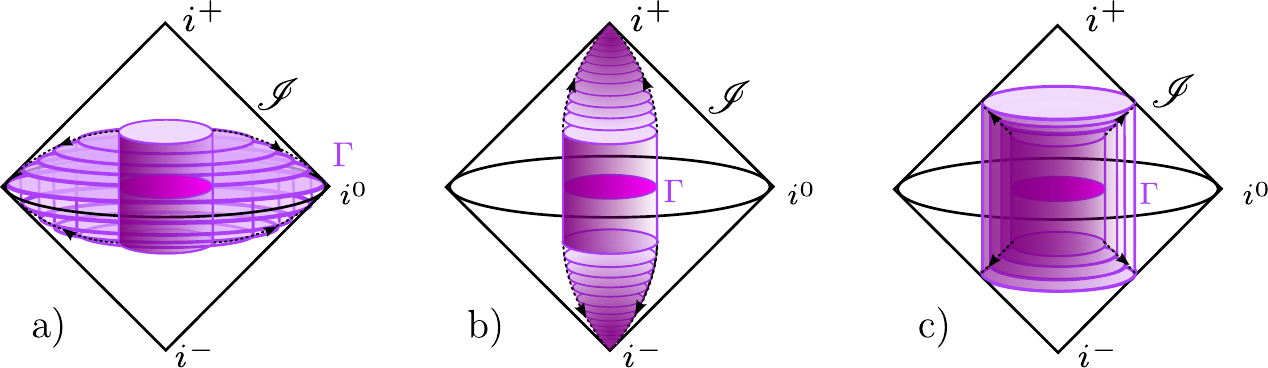}
    \captionof{figure}{\footnotesize Schematic representation of different limits for the subregion. In (a) we get a finite time evolution along $i^0$ between two complete Cauchy surfaces. In (b) the regional initial/final surfaces end up somewhere \textit{inside} $i^\pm$. The last option (c) is describing the reduced dynamics between two Milne wedges.}
    \label{fig:Soft_edges}
\end{minipage}\\

In the first case (a) we can compare the charge on $\Sigma$ with the charge on $\scri^+\cup i^+$. We get:\footnote{Due to its timelike nature, there is no matter flow through $i^0$ and thus the vacuum equations of motion $\dt \star F \approx 0$ hold.}
 \be
\int_{\p \Sigma} \rho \star F -\int_{\scri^+_-} \rho \star F
    \approx -\int_{[i^0]} \dt \rho \wedge \star F\,,
 \ee
where $[i^0]$ is the portion of the timelike asymptote between $\Sigma$ and $\scri^+_-$, i.e. $t\in (t^+, \infty)$.  The integral over this portion is determined by the global boundary conditions. Notice that in the first integral $\rho$ refers to the frame reorientations parameter, while in the second to the large gauge parameter. In this limit the two transformations, however, coincide. Although discussing this limit generally necessitates the inclusion of matter within the subregion, doing so will not alter the discussion about the solution space, the analyticity of the solution will now be only in a neighbourhood of the boundary.  In principle, we should incorporate matter into the symplectic form, but for the comparison of the value of the charges (and not their role as generators), it does not matter whether the matter is inside or outside the region of interest. In this limit, the dynamical content of the two charges is identical, as their difference is determined by a background structure -- namely, the boundary conditions on $i^0$ given in~\eqref{eq:global bcs} and~\eqref{eq:global bcs 2}. This can also be explained by the absence of radiation on $[i^0]$, leading to the charge conservation. Thus, this provides a stronger version of the matching conditions employed in celestial holography to equate the charge across $\scri^+$ and $\scri^-$.\\

Let us move on to the second limit (b), in which we keep the subregion radius finite, but we study the \textit{full} dynamics from $i^-$ to $i^+$. As anticipated at the end of subsection \ref{subsec:memory}, the \textit{infinitely long} timelike tube allows for a direct comparison between the finite-distance and infinite-distance memory modes. The absence of radial magnetic charges field at $i^\pm$ (essentially setting $\tilde A^\dr_a\big|_{t=t^\pm}=0$) makes the subregional memory mode in  \eqref{eq:subregion memory = transition} exact, leading to the definition:
\be
    \mathcal{N} := \varphi ^+ - \varphi ^- - \int_{-\infty}^\infty A_t ^\dr\big|_\Gamma \dt t\,,
\ee
where we have defined $\varphi^\pm :=\lim _{t\to \pm \infty } \varphi(t)$. This is analogous to the definition of the global soft mode, $N$, in \eqref{eq:gl_GM_SM_def}.

In the limit $t^\pm  \to \pm \infty $, equation \eqref{eq:v_scri+}, together with \eqref{eq:hphi_def}, leads to the following limiting values of the finite Goldstone mode $\varphi$: 
\begin{equation}
\begin{split}\label{eq:vpm_limit}
       &\varphi ^+  = \phi  +  \frac{1}{2}N   + \int _{S^2} \Delta _{S^2} ^{-1}  \l r_s^2  \mathcal{Q} ^+  + J^{+}  - \mathbf{Q} \r \, , \\
    &\varphi ^- =  \phi  -   \frac{3}{2}N   + \int _{S^2} \Delta _{S^2} ^{-1}  \l r_s ^2 \mathcal{Q} ^-  + J^{-}  - \mathbf{Q} \r \, ,
\end{split}
\end{equation}
where $J^\pm := \lim _{u \to \pm \infty } j(u^\pm)$. Notice that both $\varphi ^+$ and $\varphi ^-$ are expressed in terms of quantities defined on $\scri^+$, because of the specific frame we chose, that dressed everything to $\scri ^+$.

In this ``infinite time extension'' limit, the subregional vacuum transition formula in \eqref{eq:memory_relation_1} simplifies, leading to the following explicit relation between finite and asymptotic memory modes:
\begin{equation}
\label{eq:memories_relationship}
   \begin{split}
       \mathcal{N}+\int_{-\infty}^\infty A_t ^\dr\big|_\Gamma \dt t= 2 N  +\int _{S^2} \Delta _{S^2} ^{-1}  \l r_s ^2 \mathcal{Q}^+  - r_s ^2 \mathcal{Q} ^-    + J^+ - J^- \r   \,.
   \end{split}
\end{equation} 

We note that now the initial and final Cauchy surfaces of the subregion belong to $i^\pm$. No electromagnetic radiation is present there, and therefore the subregional charge densities $\mathcal{Q} ^\pm $ can only be sourced by a non-vanishing matter current $j_{\mu}$ in the complement. Therefore, in the case of free Maxwell theory we have a more straightforward relationship between $\varphi ^\pm $ and $\l \phi, \,  N \r $, since $J^\pm = \mathbf{Q}=\mathcal{Q}^\pm=0$. Indeed, the finite and asymptotic memory modes become related in a simple manner:
\be
    \text{Free Maxwell }\implies \mathcal{N}+\int_{-\infty}^\infty A_t ^\dr\big|_\Gamma \dt t=2N.
\ee
They are equal observables on the postselected phase space corresponding to the infinite timelike tube, up to a regional phase space constant determined via $A_t^{\rm dr}|_\Gamma$ which is fixed via the soft boundary conditions. This is not true if matter exists outside the tube, as equation \eqref{eq:memories_relationship} clearly shows.

We now discuss what happens if \emph{after} we take this (b) limit, we further take the large $r_s$ limit of $\Gamma$. This is a good sanity check for the definitions of $\mathcal{N}$ and $\varphi $ as we can compare their asymptotic values with the global variables $N$ and $\phi$.

In this limit, there are no solutions with matter in the complement, as the latter disappears. This leads to $\lim _{r_s \to \infty} J^\pm=0$. Next, since the variables $r_s ^2 \mathcal{Q}^\pm $ are defined on $i^\pm$, we must take their large $r_s$ limit on that surface, which gives the radial electric fields at $\scri ^+ _+ $ and $\scri ^- _-$, respectively. These are zero for the same reason, as we remind the reader that we are always only considering solutions without matter inside the subregion. 
Combining all this we have:
\begin{equation}
    \begin{split}
         \lim _{r_s \to \infty }   \frac{1}{2}\l \varphi ^+  + \varphi ^- \r &= \phi  - \frac{1}{2}N  \, , \\
            \lim _{r_s \to \infty } \mathcal{N}  &= 2N\,,
    \end{split}
\end{equation}
where to get the last line we observe that, in this limit, the integral $\int_{-\infty}^\infty A^\dr_t\big|_\Gamma \dt t$ splits into integrals along $\scri^+$, $\scri^-$ and $i^0$. The first two are zero because of the dressing condition \eqref{eq:dr_cond_ru_out}, while the latter is zero by the global boundary conditions in \eqref{eq:global bcs}.

Curiously, we see that our subregion definition of memory asymptotes to $2N$, as opposed to just $N$, which is what one might have expected. Heuristically, this can be understood from the fact that we first took the $t^\pm \to \pm \infty $ limit and \emph{then} the $r_s\to\infty$ one. This is as if $\Gamma $ has spread across both $\scri ^+ $ and $\scri ^-$. On the other hand, from the definition of $A^\dr$ in \eqref{eq:adr_ru}, it is clear that if we were to first take the $r_s \to \infty $ limit, while keeping  $u=const$, we would have contributions only from $\scri ^+ $. 
\\

The last limit (c) is particularly relevant for studying relational properties of subregions of $\scri$. For example, in the spirit of \cite{Chen:2023tvj,Chen:2024kuq}, where the authors examine the connection between entanglement entropy and soft modes from a gauge-fixed perspective. The Goldstone mode they construct in these works is the gauge-fixed counterpart of our gauge-invariant construction. Although applying our formalism to study entanglement entropy requires extending it to the quantum realm and goes beyond the scope of this paper, we can already suggest that the crucial role of the Goldstone mode, along with its frame dependence, hints at the potential frame dependence of entanglement entropy itself \cite{Hoehn:2023ehz,DeVuyst:2024pop,DeVuyst:2024uvd}.

\section{Action of Global and Finite-Distance Charges} \label{sec:charge action}

The goal of this section is to study the sector of the global phase space postselected on a choice of soft boundary conditions (cf.~section~\ref{sec:Soft_bc}) on the finite timelike boundary $\Gamma$ in order to explore and compare the action of the global and subregion charges, respectively, on the subregion phase space variables. We will show that their action is \emph{the same}, thereby proving the claim that a large gauge transformation generates a frame reorientation on the subregion variables. Importantly, we also show that there exist \emph{other} operators, different from the global charge, which also have the same action as the subregion charge, in the sense explained above. This gives a concrete realization of the claim that there exist other complement transformations, not corresponding to large gauge transformations, which, nevertheless, also generate subregional frame reorientations.

To begin with, let us recall the key players, shown in table \ref{table:key players}. The relationship between the variables on the third line is given in $\eqref{eq:v_scri+}$.

\begingroup
\renewcommand{\arraystretch}{1.9}
\begin{table}[h!]
    \centering
        \begin{tabularx}{0.95\textwidth}{|c||c|>{\centering\arraybackslash}X|}
            \hline
             & \textbf{Global theory} & \textbf{Subregion theory}\\
            \hhline{|=||=|=|}
            \textbf{Dressed field} &  $\bar{A}^\dr  \big|_{\scri ^+}$ & $A^\dr  \big|_{\Sigma }$\\
            \hline
            \textbf{Radiative field} & $\Tilde{A}^\dr  \big|_{\scri ^+}  = \bar{A}^\dr   \big|_{\scri ^+}  - \dt \phi$ &  $\Tilde{A}^\dr \big|_{\Sigma }  = A^\dr   \big|_{\Sigma }   - \dt \varphi$\\
            \hline
            \textbf{Goldstone mode} & $\phi  = \frac{1}{2} \l \phi ^+ + \phi ^-\r$ & $D ^2 \varphi   = \Tilde{\nabla} ^a A^\dr_a \big|_{\p \Sigma }$\\
            \hline
            \textbf{Memory mode} & $\p _a N    = \p _a \l  \phi ^+ - \phi ^- \r$ & $\mathcal{N}_a =  A_a ^{\dr} \big|_{t=t^+} - A_a ^{\dr }  \big|_{t=t^-}   - \p _a  \l \int  _{t^-} ^{t^+ } \, A_t ^{\dr}\big|_{\Gamma} \dt  t\r  $\\
            \hline
        \end{tabularx}
        \caption{Summary of the various quantities appearing both at finite distance and at the asymptotic boundary.}
        \label{table:key players}
\end{table}
\endgroup

\begin{figure}[h!]
    \centering

\tikzset{every picture/.style={line width=0.75pt}} 

\begin{tikzpicture}[x=0.75pt,y=0.75pt,yscale=-1,xscale=1]

\draw  [color={rgb, 255:red, 59; green, 45; blue, 131 }  ,draw opacity=1 ][line width=0.75]  (333.48,35.67) -- (481.69,182.53) -- (334.22,328.67) -- (186.01,181.8) -- cycle ;
\draw  [color={rgb, 255:red, 144; green, 19; blue, 254 }  ,draw opacity=1 ][fill={rgb, 255:red, 144; green, 19; blue, 254 }  ,fill opacity=0.19 ][line width=1.5]  (361.78,107.02) -- (359.46,246.89) .. controls (359.38,251.37) and (347.21,255) .. (332.26,255) .. controls (317.32,255) and (305.26,251.37) .. (305.34,246.89) -- (307.66,107.02) .. controls (307.73,102.54) and (319.91,98.91) .. (334.85,98.91) .. controls (349.8,98.91) and (361.85,102.54) .. (361.78,107.02) .. controls (361.7,111.51) and (349.53,115.14) .. (334.58,115.14) .. controls (319.64,115.14) and (307.58,111.51) .. (307.66,107.02) ;
\draw  [color={rgb, 255:red, 245; green, 166; blue, 35 }  ,draw opacity=1 ][fill={rgb, 255:red, 245; green, 166; blue, 35 }  ,fill opacity=0.31 ][line width=1.5]  (307.01,173.78) .. controls (307.01,169.15) and (319.03,165.39) .. (333.85,165.39) .. controls (348.67,165.39) and (360.69,169.15) .. (360.69,173.78) .. controls (360.69,178.41) and (348.67,182.17) .. (333.85,182.17) .. controls (319.03,182.17) and (307.01,178.41) .. (307.01,173.78) -- cycle ;
\draw [color={rgb, 255:red, 245; green, 166; blue, 35 }  ,draw opacity=1 ][line width=2.25]    (360.69,173.78) -- (417,119) -- (481.67,182.53) ;
\draw [color={rgb, 255:red, 245; green, 166; blue, 35 }  ,draw opacity=1 ][line width=2.25]    (186.01,181.8) -- (249,119.67) -- (307.01,173.78) ;
\draw  [color={rgb, 255:red, 245; green, 166; blue, 35 }  ,draw opacity=1 ] (239,113.67) .. controls (235.72,110.35) and (232.42,110.33) .. (229.1,113.61) -- (212.52,129.99) .. controls (207.78,134.68) and (203.77,135.36) .. (200.49,132.04) .. controls (203.77,135.36) and (203.04,139.36) .. (198.3,144.05)(200.43,141.94) -- (181.72,160.43) .. controls (178.4,163.71) and (178.38,167.01) .. (181.66,170.33) ;
\draw  [color={rgb, 255:red, 245; green, 166; blue, 35 }  ,draw opacity=1 ] (484.33,169.67) .. controls (487.59,166.33) and (487.55,163.03) .. (484.22,159.77) -- (467.71,143.65) .. controls (462.94,138.99) and (462.19,134.99) .. (465.45,131.65) .. controls (462.19,134.99) and (458.18,134.33) .. (453.41,129.67)(455.55,131.77) -- (436.9,113.55) .. controls (433.57,110.29) and (430.27,110.33) .. (427,113.66) ;

\draw (414.22,94.18) node [anchor=north west][inner sep=0.75pt]  [font=\normalsize,color={rgb, 255:red, 241; green, 155; blue, 14 }  ,opacity=1 ]  {$u'$};
\draw (364.69,173.85) node [anchor=north west][inner sep=0.75pt]  [font=\normalsize,color={rgb, 255:red, 241; green, 155; blue, 14 }  ,opacity=1 ]  {$t'$};
\draw (390.85,149.79) node [anchor=north west][inner sep=0.75pt]  [font=\normalsize,color={rgb, 255:red, 245; green, 166; blue, 35 }  ,opacity=1 ]  {$\textcolor[rgb]{0.96,0.65,0.14}{\overline{\Sigma }}$};
\draw (259.1,149.79) node [anchor=north west][inner sep=0.75pt]  [font=\normalsize,color={rgb, 255:red, 245; green, 166; blue, 35 }  ,opacity=1 ]  {$\textcolor[rgb]{0.96,0.65,0.14}{\overline{\Sigma }}$};
\draw (178.67,113.07) node [anchor=north west][inner sep=0.75pt]  [color={rgb, 255:red, 245; green, 166; blue, 35 }  ,opacity=1 ]  {$\mathcal{I}_{u'}$};
\draw (288,118.4) node [anchor=north west][inner sep=0.75pt]  [color={rgb, 255:red, 144; green, 19; blue, 254 }  ,opacity=1 ]  {$\Gamma $};
\draw (327.33,188.4) node [anchor=north west][inner sep=0.75pt]  [color={rgb, 255:red, 245; green, 166; blue, 35 }  ,opacity=1 ]  {$\Sigma $};
\draw (468.67,115.07) node [anchor=north west][inner sep=0.75pt]  [color={rgb, 255:red, 245; green, 166; blue, 35 }  ,opacity=1 ]  {$\mathcal{I}_{u'}$};

\end{tikzpicture}
    \caption{The global Cauchy surface considered in this section, drawn in orange. It is composed of 3 parts - a portion of $\scri ^+ $, called $\mathcal{I} _{u'}$ that starts from $\scri ^+ _ -$ and ends at $u=u'$, another null hypersurface that connects the $u'$ constant cut of $\scri ^+ $ with the boundary $\Gamma $ of the subregion, and finally the Cauchy surface of the subregion itself, denoted by $\Sigma$.  }
    \label{fig:deformed_CS}
\end{figure}
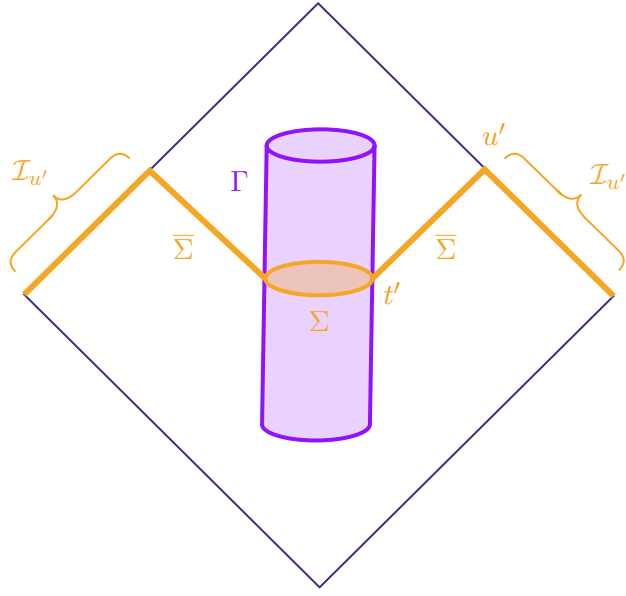

Comparison of the phase space flows of different variables requires us to embed the subregion Cauchy slice $\Sigma $ in the global one $\mathbf{\Sigma}$. In order to achieve this, and still be able to use the form of the solution around $\scri ^+ $ presented in section \ref{sec:sol_space}, we consider a deformed Cauchy slice, as illustrated in figure \ref{fig:deformed_CS}. Part of the global Cauchy slice is a portion of $\scri ^+ $ from $u=-\infty $ to $u'$, that we call $\mathcal{I} _{u'}$. Afterwards, it extends in the bulk along a null direction, specified by $u=u'$ until it reaches the subregion Cauchy slice $\Sigma $ at time $t' = u'+r_s$. We call this portion of the Cauchy slice  $\bar{\Sigma}$. The symplectic forms of the global and subregional theories, respectively, evaluated on the Cauchy slice described above, are (using \eqref{eq:omega scri}):
\begin{equation}
\begin{split}
    \mathbf{\Omega} & = \int _{\mathcal{I} _{u'}}\,\epsilon _{\scri^+}  \, \l \d \p _a \hat{\phi } \curlywedge \d D^a  \p _u \hat{\phi } + \d \p _a b \curlywedge \d D^a \p _u  b \r + \int _{\scri ^+ _-} \, \epsilon _{\scri ^+ _-} \,   \d \phi \curlywedge \d \l  D^2 N  - \mathbf{Q}  \r + \bar{\Omega } \big| _{\bar{\Sigma }}   + \Omega  \, , \\
     \Omega & = \int _{\Sigma }\, \epsilon _\Sigma  \, \d \Tilde{A} ^{\dr} _a \curlywedge \d  F^{ta}+ \int _{\p \Sigma } \,\epsilon _{\p \Sigma } \, \d \varphi  \curlywedge \d \mathcal{Q}  \, . 
\end{split} \label{eq:global and sub sympl}
\end{equation}
Once again,  the choice of phase space variables has made the symplectic form ``block anti-diagonal'' and thus  invertible. As shown in subsection \ref{subsec:sym_form}, for the subregion symplectic form this crucially depends on the soft boundary conditions.

The global charges associated to the large gauge symmetry $\phi \mapsto \phi + \alpha$ are:
\begin{equation}
    \mathbf{Q} _{\alpha }  = \int _{\scri ^+ _-}  \,  \epsilon _{S^2 }  \, \alpha \, \l D^2 N  - \mathbf{Q}\r\,,
\end{equation}
where $\alpha $ is an arbitrary function on the celestial sphere. The action of these charges on any global phase space functional is given by the Poisson bracket (with $x\in\mathbf\Sigma$):
\begin{equation}
    \{ \mathbf{Q} _{\alpha }  , \mathbf{O} (x) \} =   \int _{S^2}\,  \dt y^2 \, \sqrt{g}  \,  \alpha (y) \, \frac{\d \mathbf{O} (x)}{\d \phi (y)}\,.
\end{equation}
From this, $A^\dr = \Tilde{A} ^\dr + \dt \varphi $, equation \eqref{eq:v_scri+} and the symplectic form above we can deduce that:
\begin{equation}
\begin{split}
     &\{ \mathbf{Q} _{\alpha }, \,  \Tilde{A} ^\dr (x) \big|_{\scri ^+} \}  = \{ \mathbf{Q} _{\alpha }, \, \Tilde{A} ^\dr (x) \big|_{\Sigma } \} =0, \\
    & \{ \mathbf{Q} _{\alpha }  ,  \, \phi (x_0(x)) \}  = \{ \mathbf{Q} _{\alpha }  , \,  \varphi (x) \} = \alpha  (x_0(x)) \,,\\
    &  \{ \mathbf{Q} _{\alpha } ,  \,  \bar{A} ^\dr (x) \big|_{\scri ^+  } \}  =\{ \mathbf{Q} _{\alpha } , \,  A ^\dr (x) \big|_{\Sigma } \}  = \dt \alpha (x_0 (x)) \,,
\end{split}
\end{equation}
where $x_0(x)$ is the point on the celestial sphere mapped to the point $x$ in the bulk via the choice of Wilson line path used in our construction. The last line is what we wanted to prove and we note that the soft boundary conditions (and hence the postselection) played a crucial role in the evaluation of the brackets, because it guaranteed the independence of the different bulk and corner phase space variables.

Now restricting ourselves to the subregion phase space, we consider the subregion charges of the frame reorientation symmetry $\varphi  \mapsto \varphi  + \rho $:
\begin{equation}
    Q_{\rho  }  = \int _{\p \Sigma } \, \epsilon_{\p\Sigma} \,\rho \,  \mathcal{Q}\,.
\end{equation}
From the subregion symplectic form, we can deduce that:
\begin{equation}
    \begin{split}
        \{ Q _{\rho } , \, \varphi  \} = \rho  \, \quad  \{ Q _{\rho } , \,  \Tilde{A} ^\dr  \big|_{\Sigma } \} = 0 \,.
    \end{split}
\end{equation}
The final bracket we need to compute is the one between the charges themselves. We use the global symplectic form to get:
\begin{equation}
    \{ \mathbf{Q} _\alpha , \, Q_{\rho }  \} =0 \,.
\end{equation}

The conclusion is that the action of the global and subregion charges on the subregion variables $A^\dr\big|_\Sigma $ is the same (upon identifying $\alpha(x_0(x))=\rho(x)$ for $x\in\Gamma$), as advertised.

Finally, we show that there are other operators from the global theory that have similar effects on $ A ^\dr (x) \big|_{\Sigma } $. Let us look for example at the following operator: 
\begin{equation}
\begin{split}
    \mathbf{O} _{\alpha }(u) := \frac{1}{2}\int _{S^2 _u } \, \epsilon _{ S^2 }  \,   \alpha \,  D^2   \hat{\phi }  (u) \,, \quad u'\geq u \, .
\end{split}
\end{equation}
 This operator does \emph{not} generate large gauge transformations, since $\{\hat{\phi}(u),\phi\}=0$, as follows from the global symplectic form in \eqref{eq:global and sub sympl}. In order to find its action on $\varphi$, which depends linearly on $\hat{\phi}$, we need the bracket between $\hat{\phi}$ and itself. The symplectic form \eqref{eq:global and sub sympl} tells us that $  \{ \p _a \p _u \hat{\phi} \l u , x^a  \r  ,\,  D^b \hat{\phi } \l \bar{u} , \bar{x} ^a  \r \} = \d(u-\bar{u})\d^a _b  \d ^2 (x^a - \bar{x} ^a)$, so to get what we need we have to integrate over $u$ and antisymmetrise with respect to $u$ and $\bar{u}$. The result is: 
\begin{equation}
\begin{split}
  \{ \p _a \hat{\phi} \l u , x^a  \r  ,\,  D^b \hat{\phi } \l \bar{u} , \bar{x} ^a  \r \}  = \l  - \Theta_H(\bar{u} - u)  +  \Theta_H(u-\bar{u})   \r  \, \d _a ^b \, \d ^2 (x^a - \bar{x} ^a)  , \quad u' \geq  u, \  u' \geq  \bar{u} \,,
\end{split}
\end{equation}
where $\Theta_H(x)$ is the Heaviside step function, and the conditions $u' \geq u, \  u' \geq \bar{u}$ ensure that both operators above live on the Cauchy slice we have picked. Notice that the field $\hat{\phi} $ does not commute with itself at different points of $\mathcal{I} _{u'} $. This is a consequence of the null nature of this portion of the Cauchy slice. Intuitively  this makes sense, as commutation between observables indicates they are causally disconnected -- something that is not true when the operators are separated by a null curve. 

With the equation above and the relationship between $\varphi $ and $\phi $ from the second line of \eqref{eq:v_scri+} we can work out the bracket of interest:
\begin{equation}
    \{  \mathbf{O}_\alpha (u) 
    , \, \dt  \varphi (t', x^a ) \} = \dt \alpha (x^a ) \, , \quad  u'\geq u \, .
\end{equation}
Importantly, $\mathbf{O}_\alpha (u)$ commutes with all other subregion observables for $u'\geq u$, by causality, since they are local to the subregion. Furthermore, it leaves $A^\dr_t\big|_\Gamma$ invariant and so preserves the soft boundary conditions, thereby generating a genuine frame reorientation symmetry of the subregion theory.

The importance of the above equation is that it shows that the symmetry generators of the global theory are only a subset of the symmetry generators of the subregion theory. This statement is equivalent to what was claimed earlier in this paper -- large (asymptotic) gauge transformations are not the only global transformations yielding reorientations of soft edge frames at finite distance (even though for every reorientation $\rho$ there is a corresponding large gauge transformation $\alpha$). In fact, we see that the operator $\mathbf{O} _{\alpha }(u)$ above has the effect of adding \emph{hard radiation} modes at the cut $S^2_u$ of $\scri^+$. This is an example of a change of initial data localized at spacelike separation from the subregion.

Interestingly, the addition of \emph{soft radiation} also generates valid frame reorientations. This is a highly delocalized ``perturbation'' corresponding to shifts in the soft mode $N\mapsto N+\alpha$, for $\alpha$ a function on the celestial sphere. We see from equation \eqref{eq:v_scri+} that this has the desired effect on $\varphi$. However, finding the corresponding charge generator turns out to be challenging. The difficulty stems from the corner piece of the global symplectic form in \eqref{eq:global and sub sympl}, which involves the combination $D^2N-\mathbf{Q}$, thereby making it not as straightforward to isolate a would-be conjugate variable to $N$ alone. We leave this as an interesting open question.

While the analysis here was done for a particularly nice type of soft edge mode, built along null lines, we expect the general structure to survive for any soft edge mode: namely that corner symmetries can always be generated by, among other ways, large-gauge transformations, addition of soft modes and addition of spacelike separated hard modes. The advantage of our choice of edge mode was simply that it made this structure particularly evident via equation \eqref{eq:v_scri+}.

These examples highlight one of the key physical insights gained through our relational framework for building subregional phase spaces: subregions in gauge theory may experience a certain \emph{symmetry generator enhancement} compared to the global theory. While the set of corner symmetries may be the same as the set of asymptotic symmetries, the set of generators in the global phase space whose action coincides with the action of the corner charges on the subregional phase space may include, in general, more operators than just the asymptotic large-gauge charges.\footnote{Here, the careful use of the word ``may'' serves to remind the reader that such statements are highly dependent on the choice of edge mode frame. In particular, the general conclusion here applies to soft edge modes, reaching the asymptotic boundary. If we were to dress to, say, charged matter in the complement, then it is not even true that large-gauge transformations generate corner symmetries, so the comparison between subregion and asymptotic symmetry structures loses significance, as the two are essentially independent of each other.} This discussion sheds considerable light on the question of \emph{how} to operationally generate corner symmetries for subregions in gauge theories and we anticipate its physical implications might gain even more relevance in the case of gravity. The physical interpretation of corner symmetries in gauge theories touches base with debates in the philosophy of physics community on the ``direct empirical significance'' of gauge symmetries, \cite{Gomes:2019otw,BradingBrown,GreavesWallace}. Our conclusion has resonance with \cite{Gomes:2019otw} (see also the ``gluing theorem'' in \cite{Gomes:2019xto}), in that subregion symmetries encode relational information with the complement and cannot be detected intrinsically. However, it differs in that our symmetries are \emph{not} related to (subgroups of) gauge symmetries, but rather correspond to physical transformations of complement/asymptotic data.

\section{Conclusions and Outlook}
\label{sec:conclusions}

In this work, we have explored the intricate connection between asymptotic and finite-boundary symmetries within gauge theories, unified under the concept of soft edges. Traditional approaches to gauge theories and boundary behaviour had long suggested a relationship between these two regimes, typically viewing soft modes as an asymptotic limit of edge modes. However,  through the lens of edge modes as dynamical reference frames, we revealed that finite-distance boundaries resonate with the asymptotia without requiring an infinite-distance or infinite-volume limit.

To clarify this interplay, we focused on classical Maxwell theory, analyzing finite subregions with timelike boundaries embedded in flat spacetime. This setup allowed us to combine an analysis of the symplectic flux, phase space structure, and boundary symmetries with equations of motion, shedding light on how boundary condition choices shape gauge theory structures. A key part of our analysis was the distinction between two classes of reference frames: intrinsic and extrinsic frames. Intrinsic frames are defined solely within the subregion, lacking reorientation symmetries and are thus unable to support boundary charges. Conversely, extrinsic frames extend beyond the subregion, connecting it with the global structure and generating boundary charges, embodying the concept of \textit{soft edges}. Our discussion also clarified certain debates in the literature, highlighting how different frame and boundary conditions choices impact interpretations of edge symmetries and modes. While some works, such as those by Donnelly and Freidel \cite{Donnelly:2016auv}, emphasized the presence of boundary symmetries in the extended phase space, others like \cite{Gomes:2018dxs,Riello:2021lfl,Riello:2020zbk} had questioned these symmetries, working implicitly within an intrinsic frame context where no boundary charges can arise. Our broader perspective based on \cite{Carrozza:2021gju} bridged these varied approaches, situating each within a framework where intrinsic and extrinsic frames produce distinct observables, reconciling previously divergent conclusions. 

Both types of frames were crucial in our discussion. The extrinsic frame provides a concrete mechanism for asymptotic symmetries to influence finite-distance physics by transmitting the effects of asymptotic soft modes and charges into finite regions via finite-distance analogues. Combined with the intrinsic one, it allowed us to obtain a distinction between ‘‘radiative'' (bulk) and ‘‘soft'' (corner) sectors within the phase space, enabling a quasi-local description that applies beyond the traditional asymptotic framework. Radiative degrees of freedom, associated with photon radiation, depend solely on the subregion’s interior, while the Goldstone soft sector, embodied by the gauge-invariant pair $\{\varphi , E_\perp\}$, represents a set of cross-boundary observables, linking the subregion with its exterior.
This gauge-invariant bulk-corner split, dictated by differences between intrinsic and extrinsic reference frames, encodes frame dependence and relationality within the Goldstone mode, while the intrinsic sector remains frame-independent.

This relational framing naturally led us to the selection of boundary conditions within the new  soft class in table \ref{table:sbc_types}, enabling a finite-region realization of the infinite-dimensional charge algebras typically observed at asymptotic boundaries. In contrast to traditional Dirichlet or Neumann boundary conditions, which constrain the soft sector by fixing one of the variables in the Goldstone pair, soft boundary conditions preserve boundary symmetries and their associated charges, allowing reorientations of frames that mirror asymptotic charges in finite regions. This is a key feature of each type of soft boundary conditions -- Neumann, Dirichlet and Robin. What differentiates them is their action on the remaining radiative data, namely the magnetic field on the boundary.

Our definition of the soft sector aligns with the usual splitting seen at asymptotic boundaries, though our description adds a relational, manifestly gauge-invariant layer to previous gauge-fixed approaches. Moreover, we showed that the structure of this Goldstone pair at finite distances mirrors aspects of the asymptotic infrared triangle. A key insight from our analysis is the classical relationship between two vertices (one edge) of this triangle at finite distances, providing a novel perspective on memory effects and associated symmetries. Our construction provides a quasi-local version of the memory effect in Maxwell theory, based on finite-distance, dynamical observables. Recoverable in the limit approaching traditional asymptotic formulations, this quasi-local memory effect is operationally more accessible and applicable to finite observers. At finite distances, we also observed a correspondence between memory and vacuum transitions akin to the asymptotic regime, mediated here by the postselection parameter $A^\dr_t\big|_\Gamma$ (eq.\ \eqref{eq:memory_Gm_finite}). Furthermore, a connection exists between quasi-local and global vacuum transitions, although it is frame-dependent and ultimately relies on complementary data (such as matter contributions), making it inaccessible through quasi-local measurements alone.

Our study of edge modes as dynamical reference frames underscores the importance of frame dependence in linking asymptotic and finite-distance physics.  We observed that extrinsic frames reveal distinct subsets of gauge-invariant observables, reinforcing a subsystem-relative character in gauge theories \cite{AliAhmad:2021adn,Hoehn:2023ehz,DeVuyst:2024pop,DeVuyst:2024uvd}. This relative dependence is especially apparent in how boundary conditions shape symmetry structures, giving a practical demonstration of subsystem relativity in gauge theory: the choice of reference frame fundamentally influences the observable quantities within the subregion.

Looking ahead, extending this framework into the quantum regime, involving the third vertex of the soft triangle, remains a promising direction for future research. While scattering at finite distance is not well understood, in the last years progress has been made in understanding quantum field theories for finite regions in terms of their associated algebras of observables. Studying the properties of these algebras for subregion gauge theories, described with different dynamical reference frames, is an exciting new avenue to explore. 
In recent work \cite{Araujo-Regado:2025ejs}, we have employed the structures developed here in the context of lattice gauge theories, while extending the construction in several new directions. Firstly, we provide a quantum version of the setup, while generalizing it to arbitrary non-Abelian compact Lie structure groups. Secondly, we provide a systematic study of subregional entanglement entropies, thereby shedding light on the edge mode contribution, \cite{Ball:2024hqe, Donnelly:2015hxa, Donnelly:2016auv}. Crucially, a relational definition of entanglement entropy (extending the one introduced in \cite{Hoehn:2023ehz} to the lattice setup) based on quantum reference frames yields a distinct one from standard constructions in gauge theory. These turn out to be frame-dependent, however, avoid the non-distillable part usually appearing in entanglement entropy constructions in gauge theories \cite{Casini:2013rba,Ghosh:2015iwa}. The distinction between extrinsic and intrinsic frames turns out to be of great importance in understanding the precise relationship between the so-called ‘‘electric/magnetic center'' algebras of \cite{Casini:2013rba,Ghosh:2015iwa, Delcamp:2016eya} and the respective relational algebras. 
In fact, as shown in \cite{DeVuyst:2024pop,DeVuyst:2024uvd}, the recent explorations of gravitational entropies in subregions \cite{Chandrasekaran:2022cip,Jensen:2023yxy,Kudler-Flam:2023qfl} can be understood in precisely this sense, likewise exhibiting a quantum frame dependence of the gravitational entropy of a local subregion.

Applying this approach to more complex theories, such as gravity and continuum non-Abelian gauge theories, presents other exciting prospects. In gravity, the ‘‘soft edges formalism'' could enable the definition of unambiguous symmetry algebras of diffeomorphisms for arbitrary subregions, including those bounded by horizons, with a clearer physical interpretation of the symmetry origin. This may deepen our understanding of black hole entropy, possibly clarifying aspects of the soft hair hypothesis \cite{Hawking:2016sgy, Haco:2018ske, Haco:2019ggi} and advancing our grasp on black hole microstates. We point the reader to the recent work \cite{Pulakkat:2025eid}, in which a proposal for an intrinsic subregional phase-space in classical gravity is provided. Also, see \cite{Ball:2024gti} for recent progress on subregional phase-spaces in Yang-Mills theory.

Together, these findings endorse the perspective that asymptotic symmetries, soft modes, and memory effects naturally extend to finite-distance settings, framed by the relational and quasi-local structures of gauge theories. Our work offers new insights into the ‘‘corner program'', opening avenues for finite-region formulations in gauge theories and gravity. This holds promise for quantum gravity and quantum information theory, particularly in studying entanglement, holography, and the foundational understanding of subsystems and boundaries in field theory and gravity.

\section*{Acknowledgements}

\noindent We would like to thank Stefan Eccles, Henrique Gomes, Temple He, Alok Laddha, Aldo Riello and Jesse Woods for helpful discussions. PH would further like to thank the Fields and Strings Laboratory led by Jo\~{a}o Penedones at EPFL Lausanne for hospitality during stages of this work. This work was supported by funding from Okinawa Institute of Science and Technology Graduate University and also made possible through the support of the ID\# 62312 grant from the John Templeton Foundation, as part of the \href{https://www.templeton.org/grant/the-quantum-information-structure-of-spacetime-qiss-second-phase}{\textit{`The Quantum Information Structure of Spacetime'} Project (QISS)}.~The opinions expressed in this project are those of the authors and do not necessarily reflect the views of the John Templeton Foundation. The latex layout of the paper is due to Josh Kirklin and we thank him for permitting us to use it.

\appendix

\section{Boundary Lagrangian for Soft Boundary Conditions}
\label{app:post_DBC}
In this appendix, we provide a short illustration of the boundary Lagrangian that realizes the soft boundary conditions, following the algorithm presented in \cite{Carrozza:2021gju}. 

We start by outlining the philosophy by which we fix the boundary behaviour of the dynamical fields. We are given the action with a local and covariant boundary term $\ell$:
\begin{equation}
\label{eq:action_post}
    S = \int _{\cM} \, \star F \wedge F  + \int _\Gamma  \ell\,.
\end{equation}
The variation of the boundary component will contribute to boundary equations of motion, whose solutions include the desired boundary conditions (i.e.\ postselection) for the boundary dressed fields  $\star F |_\Gamma$ and $A^{\ra} |_\Gamma$. This is different from the viewpoint where boundary conditions are imposed \textit{a-priori} on field variations in $\delta S$, rather than emerging as equations of motion, leading to a different $\ell$.

We start by decomposing the dressed field $A^\ra|_\Gamma:= \Tilde A ^\dr +\dt \varphi$, as described in section \ref{subsec:framechange}, and furthermore separate the time and spacelike components:
\be
A^\ra\big|_\Gamma:= A^{(0)} +\Tilde A ^\dr + {\dt_{\p \Sigma} \varphi}  \,\q \text{with}\;\; A^{(0)}:=A_t^\dr \dt t\,,\;\;  \Tilde A^\dr :=\Tilde A^\dr_a \dt x^a\,,\;\; {\dt_{\p \Sigma} \varphi} = \p_a \varphi \dt x^a\,.
\ee
We will also introduce a symplectomorphism on the photon phase space $\{ h |_\Gamma, \p _r h |_\Gamma \}$. Providing the boundary Lagrangian for the soft-Robin boundary conditions, we are also collectively describing the Dirichlet and Neumann cases, taking the opportune limits. For this, we recall that both $\Tilde A^\dr_a$ and $F^{ra}$ admit a unique Hodge decomposition on each constant $r$ and $t$ slice in the subregion:
\begin{equation}
    \Tilde{A}^\dr _a  = \l \dt \varrho   + \star_{\p \Sigma} \dt h  \r _a  , \quad g _{ab} F^{rb} = \l \dt \mathfrak{p}   + \star_{\p \Sigma} \dt \mathfrak{h}  \r _a \,.
\end{equation}
Let us define the quantities:
\be
B= \p_a (\alpha h  +  \beta   \mathfrak{h}) \dt x^a \big| _\Gamma \, ,\q H= \sqrt{|g|} \p_a (\gamma  h +  \zeta  \mathfrak{h} ) \dt x^a \wedge \dt t \big| _\Gamma  \,,
\ee
with $\alpha \zeta - \beta \gamma =1$ and $g_\Gamma$ the metric induced on the boundary. With this definition, we ensure that
\be
\delta B\wedge \delta H = \delta \Tilde A ^\dr \wedge \delta \star F \big|_\Gamma\,.
\ee
Following the algorithm described in \cite{Carrozza:2021gju} to build the postselected theory, we get
\be
\ell_\text{soft} = (A^{(0)} -X^{(0)})\wedge \star F +{\dt_{\p \Sigma} \varphi} \wedge \mathcal{E} - B_0\wedge H +\f{1}{2}\l \Tilde A\wedge\star F +B\wedge H \r\,,
\ee
where
\be
X^{(0)} = X_0\, \dt t\,,\q \mathcal{E} = \sqrt{|g _\Gamma|} \p_a \varepsilon\, \dt x^a \wedge \dt t   \q \text{and}\q B_0 =  b_a\, \dt x^a
\ee
are fixed background structures. Inserting this into the action \eqref{eq:action_post}, after a few manipulations, the corresponding variational principle will give
\bsub\be
\delta S =&\delta\left[S_0 + \int_\Gamma \ell_\text{soft} \right]\\
=&- \int_\cM \delta A \wedge \dt \star F +\int_{\Sigma_1-\Sigma_2} \epsilon_\Sigma F^{it} \delta A_i  \\&+\int_\Gamma \dt (\delta \rf E_\perp) +\delta \rf\, \dt \star F +(A^{(0)} -X^{(0)})\wedge \delta \star F - (B -B_0)\wedge \delta H-\delta {\dt_{\p \Sigma} \varphi} \wedge(\p_t E_\perp -\cE)\,,\notag
\ee\esub
with the perpendicular electric field to the boundary $E_\perp :=\epsilon_{\p\Sigma} m_\mu F^{\mu t}$ and $\epsilon_{\p\Sigma}$ the volume form on the corner. 
We require stationarity up to terms localised on the initial and final Cauchy slices. This happens if \emph{both}:
\begin{itemize}
    \item the boundary equations of motion are satisfied: 
    \[\dt \star F|_\Gamma\approx0\,,\q \p _t \l \sqrt{g} F^{rt} \r  = \sqrt{g} \Tilde{\nabla}^2 \varepsilon\,;\]
    \item the boundary conditions are satisfied:
    \[A^\dr_t \,\hat\approx\,X_0\,,\q \p_a(\alpha h + \beta \mathfrak{h} )\,\hat\approx\, b_a\,.\]
\end{itemize}
The corner term, containing the normal electric field and the edge mode $\Phi$, should not be seen as living on the timelike boundary, but instead on the corner of the Cauchy slice.
Although this seems just a matter of names, as  $\p \Gamma = \p\Sigma_1 -\p\Sigma_2$, it implies that we should look at the corner contribution as part of the phase space (initial conditions) and not boundary data. This is consistent with the symplectic form presented in the main body of the paper.

\section{Lorenz Gauge and other Examples of Immaterial Frames}
\label{app:examples}

As announced in the main body of the paper, Wilson lines are not the only tool to realize immaterial frames. Specifically, for some choices of the condition $\cG[A^\dr]=0$, the corresponding dressing is not achievable with simple Wilson lines. An example of this is given by the common choice 
\be
\label{eq:Lorenz}
\nabla^\mu A_\mu^\dr =0\,,  
\ee
also referred to as \textit{Lorenz gauge} when used to gauge-fix the undressed $A_\mu$. This is known not to be a complete gauge-fixing and is usually accompanied by another condition. For example, a traditional extra choice in the edge mode literature \cite{Riello:2021lfl,Ball:2024hqe,Ball:2024xhf,Harlow:2019yfa} and in canonical electromagnetism is $A_t^\dr =0$, in conjunction yielding the \textit{Coulomb gauge} (when applied to the bare connection). Another option,  aligning with choices made in the celestial holography literature \cite{Raclariu:2021zjz,Pasterski:2017kqt,Pasterski:2016qvg,Pasterski:2021rjz,Strominger:2017zoo} is $r A_r^\dr +t A_t^\dr =0$. In both cases, the Wilson line dressing, used to set one component of $A^\dr$ to zero, comes along with a second dressing, realising the Lorenz condition \eqref{eq:Lorenz}. Let us study this in detail for the \textit{holographic gauge}, due to its relevance in celestial holography. It is characterised by the following conditions:
\be
r A^{{\text{hol}}}_r +  t A^{{\text{hol}}}_t  = 0\,,\q\q \nabla^\mu A^{{\text{hol}}}_\mu =0\,.
\ee
The first condition is achieved through an intermediate dressing with Wilson lines on which $r \propto t$ (see figure \ref{fig:Milne} for a schematic representation). Let us define
\be
\mathfrak{A}^{\text{hol}} =A - \int_{\gamma_{\text{hol}}} A\,,
\ee
where the $\gamma_{\text{hol}}$'s are the lines drawn in figure \eqref{fig:Milne}. We have, by construction,
\be
r \mathfrak{A}^{{\text{hol}}}_r +  t \mathfrak{A}^{{\text{hol}}}_t  = 0\,.
\ee

\begin{minipage}{.95\textwidth}
    \centering
    \includegraphics[width=0.2\linewidth]{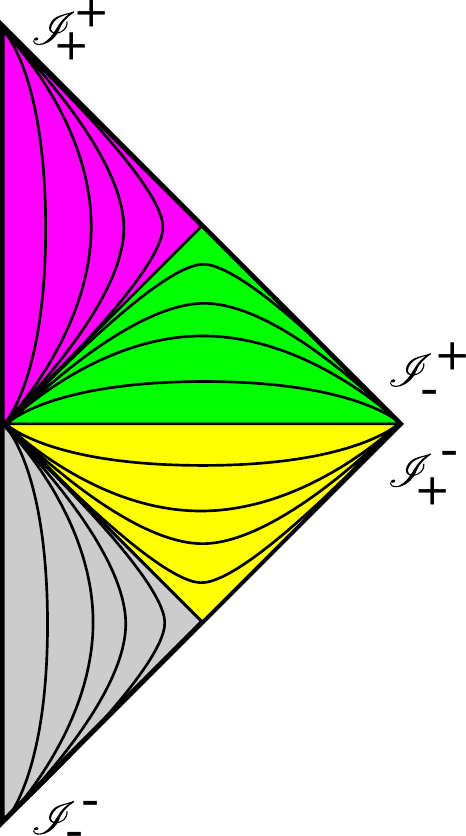}
    \captionof{figure}{\footnotesize Schematic representation of the \textit{holographic} gauge Wilson lines. Along each line we have $u=\kappa r$, and, depending on the value of $\kappa$, the anchor point is different. We can show that for $\kappa>0$, the lines go to $\scri^+_+$ (pink region); for $0>\kappa>-1$, they go to $\scri^+_-$ (green region); for $-1>\kappa>-2$, they go to $\scri^-_+$ (yellow) and when $\kappa<-2$ to $\scri^-_-$ (gray).}
    \label{fig:Milne}
\end{minipage}\\

The second condition is implemented as follows. Let us start by considering the scalar $\rf^{\text{hol}}_0$, which is a solution to:
\be\label{eq:phi_0}
\Box \rf^{\text{hol}}_0 =\nabla^\mu \mathfrak{A}^{\text{hol}}_\mu\,,\q r \p_r \rf^{\text{hol}}_0 +  t \p_t \rf^{\text{hol}}_0  = 0\,.
\ee
To solve these equations uniquely, we need to specify a suitable set of boundary conditions, on a codimension 2 surface. To see this, we insert the second equation into the first one. This maps a four-dimensional d'Alembertian to a three-dimensional Laplacian on hypersurfaces orthogonal to the Wilson lines. The Laplacian equation is solved on any of these hypersurfaces with boundary conditions on the corresponding codimension-2 boundary. An explicit complete base for the solutions of the homogeneous version of equation \eqref{eq:phi_0} is provided by the Goldstone conformal primaries \cite{Raclariu:2021zjz,Pasterski:2017kqt,Pasterski:2016qvg,Pasterski:2021rjz}. The main takeaway message is that giving the value of $\rf^{\text{hol}}_0$ on a codimension-2 surface, gives a unique solution everywhere for each configuration of the connection $A$. The frame phase ($\rf^{\text{hol}}$) is finally defined as:
\be
\rf^{\text{hol}} :=\rf^{\text{hol}}_0+ \int_{\gamma_{\text{hol}}} A \,.
\ee 

A crucial point to emphasize here, related to the equivalence between frame reorientations and large gauge transformations, mirrors the distinction between active and passive transformations in the context of diffeomorphisms.  For the example in section \ref{subsec:fr_reor}, a frame reorientation involves keeping the field configuration fixed in the complement while altering the convention for constructing $\rf$, by changing the initial value of the Wilson line. In contrast, large gauge transformations represent the active version, where we maintain the same convention (e.g.\ $\rf^A|_{i^0}=0$), but we modify the value of the fields $A$, near the attach point of the line. We argued in the main body of the paper that these two approaches are equivalent from the perspective of the subregion.

For the example here, however, we attach to different points depending on the ratio between radial and time coordinates  (see figure \ref{fig:Milne}). For instance, if $0<t<r_s$ the lines start on $\scri^+_-$, if $0>t>-r_s$ they are attached to $\scri^-_+$ and when $t=0$, they coincide with the radial ones used in Sections \ref{sec:BC} and \ref{sec:Soft_bc}. Therefore, the reorientation is associated with the value of the large gauge transformation at two different cuts of the global boundary. Nonetheless, due to the global boundary conditions, we have $\p_t \alpha=0|_{i^0}$,  indicating that the cut's location is irrelevant. A more subtle scenario arises when $t>r_{\Gamma}$ or $t>-r_{\Gamma}$, where the lines are attached to $\scri^+_+$ or $\scri^-_-$. Thanks to the Lorenz condition $\nabla^\mu A_\mu^{\text{hol}}=0$, the value of $\alpha$ can be pulled back from the future (past) of $\scri^+$ ($\scri^-$)to its past (future), meaning that, once again, the reorientation is equivalent to a large gauge transformation, determined on a single cut of the global boundary. 

A more effective way to understand this relationship is through the active version. Let us consider a general large gauge transformation with a certain profile $\alpha|_{i^0}$, such that $\p_t \alpha|_{i^0}=0$, consistent with the global boundary conditions. We leave unaffected the value of the bare $A$ at a finite distance, i.e.\ $\alpha|_\Gamma=0$, as per the definition of frame reorientations in \cite{Carrozza:2021gju}. Using the frame construction of this appendix, with e.g. the convention $\rf^{\text{hol}}|_{\scri^+_-}=0$, we first evaluate the change of $\mathfrak{A}$, which transforms as
\be
{\mathfrak{A}'}^{\text{hol}}(x)=\mathfrak{A}^{\text{hol}}(x) +\dt\left (\alpha(x_0)\right)\,, 
\ee
where $x_0$ is the anchor point for the line ending at $x\in \Gamma$. This means that the new orientation $ {\rf'}^{\text{hol}}$ is determined after solving
\be
{\rf'}_0^{\text{hol}}|_{\scri^+_-}=0\q r \p_r {\rf'}^{\text{hol}}_0 +  t \p_t {\rf'}^{\text{hol}}_0  = 0\,,\q \Box {\rf'}^{\text{hol}}_0 =\nabla^\mu {\mathfrak{A}'}^{\text{hol}}_\mu= \Box (\rf^{\text{hol}}_0 +\alpha(x_0))\,.
\ee
The corresponding solution is
\be
{\rf'}_0^{\text{hol}}(x) = \rf^{\text{hol}}_0(x) +\alpha(x_0) -\rho^{\text{hol}}(x)\,,
\ee
where $\rho^{\text{hol}}$ is the unique solution of the homogeneous equation
\be
\rho^{\text{hol}}-\alpha|_{\scri^+_-}=0 \q (r \p_r+  t \p_t)\rho = 0\,,\q \Box \rho^{\text{hol}}=0\,.
\ee
Notably, the solution to the dressing conditions from $\rho$ (the second and third equations) is determined by a single codimension 2 surface (the first equation). The difference between the two frames phases is precisely given by
\be
\rf^{\text{hol}}-\rf'^{{\text{hol}}} = \rf^{\text{hol}} -\rf'^{{\text{hol}}}_0-  \int_{\gamma_{\text{hol}}} A' =\rf_0^{\text{hol}} -\rf'^{\text{hol}}_0   +\alpha(x_0)= \rho^{\text{hol}}(x) \, . 
\ee
Unsurprisingly, the change in frame orientation exactly corresponds to a large gauge transformation that propagates into the bulk while respecting the dressing condition. The same construction applies to the Coulomb gauge, substituting the Wilson lines with timelike ones and the d'Alembertian with the Laplacian.

The argument connecting large gauge transformations and reorientations solely through Wilson lines may be somewhat misleading in the last two examples. The correct interpretation is that for every frame, the frame reorientation is equivalent to a large gauge transformation consistent with the dressing condition and determined by its asymptotic value. This is fully consistent with the equivalence between dressing and gauge-fixing. 

Here, we notice an immediate difference with the illustrative example used in the main body of the paper. The latter constrains the frame reorientation to be determined by a codimension-2 surface only after the imposition of the soft boundary condition, while the example presented here achieves this goal by construction. This has a consequence on boundary conditions because the dressing already enforces a common time profile for the Goldstone modes, hence already acting as a boundary condition. 

This distinction does not change the fundamental properties of the Goldstone mode or the conclusions drawn in the main body of the paper, but it would have added complexity to the presentation of the soft boundary conditions. For this reason, we used the radial Wilson lines as an example in the main body of the paper, despite the fact that the use of the Lorenz gauge would have significantly streamlined the discussion on the evolution of initial conditions in section \ref{subsec:in_dat_ev}. It is indeed well known that the condition \eqref{eq:Lorenz} leads to a decoupling of the equations of motion for each component of the field $A_a$ as $ \Box A^\dr_\mu =0$.

\section{Perfectly Conducting Boundary Conditions}
\label{app:BC_2}

In the main body of the paper (section \ref{sec:BC}), we argued that traditional choices of boundary conditions fail to generate a non-trivial edge sector in the phase space. In this appendix, we further examine the mechanism behind this limitation.

We recall that these two classes of boundary conditions sometimes referred to as \textit{perfeclty conducting} boils down to setting either all components of $\delta A^\dr|_\Gamma$ or $\delta \star F|_\Gamma$ to zero. The first case corresponds to the Dirichlet (or perfectly electric conducting) boundary conditions, where all components of $ A^{\ra}|_\Gamma$ are fixed, thereby fixing the tangential electric field and the perpendicular magnetic field (encoded in $F|_\Gamma)$. The term ‘‘perfectly electric conducting'' stems from the analogy with electric conductors, where the electric field is perpendicular to the surface, with no tangential component. Crucially, the condition we discuss here affects the \emph{dynamical} electric field, ensuring that its tangential component must vanish on $\Gamma$ to maintain the boundary conditions. Any non-zero value of the tangential electric field on $\Gamma$ should therefore be regarded as part of a fixed background structure. Complementarily, the Neumann case represents \textit{magnetic conducting} conditions, where the tangential magnetic field and the perpendicular electric field are fixed.

In \cite{Carrozza:2021gju}, an additional case known as \textit{Robin} boundary conditions is discussed. It represents a linear combination of Dirichlet and Neumann conditions. However, this approach does not introduce a significantly different class of boundary conditions, as it yields results similar to those of the Dirichlet case. 

In section \ref{sec:BC}, we showed that both Dirichlet and Neumann boundary conditions constrain the edge sector of the phase space, either by prohibiting reorientations or by freezing the associated charge. Since the symplectic form must be invertible, removing one degree of freedom from the phase space requires also removing its conjugate. This mechanism is quite intuitive for Neumann boundary conditions.  Specifically, these conditions freeze $ \star F |_{\p \Sigma }\sim E_\perp$, which is the generator of the frame reorientations, or in other words the charge aspect \eqref{eq:charge}. Since it becomes a constant on phase space, it cannot generate a non-trivial Hamiltonian flow which shifts the fields according to \eqref{eq:reor}. Equivalently, by removing $\star F|_{\partial\Sigma}$ from the phase space, to preserve the invertibility of $\Omega$, we must ``gauge-fix'' its symplectic conjugate, namely the edge frame $\Phi$, to a certain profile, thus eliminating the possibility of shifting it by $\Phi \mapsto \Phi - \rho$. 

For Dirichlet boundary conditions, although the mechanism initially seems different, where reorientations become \textit{meta-symmetries} \cite{Carrozza:2021gju}, the symplectic analysis leads to the same triviality in the edge sector. Fixing the Goldstone mode also makes the normal electric field a degenerate direction in phase space.

To further delve into this let us consider two different solutions of Maxwell's equation, in the radial dressing $A_r^\dr=0$, obtained by evolving the same initial data for the angular components of the connection and the strength tensor, but for the first one $a_t=0$ and $E_\perp=0$, and both non-vanishing for the second solution. By integrating \eqref{eq:config_evol} in time and using the relationship between $A_t^\dr$ and $F_{rt}$ (equation \eqref{eq:tilde fixed}), we get
\begin{align}
\label{eq:change_At&E}
    \left(A^{(1)}_a-A^{(2)}_a\right)|_{\Sigma(t)}&= \left( \int_0^t dt' \p_a \left[A^{(1)}_t(t')-A^{(2)}_t(t') \right] \right) \\
    &= \p_a \left( \int_0^t dt'\left [a_t(t',x^a)+ \int_{r_s}^r dr' \left(F^{(1)}_{rt}-F^{(2)}_{rt}\right)\right] \right) \\
    &=\p_a \left( \int_0^t dt'\left [a_t(t',x^a)+ (r-r_s) E_\perp|_{\p \Sigma(t')}\right] \right) \,.
\end{align}
To obtain the final result we have dropped the term that vanished because of the equivalence of the bulk initial conditions. We see that both $a_t$ and $E_\perp$ contribute as an exact piece to the radiative fields.

While under Dirichlet and dynamical soft boundary conditions, changing $a_t$ as in \eqref{eq:change_At&E}, acts as a meta-symmetry, mapping solutions between two different post-selected sectors, this is a time-dependent reorientation, hence a genuine gauge transformation for Neumann boundary condition. Conversely, $E_\perp$ in Dirichlet boundary conditions behaves like a superselection parameter, a physical variable, that is absent from the phase space. 

It is noteworthy that in gravitational contexts, reorientations never manifest as gauge symmetries \cite{Carrozza:2022xut}. The corresponding Neumann case, fixing the extrinsic curvature of the boundary, still retains some reorientations as physical symmetries. This is likely because gravitational waves cannot be fully shielded by boundary conditions, unlike electromagnetism, where the magnetic conducting boundary conditions achieve precisely this function. Similarly, we could argue for the presence of superselection for the Dirichlet case invoking the discontinuity of the normal electric field in idealized perfectly electric conducting devices, effectively separating this degree of freedom between the subregion and its complement.

\printbibliography

\end{document}